\pdfoutput=1
\documentclass[10pt,a4paper]{article}

\usepackage{amsmath,amsgen,latexsym}
\usepackage{amstext,amssymb,amsfonts,latexsym}
\usepackage{theorem}
\usepackage{pifont}
\usepackage{graphicx}

 \newcommand{\bs}{\bigskip}
 \newcommand{\ms}{\medskip}
 \newcommand{\n}{\noindent}
 \newcommand{\s}{\smallskip}
 \newcommand{\hs}[1]{\hspace*{ #1 mm}}
 \newcommand{\vs}[1]{\vspace*{ #1 mm}}

 \newcommand{\setempty}{\varnothing}
 
 \newcommand{\nat}{\mathbb{N}}
 
 \newcommand{\integer}{\mathbb{Z}}

 \newcommand{\co}{\mathrm{co}\mbox{-}}

 \newcommand{\CC}{{\cal C}}
 \newcommand{\FF}{{\cal F}}

 \newcommand{\KK}{{\cal K}}
 \newcommand{\LL}{{\cal L}}
 \newcommand{\NN}{{\cal N}}
 
 \newcommand{\MM}{{\cal M}}

 \newcommand{\PP}{{\cal P}}

 \newcommand{\dl}{\mathrm{L}}
 \newcommand{\nl}{\mathrm{NL}}
 \newcommand{\p}{\mathrm{P}}
 \newcommand{\np}{\mathrm{NP}}

 \newcommand{\poly}{\mathrm{poly}}

 \newcommand{\dcfl}{\mathrm{DCFL}}

\theoremstyle{plain}
\theoremheaderfont{\bfseries}
 \newtheorem{theorem}{Theorem}[section]
 \newtheorem{lemma}[theorem]{Lemma}
 \newtheorem{proposition}[theorem]{{\bf Proposition}}
 \newtheorem{corollary}[theorem]{Corollary}
 \newtheorem{fact}[theorem]{Fact.}

 {\theorembodyfont{\rmfamily}  \newtheorem{definition}[theorem]{Definition}}
 {\theorembodyfont{\rmfamily} }
 {\theorembodyfont{\rmfamily} }

 \newenvironment{proofof}[1]{\vspace*{5mm} \par \noindent
         {\bf Proof of #1.\hs{2}}}{\hfill$\Box$ \vspace*{3mm}}

 \newtheorem{yclaim}[theorem]{Claim}

 \newenvironment{proof}{\par \noindent
            {\bf Proof. \hs{2}}}{\hfill$\Box$ \vspace*{3mm}}

 \newcommand{\ceilings}[1]{\lceil #1 \rceil}
 \newcommand{\floors}[1]{\lfloor #1 \rfloor}
 \newcommand{\pair}[1]{\langle #1 \rangle}

\newcommand{\ignore}[1]{}

\newcommand{\cent}{|\!\! \mathrm{c}}
\newcommand{\dollar}{\$}

 \newcommand{\oned}{1\mathrm{D}}

 \newcommand{\onep}{1\mathrm{P}}
 \newcommand{\onebp}{1\mathrm{BP}}
 \newcommand{\onen}{1\mathrm{N}}
  \newcommand{\twou}{2\mathrm{U}}
 \newcommand{\twod}{2\mathrm{D}}
 
 \newcommand{\twon}{2\mathrm{N}}

 \newcommand{\twobp}{2\mathrm{BP}}

 \newcommand{\oneparity}{1\oplus}

 \newcommand{\oneu}{1\mathrm{U}}

 \newcommand{\twoupd}{2\mathrm{UPD}}
 \newcommand{\onedpd}{\mathrm{1DPD}}
 \newcommand{\onenpd}{\mathrm{1NPD}}
 \newcommand{\twodpd}{\mathrm{2DPD}}
 \newcommand{\twonpd}{\mathrm{2NPD}}
 \newcommand{\logdcfl}{\mathrm{LOGDCFL}}
 \newcommand{\logcfl}{\mathrm{LOGCFL}}

 \newcommand{\twof}{2\mathrm{F}}
 \newcommand{\twofct}{2\mathrm{FCT}}
 \newcommand{\lintwod}{\mathrm{lin2D}}

 \newcommand{\linonep}{\mathrm{lin1P}}
 \newcommand{\linoner}{\mathrm{lin1R}}

 \newcommand{\lintwodpd}{\mathrm{lin2DPD}}
 \newcommand{\lintwonpd}{\mathrm{lin2NPD}}

 \newcommand{\onedct}{\mathrm{1DCT}}
 \newcommand{\onenct}{\mathrm{1NCT}}
 \newcommand{\twodct}{\mathrm{2DCT}}
 \newcommand{\twonct}{\mathrm{2NCT}}
 \newcommand{\twouct}{\mathrm{2UCT}}

 \newcommand{\onedpdct}{\mathrm{1DPDCT}}
 \newcommand{\onenpdct}{\mathrm{1NPDCT}}
 \newcommand{\twodpdct}{\mathrm{2DPDCT}}
 \newcommand{\twonpdct}{\mathrm{2NPDCT}}

 \newcommand{\ptime}{\mathrm{ptime}\mbox{-}}
 \newcommand{\etime}{\mathrm{etime}\mbox{-}}
 \newcommand{\polylog}{\mathrm{polylog}}

 \newcommand{\oner}{1\mathrm{R}}

\newcommand{\nauxpdaspti}[2]{\mathrm{NAuxPDA},\!\mathrm{SPTI}( #1, #2 )}
\newcommand{\dauxpdaspti}[2]{\mathrm{DAuxPDA},\!\mathrm{SPTI}( #1, #2 )}

\begin{document}

\pagestyle{plain}
\pagenumbering{arabic}
\setcounter{page}{1}
\setcounter{footnote}{0}

\begin{center}
{\Large {\bf How Can Size, Ceiling, and Time Bounds Affect the \s\\ Complexity of Nonuniform Automata Families?}}\footnote{This is an extended and corrected version of a preliminary report  (with a slightly different title)  \cite{Yam26} appeared in the
Proceedings of the 16th International Workshop on Non-Classical Models of Automata and Applications (NCMA 2026), Rouen, France, June 29--30, 2026. Electronic Proceedings in Theoretical Computer Science (EPTCS) vol. 446, pp. 105--120, 2026.}
 \bs\ms\\

{\sc Tomoyuki Yamakami}\footnote{Affiliation: Faculty of Engineering, University of Fukui, 3-9-1 Bunkyo, Fukui 910-8507, Japan}
\bs\\

\end{center}

\sloppy
\begin{abstract}
In the past literature, families of two-way finite automata and pushdown automata having limited
resource-bounded complexity have been studied in direct connection to (mainstream)  space-bounded complexity classes equipped with Karp-Lipton style advice of limited size when all inputs given to the automata have bounded length.
We acknowledge three major factors---size, ceiling, and time---of such families, which have a significant impact on the computational/descriptional complexity of finite and pushdown automata families, where the ``size'' refers to the (stack-)state complexity, the ``ceiling'' refers to the maximum input length, and the ``time'' refers to the execution time of underlying automata.
In this line of study, we further explore these effects caused by different sizes and ceilings. We exploit automata-based reducibility and translatability to space-bounded advised Turing machines and auxiliary automata.

\s

\n{\bf Keywords.} nonuniform state complexity, promise problem,  stack-state complexity, size, ceiling, advice
\end{abstract}

\sloppy
\section{Nonuniform Automata Families: Old and New Issues}

Let us quickly review necessary background materials and overview the main contribution of this work.

\subsection{Background and Challenges of This Work}\label{sec:introduction}

In 1959, Rabin and Scott \cite{RS59} published a paper claiming that, for each fixed positive integer $n$,  each one-way nondeterministic finite automaton (or 1nfa, for short) $N_n$ of $n$ inner states can be converted into another computationally equivalent\footnote{Two machines are \emph{computationally equivalent} if their outcomes coincide on every input.}
one-way deterministic finite automaton (or 1dfa) $M_n$ of $2^{n}$ inner states.
The total number of inner states used to describe each finite automaton, known as the \emph{state complexity}, has served as a useful descriptional complexity measure for the ``size'' of the finite automaton, which expresses how large a machine's inner program is.
The result of Rabin and Scott can be rephrased in such a way that every 1nfa of size $n$ can be simulated on an appropriate 1dfa of size at most $2^n$.
This result has provided a measurable  complexity  comparison between two types of automata.
An increase of the state complexity by this simulation expresses the ``cost'' of converting one type of automaton to another type of computationally equivalent automaton. This notion has been used for  scaling the relative complexity of these two types of automata.
After the publication of their paper, researchers have been wondering if the value $2^n$ can be significantly reduced.

As for  the two-way versions of 1nfa and 1dfa, called 2nfa and 2dfa, Berman and Lingus \cite{BL77} reported that a simulation between an $n$-size  2nfa $N_n$ on an appropriate  2dfa $M_n$ of $n^{O(1)}$ inner states  has an intimate connection to the $\dl=?\nl$ question, where  $\dl$ is the deterministic log-space complexity class and $\nl$ is its nondeterministic variant.
Limiting input alphabet does not seem to help because the best known simulation of an $n$-size ``unary''  2nfa requires  at most $n^{O(\log{n})}$ inner states on an appropriate 2dfa \cite{GMP03}.

In 1978, Sakoda and Sipser \cite{SS78} studied a computational model of nonuniform families of finite automata $M_n$  in order to solve given families of promise (decision)  problems\footnote{A \emph{promise (decision) problem} is a pair of disjoint
sets over the same alphabet.} $(L_n^{(+)},L_n^{(-)})$,  indexed by natural numbers $n$.
At this point, we take a quick look at a simple, concrete example. Assuming a suitable binary encoding $\pair{G}$ of a graph $G$, let us consider the family $\LL_{bipartite} = \{(L_n^{(+)},L_n^{(-)})\}_{n\in\nat}$ of promise problems defined by $L_n^{(+)}=\{x\in\{0,1\}^*\mid |x|\leq 2^n, x=\pair{G}, \text{ $G$ is a bipartite graph}\}$ and $L_n^{(-)}=\{0,1\}^* - L_n^{(+)}$ for each index $n$.   This family $\LL_{bipartite}$ can be solved by a suitable family $\{N_n\}_{n\in\nat}$ of 2nfa's using $n^{O(1)}$ inner states.
After the paper of Sakoda and Sipser, a series of followup studies has been conducted intensively in the past literature
\cite{Gef12,Kap09,Kap12,Kap14,KP15,Yam18,Yam19a,Yam19b,Yam21,Yam23}.

Unlike single finite automata, the study of nonuniform automata families has demanded different kinds of insight.
These ``families'' of finite automata takes a similar advantage to (nonuniform)  Boolean circuit families; however, a major difference also lies in the point that, whereas  each circuit appearing in a Boolean circuit family is limited to particular input size, each finite automaton in an automata family is generally not, as in the previous example of $\{N_n\}_{n\in\nat}$ for $\LL_{bipartite}$.


The focal object of Sakoda and Sipser is in fact the nonuniform families of ``polynomial-size'' finite automata. Here, the ``size'' of a finite automaton signifies the descriptional complexity  of this automaton, more akin to the (work) space complexity measure of a Turing machine equipped with an external information source known as ``advice''.
This association contributes, as a hidden gem, to the study of automata theory by way of (mainstream) computational complexity theory.

Sakoda and Sipser  invented the now widely-used notations of $\twod$ and $\twon$ to treat collectively  promise problem families solvable by families of polynomial-size  2dfa's  and of polynomial-size  2nfa's, respectively.
The one-way analogues of $\twod$ and $\twon$ are denoted respectively by $\oned$ and $\onen$.
The aforementioned family $\LL_{bipartite}$ belongs to $\twon$ but it is not clear whether $\LL_{bipartite}$ falls in $\twod$. Sakoda and Sipser asked a fundamental question of whether those two nonuniform complexity classes coincide.

\begin{figure}[t]
\centering
\includegraphics*[height=5.0cm]{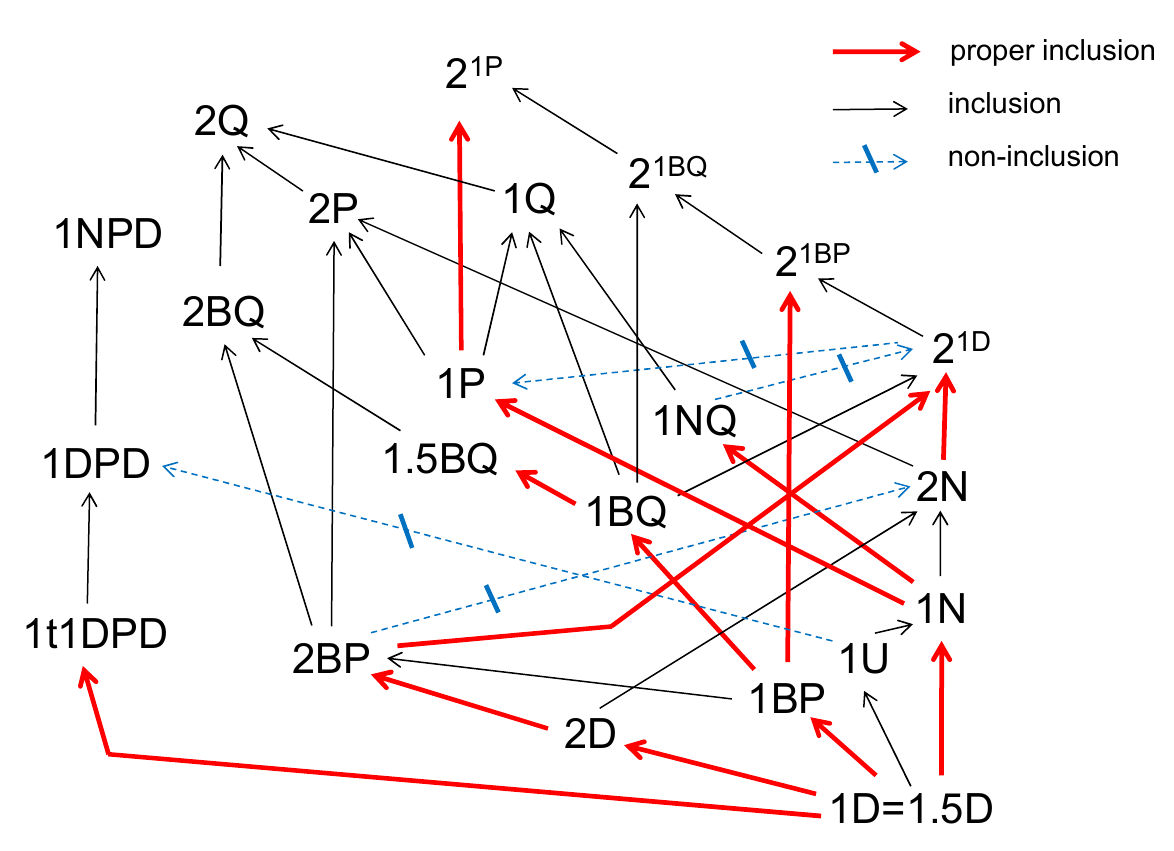}
\caption{Known inclusion relationships among basic nonuniform complexity classes in the past literature \cite{Kap09,Kap12,SS78,Yam19a,Yam22,Yam23,Yam24,Yam24b}. Red arrows, in particular, indicate the ``proper'' inclusion relationships.}\label{fig:figure-diagram}
\end{figure}

In recent literature, the study of nonuniform automata families has been significantly expanded to other types of automata, including unambiguous, co-nondeterministic, probabilistic, alternating, counting, and quantum automata \cite{Gef12,Kap09,Kap12,Yam19b}.
The nonuniform complexity classes $\onebp$ and $\onep$, for instance, stand for the classes of all families of promise problems solvable by families of polynomial-size one-way probabilistic finite automata with bounded-error and unbounded-error probabilities, respectively. Fig.~\ref{fig:figure-diagram} illustrates  some of the relationships among the complexity classes induced by those nonuniform automata families. Nonuniform counting complexity classes were further studied in \cite{Yam23}.


As a natural extension of finite automata families, Yamakami \cite{Yam21,Yam23,Yam24,Yam24b} studied the computational complexity of \emph{pushdown automata families}.
For pushdown automata families,  the ``size'' refers to its \emph{stack-state complexity}, which indicates the total number of inner states multiplied by the total number of ``pushable'' strings \cite{Yam21,Yam21b,Yam23}.
Hence,  for pushdown automata families, the term of ``polynomial-size'' refers to the polynomial stack-state complexity  in place of the polynomial state complexity. Similar to the previous notations of Sakoda and Sipser,
the notations of $\twodpd$ and $\twonpd$ (as well as $\onedpd$ and  $\onenpd$)  were introduced in \cite{Yam21,Yam23} with the use of polynomial-size pushdown automata families.
In the course of study on finite and pushdown automata families, a multi-counter extension  was further sought and studied intensively \cite{Yam24,Yam24b}.

From a different perspective, Kapoutsis \cite{Kap09,Kap12} analyzed the behaviors of  finite automata families of \emph{exponential (i.e.,  $2^{n^{O(1)}}$) size}.
To express families of promise problems solvable by exponential-size automata families, he introduced the special notations of $2^{\twod}$ and $2^{\twon}$ (resp., $2^{\oned}$ and $2^{\onen}$), compared to $\twod$ and $\twon$ (resp., $\oned$ and $\onen$), respectively.
With these notations, the aforementioned state complexity result of Rabin and Scott can be rephrased as $\onen\subseteq 2^{\oned}$.
Similarly to the $\twod=?\twon$ question of Sakoda and Sipser \cite{SS78},
it is not known whether $2^{\twod}=2^{\twon}$ holds or not.

As seen above, the ``size'' of underlying finite and pushdown automata families affects the overall performances of those machine families. In a sharp contrast, the second factor to control their behaviors is the length of input strings fed to  those underlying machines.
Therefore, it is quite beneficial to place an appropriate cap on the length of input strings fed into underlying finite and pushdown automata. Such a cap is known as a ``ceiling'' and has played a key role  in analyzing the overall behaviors of finite and pushdown automata families.
Kapoutsis \cite{Kap09,Kap12,Kap14} is the first to have focused on various ceilings.
Following the notational convention of Kapoutsis, we write $\twod/\poly$ and $\twon/\poly$ for the restrictions of $\twod$ and $\twon$ on machines of \emph{polynomial ceilings}, which indicates the maximum length of inputs being upper-bounded by a certain polynomial.  Such a restriction is also applicable to pushdown automata families. The notations of $\twodpd/\poly$ and $\twonpd/\poly$ induced respectively from $\twodpd$ and $\twonpd$ were studied in \cite{Yam21,Yam24,Yam24b}.


The third factor is the execution time-bounds, which have played an important role in studying low nonuniform complexity classes.
To emphasize such an execution time (or a runtime), the notation $\ptime\twodpd/\poly$ and $\ptime\twonpd/\poly$ were introduced in \cite{Yam24,Yam24b} respectively  from $\twodpd$ and $\twonpd$, where the prefix ``$\ptime$'' refers to ``polynomial time'' (in index $n$ and length $|x|$ of inputs $x$).

The detailed definitions of the aforementioned uniform complexity classes will be provided in Section \ref{sec:FL-FLpoly}.

\subsection{Importance of Size, Ceiling, and Time Bounds}\label{sec:size-ceiling-time}

Throughout the aforementioned studies, there are, in essence, three major
factors, which significantly affect the overall performance of each automata family:
(i) the \emph{size} of underlying machines, (ii) the \emph{ceiling} bounds of input strings given to underlying machines, and (iii) the \emph{runtime} of underlying machines.
These three  factors seem to directly relate to the complexity measures of work space, execution time, and advice size used in a standard model of Turing machines in (mainstream) computational complexity theory.  Now, we wonder how finite and pushdown automata families can be empowered by their machines'  sizes, ceilings, and execution times.
The purpose of this work is therefore to further explore how such factors contribute to the shift of the overall computational complexity of nonuniform automata families.

It was already observed that exponential-size machines behave quite differently from polynomial-size ones.
The power of exponential-size automata families were exhibited in the following few  examples: $2^{\oned}\nsubseteq \onep$, $\onep\subsetneqq 2^{\onep}$, $\mathrm{2BP}\subseteq 2^{\oned}$ \cite{Yam19a}, $\twon\subseteq 2^{\oned}$ \cite{Kap09,Kap12}, and $\mathrm{1t1DPD}\nsubseteq \onep$ \cite{Yam23}, where ``1t'' refers to ``one (stack) turn'' and $\onep$ (resp., $\twobp$) is an unbounded-error (resp., bounded-error) probabilistic variant of $\onen$ (resp., $\twon$). See Sections \ref{sec:pushdown-automata} and \ref{sec:counter} for their precise definitions.
Unfortunately, there are few studies on the case of ``superpolynomial-size'' automata families. In this work, we wish to focus on the study of superpolynomial-size automata families.

When inputs are all limited to having polynomial ceiling, the families of polynomial-size finite automata and pushdown automata, for instance, were shown to be related to the standard log-space complexity classes, $\dl$  and $\nl$, supplemented by polynomial-size advice \cite{BL77,Kap14,SS78}.
Other than polynomial ceiling bounds, Kapoutsis also considered promise problem families of \emph{superpolynomial ceilings}.   When ceiling bounds of polynomial-size automata families are expanded to exponentials, they in turn relate to loglog-space complexity classes supported by  polylog-size advice \cite{Kap14}.
In contrast, when ceiling is exponentially bounded, those finite automata families are
related to the $O(\log\log{n})$-space complexity classes, $\mathrm{LL}$ and $\mathrm{NLL}$, equipped with poly-logarithmic-size advice \cite{Kap14}.

The execution time significantly affects the computational power of two-way finite automata of large size and of two-way pushdown automata.
For instance,  although it is not yet known that $\twoupd = \twonpd$,  we can prove that $\ptime\twoupd/\poly = \ptime\twonpd/\poly$ with no hidden assumptions  \cite{Yam24,Yam24b},  where ``$\mathrm{U}$'' indicates the ``unambiguous'' computations of underlying finite automata in the promise problem setting.

Unfortunately, we still lack a whole picture of the effects of various combinations of size,  ceiling, and runtime for nonuniform families of finite and pushdown automata.
Therefore, the purpose of this work is solely dedicated to the clarification of the effects of size,  ceiling, and runtime by studying the mutual relationships among nonuniform classes of finite and pushdown automata families.

\subsection{Illustrations of Main Contributions}

In this work, we wish to answer the fundamental questions raised in Section \ref{sec:size-ceiling-time} concerning the effects of size, ceiling, and time bounds on the overall performances of finite and pushdown automata families. For this purpose,
we mostly concentrate on the combinations of the following types of automata families:
(i) size: polynomial and superpolynomial (including exponential),
(ii) ceiling: polynomial ($/\poly$) and exponential ($/\mathrm{exp}$), and
(iii) time: polynomial ($\ptime$), exponential ($\etime$), and the case of no time bound.
We point out that these factors, in particular, size and ceiling do not directly have one-to-one correspondence  to work space and advice size, as one expects.

In this work, we intend to
develop two technical tools used for proving various collapses and separations among complexity classes: (1) reducibility by two-way deterministic finite transducers (Section \ref{sec:reductions}) and (2) translations between automata families and  advised Turing machines and auxiliary automata (Section \ref{sec:translation}).  The first tool is usable in shifting our focal point from two-way machines to one-way machines. As for collapses/separations of nonuniform complexity classes, the second tool is described as
Propositions \ref{key-proposition}--\ref{second-key-proposition}, which play an essential role in translating finite and pushdown automata families to Turing machines and auxiliary automata, respectively.  Those two technical tools will be effectively used to prove most of the collapses/separations of nonuniform classes of promise problems solvable by one-way/two-way finite and pushdown automata families.
Fig.~\ref{fig:hierarchy-poly} illustrates such collapses and separations of two-way machines and
Fig.~\ref{fig:hierarchy-exp} shows corresponding relationships of classes of exponential ceilings for nonuniform complexity classes of polynomial ceilings.


It is important to remark that we use the conventional mechanism of random access for advice strings. Since advice strings of size $2^{n^{O(1)}}$ are too long for underlying polynomial-time Turing machines to read in a standard way (i.e., moving a tape head back and forth over the tape), we need to allow those machines to make a \emph{random access} to desired locations of the long input and advice strings (see Section \ref{sec:random-access}).

\paragraph{Organization of This Paper}

The rest of this work is organized as follows. The basic notions and various notations necessary for the later sections will be explained in Section \ref{sec:preparation}, including machine models of finite and pushdown automata families. Two key technical tools used for proving various containments/separations of nonuniform complexity classes will be explained in Section \ref{sec:tools}.
The effects of polynomial ceilings will be discussed in Section \ref{sec:poly-ceilings} and those of exponential ceilings will be studied in Section \ref{sec:exp-ceilings}.
The behaviors of polynomial- and superpolynomial-size automata families of polynomial ceilings will be examined separately in Sections \ref{sec:poly-size-automata} and \ref{sec:exp-size-automata} as well as Section \ref{sec:exp-ceilings}. Section \ref{sec:exp-ceilings} will treat polynomial- and exponential-size automata families of exponential ceilings.  
Throughout Sections \ref{sec:poly-ceilings}--\ref{sec:exp-ceilings}, our focal point is two-way finite and pushdown automata families. In Section \ref{sec:one-way-pushdown}, on the contrary, we will study the computational complexity of one-way finite and pushdown automata families.
In Section \ref{sec:discussion}, we will make a brief discussion and provide a pointer to the direction of future research.

\section{Foundations of This Work}\label{sec:preparation}

We will explain the basic notions and notation that the reader needs to read through the rest of this work.

\subsection{Sets, Numbers, and Alphabets}

The notation $\nat$ refers to the class of all \emph{natural numbers}, including $0$. For the set of  positive integers, we write $\nat^{+}$. Given  two integers $m$ and $n$ with $m\leq n$, the notation $[m,n]_{\integer}$ denotes the \emph{integer interval} composed of all integers between $m$ and $n$ (including $m$ and $n$) in comparison with the \emph{real interval}  $[r_1,r_2]$. We often abbreviate as $[n]$ the specific integer interval $[1,n]_{\integer}$ for each  number $n\in\nat^{+}$.
In this work, \emph{polynomials} take only nonnegative integer coefficients and \emph{logarithms} are all  taken to the base $2$. For convenience, we further set $\log{0}$ to be $0$. We use the notations $n^{O(1)}$ (resp., $\log^{O(1)}n$ and $2^{\log^{O(1)}n}$) for $O(n^k)$ (resp., $O(\log^kn)$ and $2^{O(\log^kn)}$) for an ``arbitrary'' integer $k\geq1$.
Given a set $Q$, $\PP(Q)$ denotes the \emph{power set} of $Q$, which consists of all subsets of $Q$.

An \emph{alphabet} is a nonempty finite set of ``symbols'' or ``letters''. Given such an alphabet $\Sigma$, a \emph{string over $\Sigma$} is a finite sequence of symbols in $\Sigma$. The \emph{length} of a string $x$, denoted $|x|$, is the total number of occurrences of any symbols in $x$.
The notation $\lambda$ denotes the \emph{empty string} of length $0$.
Let $\Sigma^*$ denote the collection of all strings over $\Sigma$.
For any number $n\in\nat$, the notation $\Sigma^{n}$ (resp., $\Sigma^{\leq n}$ and $\Sigma^{<n}$) denotes the subset of $\Sigma^*$ consisting only of all strings of length exactly $n$ (resp., at most $n$ and less than $n$).
Given a string $x$ of the from $x_1x_2\cdots x_n$ with $x_i\in\Sigma$ for all $i\in[n]$, the \emph{reverse} of $x$, denoted $x^R$, is the string $x_nx_{n-1}\cdots x_2x_1$.
For two languages $A$ and $B$, the \emph{concatenation} $A\cdot B$ (or more simply, $AB$)
denotes the language $\{xy\mid x\in A,y\in B\}$. In particular, when $B$ is a singleton $\{y\}$, we succinctly write $Ay$. Similarly, when $A=\{x\}$, we write $xB$.

In general, a \emph{promise (decision) problem} over an alphabet $\Sigma$ is a pair $(L^{(+)},L^{(-)})$ of disjoint subsets of $\Sigma^*$. In particular, whenever $L^{(+)}\cup L^{(-)}=\Sigma^*$, $L^{(+)}$ is simply called a \emph{language} over $\Sigma$ and $L^{(-)}$ is said to be the \emph{complement} of $L^{(+)}$.
Generally,  the \emph{complement} of $(L_n^{(+)},L_n^{(-)})$ is $(L_n^{(-)},L_n^{(+)})$
and denoted $\co(L_n^{(+)},L_n^{(-)})$.
For a family $\LL=\{(L_n^{(+)},L_n^{(+)})\}_{n\in\nat}$, the family of complements of $(L_n^{(+)},L_n^{(-)})$'s  is $\{(L_n^{(-)},L^{(+)})\}_{n\in\nat}$ and denoted $\co\LL$.
Furthermore, for a class $\CC$ of promise problem families, the notation $\co\CC$ denotes the set $\{\LL\mid \co\LL\in\CC\}$.  Any string in $L^{(+)}\cup L^{(-)}$ is succinctly called \emph{promised} or \emph{valid}.

A function $f$ on $\nat$ (i.e., from $\nat$ to $\nat$) is said to be \emph{invertible} if there exists a function from $Im(f)=\{f(n)\mid n\in\nat\}$ to $\nat$ such that $g(f(n))=n$ for all $n\in\nat$.  This function $g$ is called the \emph{inverse} of $f$ and denoted by $f^{-1}$.

A function $f$ on $\nat$ (i.e., from $\nat$ to $\nat$) is called \emph{polynomially bounded} (or p-bounded, for short) if there exists a polynomial $p$ such that $f(n)\leq p(n)$ holds for all $n\in\nat$. In contrast, $f$ is said to be \emph{polynomially honest} (or p-honest) if there is a polynomial $p$ satisfying  $n\leq p(f(n))$  for all $n\in\nat$. The composition $f\circ g$ of two functions $f$ and $g$ is defined
by $f\circ g(x) = f(g(x))$ for all inputs $x$.

A \emph{partial function} from $\Sigma^*$ to $\Gamma^*$ for two alphabets $\Sigma$ and $\Gamma$ is expressed in this work as a pair $(f,D)$ with  a set $D$ of \emph{valid} (or \emph{defined}) inputs $x$, where the outcome of $f$ on $x$, denoted $f(x)$, is defined \emph{at least} for all strings $x$ in $D$. It is important to remark that $D$ is merely a subset of $\{x\in\Sigma^*\mid \text{ $f(x)$ is undefined }\}$. This condition is essential in accordance with the definition of promise problems.

\subsection{Finite Automata, Pushdown Automata, and Finite Transducers}\label{sec:pushdown-automata}

As a foundation, this work uses the following well-known machine models.
A \emph{one-way deterministic finite automaton} (or a 1dfa, for short) $M$ is formally a septuple $(Q,\Sigma,\{\rhd,\lhd\},\delta,q_0,Q_{acc},Q_{rej})$, where $\delta$ is a transition function from $(Q-Q_{halt}) \times \check{\Sigma}$ to $Q$ with $Q_{halt} = Q_{acc}\cup Q_{rej}$ and $\check{\Sigma}=\Sigma\cup\{\rhd,\lhd\}$. Any inner state in $Q_{halt}$ is called a \emph{halting state}. When $M$ enters such an inner state, $M$ must halt.
Initially, an input string surrounded by two endmarkers ($\rhd$ and $\lhd$) is written on a read-only input tape. The value $|Q|$ is known as the \emph{state complexity} of $M$.
In contrast, a \emph{one-way nondeterministic finite automaton} (or 1nfa) uses a transition function mapping $(Q-Q_{halt}) \times\check{\Sigma}$ to $\PP(Q)$.

Every one-way finite automaton in this work must move its tape head to the right \emph{without making any $\lambda$-moves} (in accordance with the literature).
A \emph{computation path} of $M$ on input $x$ is described as a sequence of (surface) configurations of $M$ on $x$ and it is said to be \emph{accepting} (resp., \emph{rejecting}) if the path terminates in an accepting (resp., rejecting) state.
A nondeterministic machine is said to \emph{accept} (resp., \emph{reject}) an input if there exists at least one accepting computation path (resp., all computation paths are rejecting) on the input.
We say that a machine $M$ \emph{solves} a promise problem  $(L^{(+)},L^{(-)})$ if   $M$ accepts all inputs $x$ in $L^{(+)}$ and  $M$ rejects all inputs $x$ in $L^{(-)}$.
Notice that  there is no requirement for all inputs outside of $L^{(+)}\cup L^{(-)}$.

A \emph{one-way probabilistic finite automaton} (or a 1pfa) assigns certain ``probability'' to their nondeterministic choices made at every step. Here, we remark that 1pfa's are generally defined by stochastic matrices associated with symbols in $\Sigma$. See, e.g., \cite{Yam23} for those matrix-based definitions of 1pfa's.  We say that a 1pfa $M$ recognizes a language $L$ with \emph{bounded-error probability} (resp., \emph{unbounded-error probability}) if there exists a constant $\varepsilon\in[0,1/2)$ such that, for all strings $x\in L$, $M$ accepts $x$ with probability at least $1-\varepsilon$ (resp., more than $1/2$) and, for all $x\notin L$, $M$ rejects $x$ with probability at least $1-\varepsilon$ (resp., at least $1/2$).
Moreover, $M$ recognizes $L$ with \emph{one-sided-error probability} if, for all $x\in L$, $M$ accepts $x$ with probability more than $1/2$ and, for all $x\notin L$, $M$ rejects $x$ with probability $1$.


Given two alphabets $\Sigma$ and $\Theta$, a \emph{two-way deterministic finite transducer} (or a 2dft) $f$ is a 2dfa equipped with a write-once\footnote{A tape is said to be \emph{write once} if its tape head never moves to the left and, whenever the tape head writes non-blank symbol, it should move to the next blank cell.} output tape, where it writes at most one symbol in $\Theta$ on the output tape at every step. This output restriction is slightly different from that of its one-way version (or 1dft) given in \cite{Yam23}, where a 1dft is allowed to write a string in a single step.
More precisely, $f$ has the form $(Q,\Sigma,\{\rhd,\lhd\},\Theta,\delta,q_0,Q_{halt})$, where $Q_{halt}$ consists of halting states.
 A transition function $\delta$ is a map from $(Q-Q_{halt})\times \check{\Sigma}$ to $Q\times \Theta_{\lambda}\times D$, where $\Theta_{\lambda}=\Theta\cup\{\lambda\}$ and $D=\{-1,0,+1\}$ for the directions of an input tape head.

As a natural extension of 1nfa's, a \emph{one-way nondeterministic pushdown automaton} (or 1npda) $M$ is a tuple $(Q,\Sigma,\{\rhd,\lhd\},\Gamma, \delta,q_0,\bot, Q_{acc},Q_{rej})$ with a transition function $\delta: (Q-Q_{halt}) \times\check{\Sigma}_{\lambda}\times \Gamma\to \PP(Q\times\Gamma^*)$, where $\check{\Sigma}_{\lambda} = \check{\Sigma}\cup\{\lambda\}$ and  $\Gamma$ is a stack alphabet including the bottom marker $\bot$.
If $M$ further satisfies the so-called \emph{deterministic property}, which requires  $|\delta(q,\sigma,a)\cup \delta(q,\lambda,a)|\leq 1$ for all parameters $(q,\sigma,a)$'s,  then $M$ is called  a \emph{one-way deterministic pushdown automaton} (or a 1dpda).
Remember that every pushdown automaton is allowed to make its input-tape head stay still (known as a \emph{$\lambda$-move}).

The \emph{push size} $e$ of  a pushdown automaton $M$ is the maximum size of a ``pushable'' string taken at any single steps; more precisely, the maximum value $|w|$ in the set $\{|w|: (p,w)\in\delta(q,\sigma,a)\}$.  In comparison with the state complexity of a finite automaton, the \emph{stack-state complexity} of $M$ is defined to be $|Q||\Gamma^{\leq e}|$  \cite{Yam20,Yam21b}.


The \emph{stack content} refers to a consecutive sequence of symbols stored in the stack (including $\bot$), which forms a ``string'' over $\Gamma$. The \emph{stack height} is then the length of the stack content.
A \emph{(stack) turn}\footnote{In some literature, the term ``turn'' has been also used to indicate a change of the direction of an input/work tape head move. In this work, nevertheless, we do not discuss such a turn.} of a pushdown automaton $M$ is a series of actions of $M$ during the time when the ``mode'' of $M$'s  stack operation changes from increasing stack height to decreasing stack height.  A \emph{$k$-turn pushdown automaton} is a pushdown automaton that makes at most $k$ (stack) turns on each computation path on every input.


The two-way variants of the aforementioned one-way machines, which are abbreviated as 2dfa, 2nfa, 2dpda, and 2npda, allow their input-tape heads to move in all directions, including $\lambda$-moves.
Here, we remark on the runtime of two-way machines. Unlike one-way finite automata, two-way finite automata take only at most $|Q|(|x|+2)+1$ steps. See also, e.g., \cite{GMP03}.

Two machines $M$ and $N$ over the same input alphabet are \emph{computationally equivalent} if the outcomes (i.e., acceptance or rejection) of $M$ and $N$ agree on all inputs.

In Section \ref{sec:FL-FLpoly}, we further expand the aforementioned individual machines into families of machines.

\subsection{Nonuniform Families of Machines and Size, Ceiling, and Time Bounds}\label{sec:FL-FLpoly}

In what follows, we focus on a family $\LL=\{(L_n^{(+)},L_n^{(-)})\}_{n\in\nat}$  of promise (decision) problems over an alphabet, say, $\Sigma$. To solve such a promise problem family, we wish to use a family $\MM=\{M_n\}_{n\in\nat}$ of machines of the same type (that is, 1dfa's, 2nfa's, 2dpda, etc.).
For these two families $\LL$ and $\MM$, we say that $\MM$ \emph{solves} $\LL$ if, for each index $n\in\nat$, $M_n$ solves $(L_n^{(+)},L_n^{(-)})$.

As remarked in Section \ref{sec:introduction}, we particularly concentrate on the ``size'',  ``ceiling'', and ``time'' of finite and pushdown automata families throughout this work.

Given a family $\MM=\{M_n\}_{n\in\nat}$ of finite automata (resp., pushdown automata) $M_n$ with sets $Q_n$ of inner states (resp., $Q_n$ and stack alphabets $\Gamma_n$ with push sizes $e_n$), the \emph{state complexity} (resp., \emph{stack-state complexity}) of $\MM$ is generally a function on $\nat$ defined as $|Q_n|$ (resp., $|Q_n||\Gamma_n^{\leq e_n}|$) for any $n\in\nat$.

\begin{definition}\label{size-def-01}
Given a univariate function $f$ on  $\nat$, a family $\MM=\{M_n\}_{n\in\nat}$ of finite automata (resp., pushdown automata) with sets $Q_n$ of inner states (resp., $Q_n$ aa well as sets $\Gamma_n$ of stack alphabets with push sizes $e_n$) is said to have \emph{$f(n)$ size} if $f$ satisfies $|Q_n|\leq f(n)$ (resp., $|Q_n||\Gamma_n^{\leq e_n}|\leq f(n)$) for all $n\in\nat$.
\end{definition}

Conventionally, when $f$ is a polynomial, $\MM$ is said to have \emph{polynomial size}.  Sakoda and Sipser \cite{SS78} considered a family $\MM$ of polynomial-size finite automata.


The notation $\oned$ denotes the class of all families of promise problems solvable by families of polynomial-size 1dfa's. Similarly, the notation $\onen$ is defined for its nondeterministic variant. We also define $\onebp$ and $\onep$ by demanding all underlying 1pfa's to make \emph{bounded-error probability} and \emph{unbounded-error probability}, respectively.
When 1pfa's make  \emph{one-side-error probability}, we obtain the class  $\oner$.
With the use of pushdown automata families  instead, $\onedpd$ (resp., $\onenpd$) denotes  the class of all families of promise problems solvable by families of polynomial-size 1dpda's (resp., 1npda's).

In a similar fashion, we define $\twod$, $\twon$, $\twodpd$, and $\twonpd$ using 2dfa's, 2nfa's, 2dpda's, and 2npda's, respectively, as underlying machines.
When the behaviors of 1dpda's are limited to making only at most $k$ turns, we use the notation $k\mathrm{t1DPD}$ instead of $\onedpd$.

Our target in this work includes ``superpolynomial-size'' automata families. 
A family $\MM$ of finite/pushdown automata is said to have  \emph{exponential size}  (resp., \emph{linear exponential size}), the size bound is of the form $2^{p(n)}$ for a polynomial $p$ (resp., $2^{cn+d}$ for two constants $c,b>0$).
In  the case of  ``exponential-size'' families of finite automata, for instance,
the notations of $2^{\twod}$ and $2^{\twon}$ were invented by Kapoutsis \cite{Kap09,Kap12}.

To express ``linear exponential size'' families, we introduce the notion
$2^{\mathrm{lin}\twod}$ (resp., $2^{\mathrm{lin}\twon}$) by using
linear-exponential-size 2dfa's (resp., 2nfa's).
In a similar manner,  the notation $2^{\linoner}$ is introduced using one-sided-error 1pfa's of linear-exponential size.
Now, we wish to generalize these notations using arbitrary univariate functions $f$ on $\nat$.

\begin{definition}\label{size-def}
We consider a function $f$ of a ``simple form'', such as $f(n)=n$, $f(n)=\log^kn$, and $f(n)=2^{n^k}$. Now, we introduce the generic notation of $2^{f(\twod)}$ (resp., $2^{f(\twon)}$)
by simply taking families of 2dfa's (resp., 2nfa's) of $2^{O(f(n^{O(1)}))}$  size.
Moreover, we use the notation $2^{g(\mathrm{lin}\twod)}$ (resp., $2^{g(\mathrm{lin}\twon)}$) for $2^{O(g(n))}$-size 1dfa's (resp., 1nda's).
As for pushdown automata families, we also introduce the notation $2^{f(\twodpd)}$ (resp., $2^{f(\twonpd)}$) using families of $2^{O(f(n^{O(1)}))}$-size  2dpda's (resp., 2npda's) and $2^{g(\mathrm{lin}\twodpd)}$ (resp., $2^{g(\mathrm{lin}\twonpd)}$) using families of $2^{O(g(n))}$-size 2dpda's (resp., 2npda's).
To argue one-way automata, we  simply replace $\twod$, $\twodpd$, etc. in the above notations by $\oned$, $\onedpd$, etc., respectively.
\end{definition}

As more concrete examples of Definition \ref{size-def},
if we take functions $f(n)=n$ and $f(n)=\log^k{n}$ for $2^{f(\twod)}$, then we obtain the complexity classes $2^{\twod}$ and $2^{\log^k{\twod}}$, respectively.
By taking functions $g(n)=\sqrt[k]{n}$ and $g(n)=n^{\varepsilon}$ for $2^{g(\mathrm{lin}\twod)}$,
we obtain $2^{\sqrt[k]{\mathrm{lin}\twod}}$ and $2^{\mathrm{lin}\twod^{\varepsilon}}$ for each integer $k\geq2$ and a real number $\varepsilon\in[0,1)$.
Moreover, in the case of $g(n)=n^{O(1)}$, because $2^{\log{g(n)}}=g(n)$,  the special notation $2^{\log{g(\mathrm{lin}\twod)}}$ (resp., $2^{\log{g(\mathrm{lin}\twodpd)}}$, etc.) expresses $\twod$ (resp., $\twodpd$, etc.).
We ``loosely'' expand the above notations to functions of more general form. 
For instance, if we take a function $f(n)=3\log^2n+4$, then we understand the notation $2^{f(\twod)}$ as $2^{\log^2\twod}$ by redefining $f$ as $f(n^k)=O(\log^2n)$.

In the study of nonuniform families of promise problems,  another important concept is a  \emph{ceiling} which upper-bounds the length of input strings given to each promise problem.

\begin{definition}
Let $f$ denote any function on $\nat$.
A family $\LL=\{(L_n^{(+)},L_n^{(-)})\}_{n\in\nat}$ of promise problems over an alphabet $\Sigma$ is said to have \emph{$f(n)$ ceiling} if, for all numbers $n\in\nat$, $f(n)$ is an upper bound of the length of valid inputs, namely,  $L_n^{(+)}\cup L_n^{(-)}\subseteq \Sigma^{\leq f(n)}$. In particular, we say that $\LL$ has \emph{polynomial ceiling} (resp., \emph{exponential ceiling}) if there exists a polynomial $p$ such that  $L_n^{(+)}\cup L_n^{(-)}\subseteq \Sigma^{\leq p(n)}$ (resp., $L_n^{(+)}\cup L_n^{(-)}\subseteq \Sigma^{\leq 2^{p(n)}}$) for all numbers $n\in\nat$.
\end{definition}

Kapoutsis \cite{Kap09,Kap12} again invented the notation $\twod/\poly$ (resp., $\twod/\mathrm{exp}$) for the restriction of $\twod$ onto families of promise problems having polynomial (resp., exponential) ceilings. Similarly, we can introduce the notations, such as $\onedpd/\poly$ and $\twonpd/\mathrm{exp}$.
For the sake of later convenience,  $\mathrm{ALL}/\poly$ (resp., $\mathrm{ALL}/\mathrm{exp}$) expresses the collection of all families of promise problems with polynomial (resp., exponential) ceilings.


The ``size'' and ``ceiling'' bound each underlying automata family from a physical appearance viewpoint. Another important resource bound of a two-way (pushdown) automaton is ``execution time'', which limits the ``performance'' of such automaton. In the setting of automata families, such time bounds  were discussed in \cite{Yam21,Yam22,Yam24,Yam24b}.

\begin{definition}
Let $t$ denote any  function from $\nat^2$ to $\nat$.
A family $\MM=\{M_n\}_{n\in\nat}$ of 2nfa's (resp., 2npda's, 2dfa's, 2dpda's) working over an input alphabet $\Sigma$ is said to \emph{run in $t(n,|x|)$ time} if, for any $n\in\nat$ and $x\in\Sigma^*$, the runtime of $M_n$ on the input $x$ is at most $t(n,|x|)$.
In particular,
when those machines run in \emph{polynomial time} (resp., \emph{exponential time}) in both $n$ and $|x|$, we use the special prefix of ``ptime-'' (resp., ``etime-'') as in, e.g., $\mathrm{ptime\mbox{-}}2^{\twod}$ (resp., $\etime2^{\log^k{\twodpd}}$).
\end{definition}

It is important to remark that a smaller time bound of an automaton may prevent it from using all inner states on each computation path of it on every input. For instance, a finite automaton with exponentially many inner states running in polynomial time cannot access more than  polynomially many inner states on each computation path.

Remember that the runtime bound of each $M_n$ in a family $\MM$ depends not only $|x|$ but also $n$.
It is important to remark that, as shown in \cite{GMP03}, underlying 2dfa's and 2nfa's can be all treated to have  ``polynomial'' runtime.
By sharp contrast, various runtime restrictions of 2dpda's and 2npda's are crucial for the computational complexity of solvable promise problem families.

The class $\mathrm{1F}$ was introduced in \cite{Yam23} as the collection of all nonuniform families of partial functions computable by families of polynomial-size 1dft's.
This work further introduces its two-way version, which we call $\twof$.
More precisely, a family $\{(f_n,D_n)\}_{n\in\nat}$ of partial functions from $\Sigma^*$ to $\Gamma^*$ for two alphabets $\Sigma$ and $\Gamma$ is in $\twof$ if there exist a polynomial $p$ and a family $\{M_n\}_{n\in\nat}$ of polynomial-size 2dft's such that, for any index $n\in\nat$ and any string $x\in D_n$, $M_n$ takes $\rhd\!{x}\!\lhd$ on its input tape and produces $\rhd\!{f_n(x)}$ on its write-once output tape in time $p(n,|x|)$ until $M_n$ finally enters a halting (inner) state.
As before, we call these 2dft's by ``polynomial-time 2dft's.''
This runtime bound naturally  forces the length of output strings to be polynomially bounded.

\subsection{Counter Automata and Counter Pushdown Automata}\label{sec:counter}

Since finite (and pushdown) automata have no work tapes, it is useful to append multiple ``counters'' to keep track of a certain types of information as internal data.

In the setting of nonuniform (stack-)state complexity, the power of multiple counters was discussed in \cite{Yam24,Yam24b}.
When a stack uses its stack alphabet $\Gamma$ consisting only of a single non-bottom symbol, say, ``$1$'' (i.e., $\Gamma=\{1,\bot\}$), it is specifically called a  \emph{counter}. A \emph{$k$-counter automaton} (resp., a \emph{$k$-counter pushdown automaton}) is a finite automaton (resp., a pushdown automaton) equipped with $k$ counters. See, e.g., \cite{Yam23b,Yam24,Yam24b} for their basic behaviors.

Consider the situation at which $k$ counters are given to an underlying machine family for an arbitrary number $k\geq1$.
We use the notations of $\onedct_k$, $\onenct_k$, $\onedpdct_k$, and $\onenpdct_k$, where the suffix ``$\mathrm{CT}_k$'' refers to the use of $k$ counters.
In the case of two-way head moves, we replace the prefix ``$1$'' by ``$2$'', such as $\twodct_k$, $\ptime\twofct_k$, $\twodpdct_k$, etc. In particular, when $k$ equals $1$, we tend to drop the subscript ``$k$''.

In this work, we further provide deterministic finite transducers with $k$ counters and introduce the complexity classes $\twofct_k$ and $\ptime\twofct_k$.


\begin{fact}{\rm \cite{Yam24,Yam24b}}\label{counter-collapse}
(1) For any $k\geq4$, $\twodct_k=\twodct_4$.
(2) For any $k\geq3$, $\twodpdct_k=\twodpdct_3$.
(3) $\twodct_4/\poly = \twod/\poly$ and $\twodpdct_3/\poly = \twodpd/\poly$. The nondeterministic cases also hold.
\end{fact}

\begin{fact}{\rm \cite{Yam24,Yam24b}}\label{fact:counter}
$\twod\neq \twodct$, $\twou\neq \twouct$, and $\twon\neq \twonct$.
\end{fact}

From Fact \ref{fact:counter}, we can claim the following separation because the equality $\twof= \twofct$ instantly implies $\twod=\twodct$ by taking $\{0,1\}$-valued partial function families.

\begin{corollary}
$\twof\neq \twofct$.
\end{corollary}

\subsection{Turing Machines and Auxiliary Pushdown Automata with Advice}\label{sec:Turing-machine}

Throughout this work, we use the computation models of \emph{Turing machines} (or TMs) as well as Cook's \cite{Coo71} model of \emph{auxiliary pushdown automata} (or aux-pda's). We briefly explain these machines.

A \emph{nondeterministic Turing machine} (or an NTM, for short) is of the form $(Q,\Sigma,\{\rhd,\lhd\},\Gamma,\delta,q_0,Q_{acc},Q_{rej})$ with a transition function $\delta:(Q-Q_{halt})\times \check{\Sigma}\times \Gamma \to \PP(Q\times \Gamma\times D_1\times D_2)$, where $D_1=D_2=\{-1,0,+1\}$. Note that $-1$, $0$, and $+1$ respectively mean ``move to the left'', ``stay still'', and ``move to the right''. In accordance with automata models, in this work, we assume that an NTM has a read-only input tape and a rewritable work tape.
An input string surrounded by two endmarkers $\rhd$ and $\lhd$ is
initially written on this input tape. All tape cells are indexed by natural numbers, starting with the leftmost cell that contains $\rhd$.
The \emph{time complexity} $t(n)$ means the maximum number of steps taken by the NTM  on any single computation paths on any inputs of length $n$.
In contrast, the \emph{space complexity} $s(n)$ means the maximum number of work-tape cells used by the NTM (as well as an naux-pda) on any computation paths on any inputs of length $n$.
The notation $\mathrm{NSPACE,\!TIME}(s(n),t(n))$ denotes the class of all languages recognized by NTMs $M$ within time $t(n)$ using space at most $s(n)$, where $s,t$ are resource bounding functions on $\nat$. Its deterministic variant is denoted by $\mathrm{DSPACE,\!TIME}(s(n),t(n))$.     

Cook's class  $\mathrm{SC}^k$ (resp., $\mathrm{NSC}^k$) coincides with $\mathrm{DSPACE,\!TIME}(O(\log^{O(1)}{n}),n^{O(1)})$ (resp., $\mathrm{NSPACE,\!TIME}(O(\log^{O(1)}{n}),n^{O(1)})$) for each index $k\geq1$ \cite{Coo71}.
Moreover, the standard log-space complexity class  $\dl$ (resp., $\nl$) is the same as $\mathrm{SC}^1$ (resp., $\mathrm{NSC}^1$).
It is known that $\dl\subseteq \nl\subseteq \p$, $\mathrm{SC}^k\subseteq \mathrm{SC}^{k+1} \subseteq \p$, and $\mathrm{NSC}^k\subseteq \mathrm{NSC}^{k+1} \subseteq \np$ for any index $k\geq1$.

An \emph{auxiliary pushdown automata} (or aux-pda, for short) is a two-way pushdown automaton equipped with a rewritable auxiliary (work) tape used as an extra memory device. For simplicity, we abbreviate  a \emph{nondeterministic} (resp., \emph{deterministic}) \emph{auxiliary pushdown automaton} as an \emph{naux-pda} (resp., a \emph{daux-pda}).
We introduce a general notation of $\nauxpdaspti{s(n)}{t(n)}$ (resp., $\dauxpdaspti{s(n)}{t(n)}$) for the complexity class induced by the use of naux-pda's (resp., daux-pda's) running within time $t(n)$ using space at most $s(n)$.
The class $\logdcfl$ and $\logcfl$ coincide with $\dauxpdaspti{O(\log{n})}{n^{O(1)}}$ and $\nauxpdaspti{O(\log{n})}{n^{O(1)}}$, respectively.


We further equip Karp-Lipton style advice to underlying machines (i.e., TMs and aux-pda's). An \emph{advised DTM} (resp., an \emph{advised daux-pda}) is a DTM (resp., a daux-pda) supplemented by a read-only \emph{advice tape}, in which an advice string is written, depending only on the length of inputs. An advice string is provided by an \emph{advice function} $h:\nat\to\Delta^*$ for an advice alphabet $\Delta$. On each input $x$, a machine reads such an advice string $h(|x|)$.
Remember that the use of advice tape does not affect the space complexity of machines.

By making use of $h(n)$-size advice strings, we introduce
$\mathrm{DSPACE,\!TIME}(s(n),t(n))/h(n)$ and $\dauxpdaspti{s(n)}{t(n)}/h(n)$ from $\mathrm{DSPACE,\!TIME}(s(n),t(n))$ and $\dauxpdaspti{s(n)}{t(n)}$, respectively.
Here,  the suffix ``$/h(n)$'' refers to the use of Karp-Lipton style advice of size $h(n)$.  The nondeterministic versions can be similarly defined.
Similarly,  we define  $\mathrm{SC}^k/h(n)$ (resp., $\mathrm{NSC}^k/h(n)$) by providing $h(n)$-size advice strings to underlying machines inducing $\mathrm{SC}^k$ (resp., $\mathrm{NSC}^k$).  Conventionally, we write ``$/\poly$'' (resp., ``$/2^{\polylog}$'') for ``$/n^{O(1)}$'' (resp., ``$/2^{O(\log^{O(1)})}$'').

\subsection{Random Access Mechanism}\label{sec:random-access}

In the standard models of TMs and aux-pda's, all tape heads move  back and forth along tapes. When a string written on a read-only tape is extremely long, it is impossible to  access
any intended tape cells within a short execution time. To resolve this difficulty, we introduce the so-called \emph{random access mechanism}, by which we make a \emph{random access}
(instead of moving a tape head sequentially back and forth) to symbols written on the tape when a machine tries to access those target symbols in such a long string.

In this work, we equip each advised machine with the following random-access  mechanism. In this work, we apply this mechanism to \emph{all}
read-only tapes (i.e., an input tape as well as an advice tape).
Whenever such a machine intends to access the $i$th tape cell of the read-only tape, it first produces the binary number expressing ``$i$'' on an \emph{index tape} in binary and then enters a designated ``query'' state so that the content of the $i$th tape cell is automatically retrieved and written directly on an ``answer'' tape.
The machine then reads the symbol on this answer tape (by moving its tape head from left  to right) to proceed its computation.

\section{Two Key Supporting Technical Tools}\label{sec:tools}

In Sections \ref{sec:poly-ceilings}--\ref{sec:exp-ceilings}, we will discuss the descriptional/computational complexity issues of two-way finite and pushdown automata families. For this purpose, at this point, we wish to develop two key supporting technical tools necessary for proving various containment/separation results of nonuniform complexity classes, as shown in Fig.~\ref{fig:hierarchy-poly} and \ref{fig:hierarchy-exp}. The first technical tool is ``reducibility'' conducted by 2dft's and the second technical tool is a translation between families of finite automata (respect., pushdown automata) and families of advised TMs (resp., aux-pda's).  

\subsection{Reductions by Two-Way Deterministic Finite Transducers}\label{sec:reductions}

In computational complexity theory, reductions have played an crucial role in comparing the computational complexity of two separate computational problems. With a similar spirit, we
wish to utilize the notion of ``reductions'' between two families of promise problems.
Earlier, Sakoda and Sipser \cite{SS78} discussed ``homomorphic reducibility,'' which was later used by Kapoutsis \cite{Kap09,Kap12,Kap14} as a reduction tool to identify the most difficult promise problem families.
In this work, we further introduce a notion of more powerful reductions.
We here stress that the reduction notion is interesting in its own right in the study of nonuniform automata families and it surely requires further intensive research to prove its usefulness.

\begin{definition}
Let $\LL=\{(L_n^{(+)},L_n^{(-)})\}_{n\in\nat}$ and $\KK=\{(K_n^{(+)},K_n^{(-)})\}_{n\in\nat}$ respectively denote two families of promise problems over alphabets $\Sigma$ and $\Gamma$. We say that $\LL$ is \emph{ptime-2FCT$_k$ many-one reducible} (or \emph{ptime-2FCT$_{k}$-m-reducible}, for short) to $\KK$ (denoted $\LL\leq^{\ptime\twofct_k}_{m}\KK$) if there exist a family $\{(f_n,D_n)\}_{n\in\nat}$ of partial functions in $\ptime\twofct_k$ and another p-bounded, p-honest function $d$ on $\nat$ that satisfy the following three conditions for any $n\in\nat$ and for any $x\in\Sigma^*$:  (i) $L_n^{(+)}\cup L_n^{(-)}\subseteq D_n$, (ii) $x\in L_n^{(+)}$ implies $f_n(x)\in K_{d(n)}^{(+)}$, and (iii) $x\in L_n^{(-)}$ implies $f_n(x)\in K_{d(n)}^{(-)}$.
When neither counters nor time bounds are required for reduction machines, we instead write $\LL\leq^{\twof}_{m}\KK$.
\end{definition}

The above reductions can  serve to claim a ``completeness'' of a certain family of promise problems for a target complexity class.

\begin{definition}
Given a family $\LL$ of promise problems, the special notation $\leq^{\ptime\twofct_k}_{m}\!(\LL)$ denotes the class of all families of promise problems that are 2FCT$_{k}$-m-reducible to $\LL$.
Moreover, for a class $\CC$ of families of promise problems, $\leq^{\ptime\twofct_k}_{m}\!(\CC)$ expresses the union $\bigcup_{\LL\in\CC}\leq^{\ptime\twofct_k}_{m}\!(\LL)$. We also define the classes $\leq^{\twof}_{m}\!(\LL)$   and  $\leq^{\twof}_{m}\!(\CC)$ if neither counters nor time bounds are needed.
\end{definition}

Regarding the power of counters, the following statements are proven in \cite{Yam24,Yam24b}.
(1) For any $k\geq4$, $\twodct_k=\twodct_4$.
(2) For any $k\geq3$, $\twodpdct_k=\twodpdct_3$.
(3) $\twodct_4/\poly = \twod/\poly$ and $\twodpdct_3/\poly = \twodpd/\poly$.  Those results   (1)--(3)  also hold for the nondeterministic variants.
Moreover, it is shown in \cite{Yam24,Yam24b} that
$\twod\neq \twodct$, $\twou\neq \twouct$, and $\twon\neq \twonct$, where ``U'' indicates ``unambiguous'' computation. Concerning $\ptime\twofct_k$-m-reductions, the following equations hold.

\begin{proposition}\label{reduction-closure}
(1) $\ptime\twodpdct_{3}  = \: \leq^{\ptime\twofct_2}_{m}\!(\ptime\twodpdct_{3})
= \: \leq^{\ptime\twofct_2}_{m}\!(\onedpdct_3)$.
(2) $\ptime\twodct_{4} = \: \leq^{\ptime\twofct_2}_{m}\!(\ptime\twodct_{4})
=  \: \leq^{\ptime\twofct_2}_{m}\!(\onedct_4)$.
The same equalities hold for nondeterministic computations.
\end{proposition}

We remark that the inclusions $\ptime\twodpdct_3\subseteq \: \leq^{\ptime\twofct_2}_{m}\!(\onedpdct_3)$ and $\ptime\twonpdct_3 \subseteq \: \leq^{\ptime\twofct_2}_{m}\!(\onenpdct_3)$ are reminiscent of \cite[Lemma 3]{Sud78}, in which the language recognized by a polynomial-time two-way pushdown automaton can be $\dl$-m-reducible to an appropriate context-free language.

\vs{-2}
\begin{proofof}{Proposition \ref{reduction-closure}}
(1) Here, we only need to prove that (i)  $\ptime\twodpdct_3 \subseteq \: \leq^{\ptime\twofct_2}_{m}\!(\onedpdct_3)$ and (ii)  $\leq^{\ptime\twofct_2}_{m}\!(\ptime\twodpdct_3) \subseteq \: \ptime\twodpdct_3$ because the inclusion $\leq^{\ptime\twofct_2}_{m}\!(\onedpdct_3) \subseteq \: \leq^{\ptime\twofct_2}_{m}\!(\ptime\twodpdct_3)$ is obvious.

(i) Let us take a promise problem family $\LL=\{(L_n^{(+)},L_n^{(-)})\}_{n\in\nat}$  in $\ptime\twodpdct_3$ and let $\MM=\{M_n\}_{n\in\nat}$ denote a family of polynomial-size 3-counter 2dpdcta's solving $\LL$ in polynomial time.
Take a function $t(n,|x|)=(n|x|)^c+c$ for a positive constant $c\in\nat^{+}$ for which $t(n,|x|)$ upper-bounds the runtime of $M_n$ on any input $x$. We then aim at constructing another promise problem family $\KK= \{(K_n^{(+)},K_n^{(-)})\}_{n\in\nat}$ in $\onedpdct_3$ and also a function family $\FF= \{(f_n,D_n)\}_{n\in\nat}$ in $\ptime\twofct_2$ that reduces $\LL$ to $\KK$.

Firstly, we set up $f_n$ to work in the following way. On input $x$, we move a tape head back and forth between $\rhd$ and $\lhd$ to produce $(\cent x \dollar x^R )^{t(n,|x|)}$ on the output tape, where $\cent$ and $\dollar$ are two distinguished separators.
This task is done as follows.
For convenience, we refer  to each part $\cent x\dollar x^R$ of $(\cent x\dollar x^R)^{(n|x|)^c+c}$  as a \emph{block}.
Initially, we generate $1^{n^c|x|}$ in a counter, say, CT$_1$ by sweeping the input tape between $\rhd$ and $\lhd$.
By decrementing CT$_1$ one by one, we add $1^{|x|}$ to another counter, say, CT$_2$ by sweeping the input tape to
generate $1^{n^c|x|^2}$ in CT$_2$. We then move the content of CT$_2$ to CT$_1$ and repeat the above procedure until CT$_2$ is composed of $1^{n^c|x|^c}$. Additionally, we add $1^{c}$ to CT$_2$ to produce  $1^{t(n,|x|)}$ in CT$_2$. We then define  $D_n=L_n^{(+)}\cup L_n^{(-)}$.
This procedure places $\{(f_n,D_n)\}_{n\in\nat}$ into $\ptime\twofct_2$.

Secondly, we wish to define a new 1npdcta $N_n$ that can simulate $M_n$ as follows. Given input $\rhd\! (\cent x \dollar x^R )^{t(n,|x|)}\! \lhd$, we try to run $M_n$ on $x$.
Notice that a single step of $M_n$ constitutes the following actions: changing inner state $q$ to $p$, changing a top stack symbol $a$ to $w$, and moving an input tape head in direction $d\in\{-1,0,+1\}$. To simulate this single step of $M_n$,
we take the same actions except for the tape head movement.
(a) Assume that we are scanning the $k$th tape cell of $x$ in a certain block $\cent x\dollar x^R$. If $d\in\{0,+1\}$, then we stay in the area of $x$ in the current block $\cent x\dollar x^R$ and move the tape head as $M_n$ does. Whenever $d=-1$, however, we do the following. We first write a special symbol, say, $\#$ in the stack. We then push $1$ as the tape head moves toward the next $\dollar$ symbol. When the tape head reaches $\dollar$, we continue scanning $x^R$ rightwards by popping $1$ from the stack. When we see $\#$ in the stack, we reach the $(|x|-k+1)$-th tape cell of $x^R$ in this block $\cent x\dollar x^R$. (b) In the case where we are scanning a symbol of $x^R$ in a block $\cent x\dollar x^R$, we take a similar movement by swapping the role of $d=+1$ and $d=-1$.

The obtained $N_n$ clearly simulates $M_n$. Now, we define the desired promise problem $(K_n^{(+)},K_n^{(-)})$ by setting $K_n^{(+)}= L(N_n)$ and $K_n^{(-)}= \overline{L(N_n)}$. By the definition, $\FF$ reduces $\LL$ to $\KK$; namely, for any input $x$, $x\in L_n^{(r)}$ implies $f_n(x)\in K_n^{(r)}$ for each $r\in\{+,-\}$.

(ii) We denote by  $\LL=\{(L_n^{(+)},L_n^{(-)})\}_{n\in\nat}$ an arbitrary family of promise problems in $\leq^{\ptime\twofct_2}_{m}\!(\ptime\twodpd)$. Let $\KK=\{(K_n^{(+)},K_n^{(-)})\}_{n\in\nat}$ be any family in $\ptime\twodpdct_3$ solved by an appropriate family $\{M_n\}_{n\in\nat}$ of polynomial-size 3-counter 2dpdcta's.  We then take a partial function family $\FF=\{(f_n,D_n)\}_{n\in\nat}$ in $\ptime\twofct_2$ that reduces $\LL$ to $\KK$. We intend to construct a polynomial-size 2dpdcta family $\{N_n\}_{n\in\nat}$ to solve $(L_n^{(+)},L_n^{(-)})$ as follows, where $L_n^{(+)}=\{x\mid x\in D_n, f_n(x)\in K_n^{(+)}\}$ and $L_n^{(-)}=\{x\mid x\in D_n, f_n(x)\in K_n^{(-)}\}$.

On input $x$, $N_n$ scans $x$ back and forth along an input tape and produces each symbol of $f_n(x)$ one by one onto an output tape using 2 counters. For each output symbol of $f_n(x)$, $N_n$ simulates a single step of $M_n$ using additional 3 counters. This is possible because $M_n$'s input tape head moves in one direction. Clearly, this machine $N_n$ is a 2dpdcta solving $(L_n^{(+)},L_n^{(-)})$ using 5 counters.  Hence, $\LL$ belongs to $\ptime\twodpdct_5$. By Fact \ref{counter-collapse}(2), we can reduce the number of counters down to $3$. This concludes that $\LL$ is in $\ptime\twodpdct_3$.

(2) This is similar in principle to (1) with the use of Fact \ref{counter-collapse}(1) except that we do not need to simulate stack operations.
\end{proofof}

In Proposition  \ref{reduction-closure}, if inputs given to underlying machines $M_n$ and $f_n$ are p-bounded in $n$, then we can eliminate the full use of ``counters'', even from reduction functions.

\begin{corollary}\label{twod-poly-twodpd}
(1) $\twod/\poly  = \; \leq^{\twof}_{m}\!(\oned/\poly)
= \: \leq^{\twof}_{m}\!(\twod/\poly)$.
(2) $\ptime\twodpd/\poly  = \; \leq^{\twof}_{m}\!(\onedpd/\poly)
= \; \leq^{\twof}_{m}\!(\ptime\twodpd/\poly)$.
The statement (1) and (2) hold also for the nondeterministic case.
\end{corollary}

\begin{proof}
The polynomial ceiling restriction guarantees the existence of a fixed polynomial $p$ satisfying  $L_n^{(+)}\cup L_n^{(-)}\subseteq \Sigma^{\leq p(n)}$ for all $n\in\nat$. From this fact,  the description of $f_n$ given in the proof of Proposition \ref{reduction-closure} can be replaced by $f'_n(x)=(\cent x\dollar x^R)^{n^cp(n)}$. Since its exponent $n^cp(n)$ does not depend on $|x|$,  we do not need any use of counters to compute $f'_n(x)$.
This places the obtained reduction function family $\{(f_n,D_n)\}_{n\in\nat}$ into $\twof$.

Next, we consider underlying counter machines for families $\LL$ and $\KK$ in the proof of Proposition \ref{reduction-closure}. The polynomial ceiling restriction also eliminates any counters from those underlying counter machines. Therefore, (1) and (2) follow instantly.
\end{proof}

An instant application of Corollary \ref{twod-poly-twodpd} concludes that  $\oned/\poly=\onen/\poly$ implies $\twod/\poly=\twon/\poly$, where the last equality is unknown to hold without any assumptions.  This comes from the fact that $\leq^{\twof}_{m}\!(\oned/\poly) = \twod/\poly$ and $\leq^{\twof}_{m}\!(\onen/\poly) = \twon/\poly$.  See also Proposition \ref{one-way-case-collapse}(1).

Beyond the above corollary, it is not yet clear that
$\ptime\twodpd =  \; \leq^{\twof}_{m}\!(\ptime\twodpd)$
and $\twod =  \; \leq^{\twof}_{m}\!(\twod)$ hold.
In a similar vein to Proposition \ref{reduction-closure}, we further assert the following closure property for automata families of superpolynomial size. 

\begin{proposition}\label{closure-log-k}
$\ptime2^{\log^k\twodct_{4}}   = \; \leq^{\ptime\twofct_2}_{m}\!(\ptime2^{\log^k\twodct_{4}}) = \; \leq^{\ptime\twofct_2}_{m}\!(2^{\log^k\onedct_{4}})$.
\end{proposition}

\begin{proof}
The proof of this proposition is technically similar to that of Proposition  \ref{reduction-closure}.
Firstly, it is obvious that $\leq^{\ptime\twofct_2}_{m}\!(2^{\log^k\onedct_{4}}) \subseteq  \: \leq^{\ptime\twofct_2}_{m}\!(\ptime2^{\log^k\twodct_{4}})$.

Secondly, we intend to show that $\ptime2^{\log^k\twodct_4} \subseteq \; \leq^{\ptime\twofct_2}_{m}\!(2^{\log^k\onedct_4})$.
As in the proof of Proposition \ref{reduction-closure}, we start with an arbitrary promise problem family $\LL=\{(L_n^{(+)},L_n^{(-)})\}_{n\in\nat}$  over an alphabet $\Sigma$
in $\ptime2^{\log^k\twodct_4}$.
We then design a reduction function $f_n$ to produce from each string $x\in\Sigma^*$ another string of the form $(\cent x\dollar x^R)^{t(n,|x|)}$, where $t(n,|x|)$ is a runtime bound of 2dcta $M_n$ on input $x$.  We set $D_n=L_n^{(+)}\cup L_n^{(-)}$.
Another $4$-counter 1dfcta $N_n$ uses each block $\cent{x}\dollar{x^R}$ of an input string $f_n(x)$ to simulate one step of  $M_n$ as in the proof of Proposition \ref{reduction-closure}.
The state complexity of $N_n$ is upper-bounded by $c\cdot sc(M_n)$ for an absolute constant $c>0$.

Thirdly, we wish to show that $\leq^{\ptime\twofct_2}_{m}\!(\ptime2^{\log^k\twodct_4}) \subseteq \ptime2^{\log^k\twodct_{4}}$.
Let $\LL=\{(L_n^{(+)},L_n^{(-)})\}_{n\in\nat}$ denote any family in $\leq^{\ptime\twofct_2}_{m}\!(\ptime2^{\log^k\twodct_4})$ and take a family $\FF=\{(f_n,D_n)\}_{n\in\nat}$ in $\ptime\twofct_2$ that reduces $\LL$ to another family $\KK=\{(K_n^{(+)},K_n^{(-)})\}_{n\in\nat}$ in $\ptime2^{\log^k\twodct_{4}}$. Moreover, we take a family $\{M_n\}_{n\in\nat}$ of $2^{O(\log^kn)}$-size $4$-counter 2dfcta's solving $\KK$ in polynomial time. Notice that an input to $M_n$ is $f_n(x)$.
We intend to construct a family $\{N_n\}_{n\in\nat}$ of $4$-counter 2dpdcta's to solve $\LL$ directly. Let $x$ denote any input given to $(L_n^{(+)},L_n^{(-)})$.
We use a counter, say, CT$_1$ to keep track of the location of $N_n$'s tape head, say, $l$.
If $M_n$ moves its tape head to the right, then we apply $f_n$ to $x$ to produce symbols of $f_n(x)$ one by one from left to right and find the symbol $a_{l+1}$ written at location $l+1$ by decrementing CT$_1$.  We simultaneously increment another counter, CT$_2$, to recover $l$ and then add $1$ to hold $l+1$.
We then run $M_n$ on $a_{l+1}$ using $4$ more counters. We then swap between the contents of CT$_1$ and CT$_2$ to continue the simulation.
On the contrary, if $M_n$ moves its tape head to the left, then we decrement the counter by one and run $f_n$ on $x$ to find the symbol $a_{l-1}$ at location $l-1$ using the counters. We then run $M_n$ on $a_{l-1}$. This procedure makes $N_n$ solve $(L_n^{(+)},L_n^{(-)})$ correctly. This machine $N_n$ requires $2^{O(\log^kn)}$ inner states and $6$ counters. As shown in Fact \ref{counter-collapse}, we can reduce the number of counters down to $4$. The polynomial time-bound comes from the fact that $f_n$ and $M_n$ have polynomial runtimes. Therefore, $\LL$ belongs to $\ptime2^{\log^k\twodct_{4}}$.
\end{proof}

Similarly to Corollary \ref{twod-poly-twodpd}, we obtain from Proposition \ref{closure-log-k} the following corollary, in which the use of counters and the runtime bounds in the proposition are eliminated.

\begin{corollary}\label{log-k-2D-poly}
$2^{\log^k\twod}/\poly
=\: \leq^{\twof}_{m}\!(2^{\log^k\twod}/\poly) =\: \leq^{\twof}_{m}\!(2^{\log^k\oned}/\poly)$.
\end{corollary}

\subsection{Translation between Automata Families and Advised Turing Machines}\label{sec:translation}

The second technical tool is to utilize close connections between finite automata families and advised Turing machines and also between pushdown automata families and advised auxiliary pushdown automata.
Such connections were first discussed by Berman and Lingas \cite{BL77} and by Sakoda and Sipser \cite{SS78} and have been in a central subject of recent research.

These connections lead us to well-known collapses/separations of (mainstream) advised complexity classes so that we can exploit them to demonstrate corresponding collapses/separations.of classes of automata families.
In this subsection, we wish to present two important but technical propositions, which will be frequently used in later sections.


To simplify the subsequent arguments, we intend to reduce the number of nondeterministic choices made at every step at the cost of increasing the state complexity. Given an arbitrary integer $k\geq2$, we say that a 2nfa (as well as a 2npda) $N$ is of \emph{$k$-choice} if $N$ makes at most $k$ nondeterministic choices at every step.

\begin{lemma}\label{two-choice-change}
Given an $n$-size 2nfa $N$, there exists a computationally equivalent 2-choice 2nfa $\tilde{N}$ of size at most $5n^2$.
\end{lemma}

\begin{proof}
For a given 2nfa $N$, let us consider its transition function  $\delta:(Q-Q_{halt})\times\check{\Sigma}\to \PP(Q\times D)$. We wish to define another 2nfa $\tilde{N}$ by setting $\tilde{Q}$ and $\tilde{\delta}$ as follows.
Assume that, for any pair $(q,\sigma)$, $\delta(q,\sigma)$ has the form $\{(p_1,d_1), (p_2,d_2), \ldots,(p_m,d_m)\}$ with $m\leq |Q|$. We set $\tilde{m}=\ceilings{\log{m}}$ and we prepare the new set $\Theta_{\tilde{m}}=\{0,1\}^{\leq \tilde{m}}$.

Let $\tilde{Q}= Q\cup \{(q,z)\mid q\in Q, z\in \Theta_{\tilde{m}}\}$. We set $\tilde{\delta}(q,\sigma)=\{((q,\lambda),0)\}$, $\tilde{\delta}((q,z),\sigma) = \{((q,z0),0),((q,z1),0)\}$ for any $z$ with $|z|<\tilde{m}$, and $\tilde{\delta}((q,z),\sigma)= \{(p_{\mu(q,\sigma,z)},d_{\mu(q,\sigma,z)})\}$ for any $z$ with $|z|=\tilde{m}$, where $\mu(q,\sigma,z)$ denotes the number $k$ satisfying that $z$ is the $k$th string in $\{0,1\}^{\tilde{m}}$ and $k\leq m=|\delta(q,\sigma)|$ if such a number exists; otherwise, it is undefined. It is straightforward to show that $\tilde{N}$ simulates $N$.

Obviously, $\tilde{N}$ makes 2-choices. As for the state complexity $|\tilde{Q}|$ of $\tilde{N}$, we conclude that $|\tilde{Q}|\leq |Q|+|Q|\cdot 2^{\log{|Q|}+2}\leq 5|Q|^2$.
\end{proof}


Hereafter, we wish to prove the two key technical lemmas. The first key lemma is given in the following. We introduce a special notation. Given an alphabet $\Sigma$ and a designated separator $\#\notin\Sigma$, the notation $[\Sigma,\#]$ denotes the set $\{yx\mid y\in\Sigma_{\#}^*\#\cup\{\lambda\},x\in\Sigma^*\}$, 
where $\Sigma_{\#}=\Sigma\cup\{\#\}$. A subset $K$ of $[\Sigma,\#]$ is said to be \emph{good} if the following statement is true: for any $m\in\nat$,  $y_0\in\Sigma_{\#}^*\#\cup\{\lambda\}$,  and $x\in\Sigma^*$,  if $y_0x\in K$ and $|y_0x|=m$, then $yx\in K$ holds for all $y\in\Sigma_{\#}^*\#\cup\{\lambda\}$ with $|yx|=m$. 

\begin{lemma}\label{key-proposition-lemma}
Let $f,g,s,d$ be any functions on $\nat$   and $t,u$ from $\nat^2$ to $\nat$ with $f(n)\geq1$ for all $n\in\nat$.  Let $\Sigma$ be any alphabet with $\#\notin\Sigma$.
Assume that $t,u$ are nondecreasing and that $f$ is invertible.
When execution time-bounds do not need to specify, we should ignore $t$ and $u$ in the following (1)--(3).

(1) Let $\LL$ denote a family $\{(L_n^{(+)},L_n^{(-)})\}_{n\in\nat}$ of promise problems over $\Sigma$. Assume that $\LL$ has $f(n)$-ceiling.
 If $\LL$ is solved by a family $\{N_n\}_{n\in\nat}$ of $g(n)$-size 2-choice 2nfa's (resp., 2npda's) in $u(n,|x|)$ time for inputs $x$, then
there exists
an NTM (resp., an naux-pda) $N$ such that $N$ recognizes the language $K_{\LL}$ in  time $O(r(|w|,|x|))$ using space $O(\log{e(|w|)})$  with the help of $O(e(|w|)^2)$-size advice for inputs $w$ of the form $yx$ in $[\Sigma,\#]$, where $e(|w|) = g\circ f^{-1}(|w|)$, $r(|w|,|x|)= u(f^{-1}(|w|),|x|)(|x|+\log|w|+\log{e(|w|)})$, and  $K_{\LL}  =\{w\mid w=yx\in[\Sigma,\#], \exists n\in\nat[ |w|=f(n)\wedge x\in L(N_n)] \}$.

(2) Assume that $K$ is a good subset of $\{w\mid w=yx\in[\Sigma,\#], \exists n(|w|=f(n))\}$. If $K$ is recognized by an NTM (resp., an naux-pda) $N$ in $t(f^{-1}(|w|),|x|)$ time using $s(|w|)$ space with $d(|w|)$-size advice for inputs $w$ of the form $yx$ in $[\Sigma,\#]$, then  there exists a family $\{N_n\}_{n\in\nat}$ of $e'(n)$-size 2nfa's (resp., 2npda's) solving $\LL_K=\{(L_n^{(+)},L_n^{(-)})\}_{n\in\nat}$ in  time $O(r'(n,|x|))$ for inputs $x$, where $e'(n) = 2^{O(s\circ f(n))}+O(d\circ f(n))$, $r'(n,|x|)= t(n,|x|)$,  and $\LL_K$ satisfies that, for any $n\in\nat$, $L_n^{(+)} = \{x\mid [|x|<  f(n) \wedge 1^{f(n)-|x|-1}\# x\in K] \vee [|x|=f(n) \wedge x\in K]\}$ and $L_n^{(-)} = \{x\mid [|x|<  f(n) \wedge 1^{f(n)-|x|-1}\# x\in \overline{K}]\vee [|x|=f(n)\wedge x\in \overline{K}]\}$.

(3) The above statements (1)--(2) also hold in the deterministic case.
\end{lemma}

A precursor to the following argument was given in \cite{Yam19a,Yam21,Yam24,Yam24b}, where \cite{Yam24b} particularly provides a standard, general framework to these arguments in terms of \emph{parameterized (decision) problems}.   In the following proof, we do not rely on the use of such parameterized problems.

\vs{-2}
\begin{proofof}{Lemma \ref{key-proposition-lemma}}
($a$)  In what follows, we deal with the case of 2nfa's and NTMs. The other case of 2npda's and naux-pda's will be discussed in ($b$).

(1)
Let us consider a promise problem family $\LL=\{(L_n^{(+)},L_n^{(-)})\}_{n\in\nat}$ over an alphabet $\Sigma$, which has $f(n)$-ceiling restriction (namely,  $L_n^{(+)}\cup L_n^{(-)} \subseteq \Sigma^{\leq f(n)}$ for all $n\in\nat$).
Assume that a family $\NN=\{N_n\}_{n\in\nat}$ of $g(n)$-size 2nfa's $N_n$ solves $\LL$ in time $u(n,|x|)$ for inputs $x$.

Each 2nfa $N_n$ has  $(Q_n,\Sigma,\{\rhd,\lhd\},\delta_n,q_{0,n},Q_{acc,n},Q_{rej,n})$. All transitions of the transition function $\delta_n$ can be identified as  $A_n=\{(q,\sigma,p,l)\mid (p,l)\in\delta_n(q,\sigma)\}$, which is a subset of $Q_n\times \check{\Sigma}\times Q_n\times D$, where $\check{\Sigma} = \Sigma\cup\{\rhd,\lhd\}$ and $D=\{1,0,+1\}$.
We remark that, since each $N_n$ is a 2nfa and  $|x| \leq f(n)$, it runs in time linear-polynomial in $(|Q_n|,|x|)$, say, $c|Q_n||x|+c$ for an appropriate constant $c>0$. Therefore, whenever $u(n,|x|)>c|Q_n||x|+c$,  the time bound $u(n,|x|)$ of $N_n$ can be reset to $c|Q_n||x|+c$.  Unfortunately, this remark is not applicable for 2npda's.
For any string $w$ of the form $yx$ in $[\Sigma,\#]$, if $x$ is a valid string in $L_n^{(+)}\cup L_n^{(-)}$, then we obtain $|w|=f(n)$.
When $g(n)>u(n,|x|)$, in particular, $N_n$ accesses only at most $u(n,|x|)$ inner states and at least $g(n)-u(n,|x|)$ inner states are not used on \emph{each} computation path of $N_n$ on $x$.  However, this does not mean that, since $N_n$ is nondeterministic, those $g(n)-u(n,|x|)$ inner states are redundant and thus completely eliminated from $Q_n$.

Let us define $K_{\LL}  =\{w\mid w=yx\in[\Sigma,\#], \exists n\in\nat[ |w|=f(n)\wedge  x\in L(N_n)] \}$, and thus $\{w\mid w=yx\in[\Sigma,\#], \exists n\in\nat[ |w|=f(n)\wedge  x\in \overline{L(N_n)}] \} \subseteq \overline{K}$ follows.
Now, we consider an NTM $N$ to recognize the language $K_{\LL}$. Here is a concise description of $N$, which takes an input $w$ of the form $yx$ and advice string $h(|w|)$ consisting of the information on $f^{-1}(|w|)$ and $N_{f^{-1}(|w|)}$, that is, $h(|w|)=\pair{\pair{f^{-1}(|w|)},\pair{N_{f^{-1}(|w|)}}}$ if $f^{-1}(|w|)$ exists and $h(|w|)=\pair{\lambda}$ otherwise.
On the input $w=yx$ in $[\Sigma,\#]$, we retrieve the value $f^{-1}(|w|)$ ($=n$) from $h(|w|)$
if $f^{-1}(|w|)$ exists. Otherwise, we reject the input immediately.
Assuming the existence of $f^{-1}(|w|)$, we retrieve $\pair{N_n}$ from $h(|w|)$ and run $N_n$ on $x$ within $u(n,|x|)$ steps.
To simulate each move of the tape head of $N_n$, we use $N$'s own input tape with its associated tape head. We also use $N$'s work tape to hold the information on $N_n$'s current inner state. We trace  the changes of  $N_n$'s inner states by modifying the content of the work tape by moving its tape head back and forth.
Notice that $N$ needs to make random access to its advice string $h(|w|)$, particularly in case of $g(n)>u(n,|x|)$. Notice that we do not need to check if $|w|=f(n)$ because the existence of $f^{-1}(|w|)$ ensures this equality.
This description indicates that $N$ correctly recognizes $K_{\LL}$.

Next, we estimate upper bounds of various resources used by $N$.
Since we require the run of $N_n$ on the machine $N$,  we need to encode $N_n$ into a string, expressed as $\pair{N_n}$, which contains the information on all tuples  $(q,\sigma,p,l)$ in $A_n$.
Note that the length of $\pair{N_n}$ is at most $O(|Q_n||\Sigma| |Q_n||D|) = O(|Q_n|^2) = O(g(n)^2)$, which equals $O(g(f^{-1}(|w|))^2)$ since $|w|=f(n)$. 
If $h(|w|) = \pair{\pair{f^{-1}(|w|)},\pair{N_{f^{-1}(|w|)}}}$, then it follows that $|h(|w|)|\leq O(|\pair{f^{-1}(|w|)}|+|\pair{N_{f^{-1}(|w|)}}|) \leq O(g(f^{-1}(|w|))^2)$ since $g(n)\geq1$.
The work tape of $N$ holds a current inner state of $N_n$, and thus the space bound of $N$ is $O(\log|Q_n|)$, which equals $O(\log{g(n)}) = O(\log(g(f^{-1}(|w|))))$.
The runtime of $N$ is composed of the following four items.
(i) The time for reading the input string $w$ via random access. Notice that, by the use of the random access mechanism, we do not need to move its tape head back and forth over he whole input string $w$ to read.
Note that the number of accesses to $w$ is upper-bounded by $u(n,|x|)$ and that the time to write each index on an index tape requires $O(\log|w|)$ steps.   This requires $O(u(n,|x|)\log|w|)$.
(ii) The time to check if $w$ has the form $yx$ in $[\Sigma,\#]$. This process needs $O(|x|)$ steps.
(iii) The time for reading the advice string $h(|w|)$ on the advice tape via random access.
Similarly to (i), we do not need to access the whole advice string $h(|w|)$ by moving its tape head back and forth to read it. Since the number of accesses to $h(|w|)$ is upper-bounded by $u(n,|x|)$ and writing each index requires $O(\log|h(|w|)|)$ steps.   Hence, this process requires $O(u(n,|x|) \log{|h(|w|)|})$. In comparison, if we do not use the random access mechanism, then we need $O(u(n,|x|) |h(|w|)|)$ execution time.
(iv) The time for simulating each movement of $N_n$. This requires $O(u(n,|x|))$ steps.
(v) The time for tracking the changes of $N_n$'s inner states on the work tape. This requires $O(u(n,|x|)\log{|Q_n|})$. The sum of the execution times for those items is $O(u(n,|x|)(|x|+\log|w|+ \log{|h(|w|)|}+\log|Q_n|))$, which equals $O(u(f^{-1}(|w|),|x|) (|x|+\log|w|+\log{g(f^{-1}(|w|))}))$.

(2)
Given a good subset $K$ of $\{w\mid w=yx\in[\Sigma,\#], \exists n\in\nat[|w|=f(n)]\}$, 
let us take an NTM $N$ and an advice function $h$ such that $N$ accepts (resp., rejects) an input of the form $(w,h(|w|))$ with $w=yx\in[\Sigma,\#]$ in time $t(f^{-1}(|w|),|x|)$ using space $s(|w|)$ whenever $w\in K$ (resp., $w\notin K$), where $h$ satisfies $|h(i)|\leq d(i)$ for all $i\in\nat$.
Recall that this NTM has a read-only input tape, a rewritable work tape, and a read-only advice tape in which an advice string is written.

We then consider a family  $\LL_K =\{(L_n^{(+)},L_n^{(-)})\}_{n\in\nat}$ over $\Sigma$ defined by
$L_n^{(+)} = \{x\in\Sigma^*\mid [|x|<  f(n) \wedge 1^{f(n)-|x|-1}\# x\in K] \vee [|x|=f(n) \wedge x\in K]\}$ and $L_n^{(-)} = \{x\in\Sigma^*\mid [|x|<  f(n) \wedge 1^{f(n)-|x|-1}\# x\in \overline{K}]\vee [|x|=f(n)\wedge x\in \overline{K}]\}$  for each index $n\in\nat$.
We then define a family $\{N_n\}_{n\in\nat}$ of 2nfa's for $\LL_K$ as follows.
For each machine $N_n$, we need to encode the advice string $h(|w|)$ into $N_n$'s inner states. On input $x$, we generate $w=1^{f(n)-|x|-1}\# x$  if $|x|<f(n)$ and $w=x$ if $|x|=f(n)$.
We then run $N$ on the input $(w,h(|w|))$ using space $s(|w|)$.
It is easy to see that $N_n$ solves $(L_n^{(+)},L_n^{(-)})$. Notice that $|w|=f(n)$.  The $s(|w|)$-size work tape content of $N$ can be simulated using $2^{O(s(|w|))}$ inner states.
We further need to remember the value $f(n)$ and all symbols in $h(f(n))$  using $O(f(n))$ inner states and $O(|h(f(n))|)$ inner states, respectively. The state complexity of $N_n$ is thus upper-bounded by
$2^{O(s(f(n)))}+O(d(f(n)))$.
As for the runtime of $N_n$ on input $x$, we note that the simulation of $N$ does not require its tape head to sweep its work and advice tapes.  Thus, it requires $O(t(n,|x|))$.  Hence, it suffices to set $r'(n,|x|)$ to be $t(n,|x|)$.

(3) The deterministic case is treated in a similar way.

($b$) Next, we briefly explain how to amend the above argument to deal with the case of 2npda's and naux-pda's.
We modify 2nfa's and NTMs in ($a$) by supplementing ``stacks''. Since those stacks are mutually simulatable, we do not need to encode the content of each stack into inner states as well as an  advice string. We refer to the proof of Lemma \ref{second-key-proposition-lemma} for more details.
\end{proofof}

The first key lemma (Lemma \ref{key-proposition-lemma}) intends to translate finite automata (as well as pushdown automata) into NTMs (as well as naux-pda's). Conversely, the following second key proposition translates NTMs (and naux-pda's) to finite automata (and pushdown automata),  compensating the first key lemma.

\begin{lemma}\label{second-key-proposition-lemma}
Let $f,g,s,d$ be any functions on $\nat$  and $t,u$ from $\nat^2$ to $\nat$ with $f(n),d(n)\geq1$ for all $n\in\nat$. Let $\Sigma$ be any alphabet.  Assume that $t$ and $u$ are  nondecreasing and that $f$ is invertible.
Whenever execution time-bounds do not need to specify, we should ignore $t$ and $u$.

(1) Let $K$ denote any language over $\Sigma$. If $K$ is recognized by an NTM (resp., an naux-pda) $N$ in $t(f^{-1}(|x|),|x|)$ time using at most $s(|x|)$ space with $d(|x|)$-size advice for inputs $x$, then there exists a family $\{N_n\}_{n\in\nat}$ of $e(n)$-size $O(1)$-choice 2nfa's (resp., 2npda's)  solving $\LL_K =\{(L_n^{(+)},L_n^{(-)})\}_{n\in\nat}$ in $O(r(n,|x|))$ time for inputs $x$, where  $e(n)=2^{O(s\circ f(n))}+O(d\circ f(n))$, $r(n,|x|)= t(n,|x|)$, and  $\LL_K$ satisfies that, for any $n\in\nat$, $L_n^{(+)}=\{ x\mid |x| = f(n), x\in K\}$ and  $L_n^{(-)}=\{x\mid |x| = f(n), x\in \overline{K}\}$.
This family $\LL$ has $f(n)$-ceiling.

(2) For any $f(n)$-ceiling family $\LL=\{(L_n^{(+)},L_n^{(-)})\}_{n\in\nat}$ over $\Sigma$, if $\LL$ is solvable by a family $\{N_n\}_{n\in\nat}$ of $g(n)$-size 2nfa's (resp., 2npda's) in $u(n,|x|)$ time for inputs $x$, then there exists  an NTM (resp., an naux-pda) $N$ that recognizes the language $K_{\LL}$ in $O(r'(f^{-1}(|x|),|x|))$ time using $O(\log{e'(|x|)})$ space with the help of $O(e'(|x|)^2)$-size advice for inputs $x$, where $e'(|x|)= g\circ f^{-1}(|x|)$,   $r'(n,|x|)= u(n,|x|)(\log|x|+ \log{e'(|x|)})$, and $K_{\LL}= \{x\mid n\in\nat, |x|=f(n), x\in L(N_n)\}$.

(3) The statements (1)--(2) also hold in the deterministic case.
\end{lemma}

\begin{proof}
The overall arguments are similar in principle to the ones in the proof of Lemma \ref{key-proposition-lemma}. In what follows, we mostly provide the proof for the case of 2npda's and naux-pda's.

(1) Take any language $K$ over an alphabet $\Sigma$ and an naux-pda $N$ that recognizes $K$ in time $t(f^{-1}(|x|),|x|)$ using space at most $s(|x|)$ with $d(|x|)$-size advice for inputs $x$.
We consider the family $\LL_K=\{(L_n^{(+)},L_n^{(-)})\}_{n\in\nat}$ of promise problems of the form $(L_n^{(+)},L_n^{(-)})$ with  $L_n^{(+)} = \{x\mid |x|=f(n), x\in K\}$ and $L_n^{(-)}= \{x\mid |x|=f(n), x\in \overline{K}\}$ for all $n\in\nat$.

Our goal is to solve $\LL_K$ using an appropriate 2npda family.
Let us consider the following 2npda $N_n$.
On input $x$, check that $|x|=f(n)$ for a certain number $n\in\nat$. If not, immediately reject the input; otherwise, run $N$ on $x$.
Since $N$ uses at most $s(|x|)$  work cells and an advice string of at most $d(|x|)$ size to simulate $N$,  $N_n$ needs to remember the value $f(n)$ and a $d(f(n))$-size advice string in the form of inner states. Moreover, $N_n$ uses its inner states to express the  work-tape content of $N$ and to trace the changes of the work-tape content by updating the corresponding inner states.
Notice that $N_n$ can use its own stack to maintain the content of $N$'s stack.

The stack-state complexity of $N_n$ is therefore $O(f(n))+ O(d(f(n))) +  2^{O(s(f(n)))}$, which equals $2^{O(s(f(n)))}+O(d(f(n)))$ since $d(n)\geq1$. 
Since $N_n$ has no work or advice tapes, it does not require the time to sweep or make random access to those tapes of $N$.
The runtime of $N_n$ on input $x$ is therefore at most $O(t(n,|x|))$.

(2) From a given family $\LL$ of promise problems and a family $\{N_n\}_{n\in\nat}$ of 2npda's solving $\LL$, we define the desired naux-pda $N$ and its  advice function $h$ as follows. We assume that each  $N_n$ has the form $(Q_n,\Sigma,\Gamma_n,\delta_n,q_{0,n}, \bot, Q_{acc,n},Q_{rej,n})$ with push size $e_n$. Since $N_n$ has $g(n)$ size, $|Q_n||\Gamma_n^{\leq e_n}|\leq g(n)$ follows.

We first ``encode'' the machine $N_n$ into a string $\alpha_n$. For this purpose, assuming an appropriate, efficient encoding scheme, it suffices to  encode $N_n$'s transition function
$\delta_n:(Q_n-Q_{halt,n}) \times \check{\Sigma} \times\Gamma \to \PP(Q_n\times \Gamma^{\leq e_n}\times D)$ by identifying its domain and range pair with a subset of $(Q_n-Q_{halt,n})\times\check{\Sigma}\times \Gamma_n \times Q_n \times \Gamma_n^{\leq e_n}\times D$, where $D=\{-1,0,+1\}$.
Thus, $|\alpha_n|$ is upper-bounded by $O(|Q_n||\check{\Sigma}| |\Gamma_n| |Q_n| |\Gamma_n^{\leq e_n}| |D|) = O((|Q_n||\Gamma^{\leq e_n}|)^2)$, which equals $O(g(n)^2) = O(g(f^{-1}(|x|))^2)$.
Moreover, we need to encode the value $f^{-1}(|x|)$ into another string $\beta_{|x|}$. Note that $|\beta_{|x|}|\leq O(f^{-1}(|x|)) \leq O(g\circ f^{-1}(|x|)^2)$.
We then define an advice function $h$ by setting $h(|x|)$ to be $\pair{\alpha_n,\beta_{|x|}} = \pair{\alpha_{f^{-1}(|x|)},\beta_{|x|}}$ since $|x|=f(n)$.
It thus follows that $|h(|x|)|\leq O(|\alpha_{n}|+|\beta_{|x|}|) \leq O(g\circ f^{-1}(|x|)^2+f^{-1}(|x|))  = O(g\circ f^{-1}(|x|)^2)$.

Finally, we wish to define the desired naux-pda $N$, which works as follows. 
On input $x$, we  retrieve (``part'' of) the strings $\alpha_{f^{-1}(|x|)}$ and $\beta_{|x|}$ from the advice string $h(|x|)$.
Since $n=f^{-1}(|x|)$, the string $\alpha_{f^{-1}(|x|)}$ encodes $N_n$, and thus it is possible to run $N_n$ on $x$. The machine $N$ simulates $N_n$ by tracing the changes of inner states using its work tape.
Here, it is important to remark that $N$ is sufficient to retrieve only transitions of $N_n$ actually used on each computation path of $N_n$ in order to simulate $N_n$ on $x$.
Each inner state of $N_n$ requires $O(\log|Q_n|)$ bits to express on the work tape.
Hence, $N$'s space complexity is bounded by $O(\log{g(n)})=O(\log{g(f^{-1}(|x|))})$.
The runtime of $N$ is also calculated from the following four items:
(i) the time for accessing the input string via random access,
(ii) the time for accessing the advice string via random access,
(iii) the time for simulating $N$'s move, and
(iv) the time for updating the work tape.
Those four items respectively require time at most $O(t(n,|x|)\log|x|)$, $O(t(n,|x|)\log{|h(|x|)|})$,  $O(t(n,|x|))$,  and $O(t(n,|x|)\log{g(n)})$.
Thus, the overall runtime of $N$ is at most $O(t(f^{-1}(|x|),|x|)(\log|x|+ \log{g(f^{-1}(|x|))}))$.

(3) The deterministic case is treated in a similar way.
\end{proof}


For later arguments, we intend to rephrase the first and the second key lemmas in a more practical, readable form. The first key lemma is rephrased as follows.

\begin{proposition}\label{key-proposition}
Let $f,g,s,d$ denote any functions on $\nat$ and $t,u$ from $\nat^2$ to $\nat$. Assume that $u,t$ are nondecreasing and that $f$ is invertible. Let $\Sigma$ be any alphabet with $\#\notin\Sigma$.
Whenever execution time-bounds do not need to specify, we should ignore $u$ and $t$.

(1) Let $\LL=\{(L_n^{(+)},L_n^{(-)})\}_{n\in\nat}$ be any family of promise problems over $\Sigma$. If $\LL\in u(n,|x|)\mathrm{time}\mbox{-}2^{\log{g(\mathrm{lin}\twon)}}/f(n)$ witnessed by a family $\{N_n\}_{n\in\nat}$ for inputs $x$, then  $K_{\LL}\in \mathrm{NSPACE,\!TIME}(O(\log{e(|w|)}),O(r(|w|,|x|)))/O(e(|w|)^2)$ for inputs $w=yx$ in $[\Sigma,\#]$, where $e(|w|)=g\circ f^{-1}(|w|)$, $r(|w|,|x|)= u(f^{-1}(|w|),|x|)(|x|+\log|w|+\log{e(|w|)})$,  and
$K_{\LL}  =\{w\mid w=yx\in[\Sigma,\#],  \exists n\in\nat[ |w|=f(n)\wedge  x\in L(N_n)] \}$.

(2) Let $K$ be any good subset of  $\{w\mid w=yx\in[\Sigma,\#], \exists n\in\nat[|w|=f(n)]\}$.
If $K\in \mathrm{NSPACE,\!TIME}(s(|w|),t(f^{-1}(|w|),|x|))/d(|w|)$ for inputs $w=yx$ in $[\Sigma,\#]$, then the family $\LL_K =\{(L_n^{(+)},L_n^{(-)})\}_{n\in\nat}$ satisfies that $\LL_K\in O(r'(n,|x|))\mathrm{time}\mbox{-}2^{\log{e'(\mathrm{lin\twon})}}/f(n)$ for inputs $x$, where $e'(n)=2^{O(s\circ f(n))} + O(d\circ f(n))$, $r'(n,|x|)=t(n,|x|)$,  and $\LL_K$ satisfies that, for every $n\in\nat$,
$L_n^{(+)} = \{x\mid [|x|<  f(n) \wedge 1^{f(n)-|x|-1}\# x\in K] \vee [|x|=f(n) \wedge x\in K]\}$ and $L_n^{(-)} = \{x\mid [|x|<  f(n) \wedge 1^{f(n)-|x|-1}\# x\in \overline{K}]\vee [|x|=f(n)\wedge x\in \overline{K}]\}$.

(3) It is possible to replace $\mathrm{lin}\twon$ and $\mathrm{NSPACE,\!TIME}(\cdot,\cdot)$ in (1)--(2) by $\mathrm{lin}\twonpd$ and $\nauxpdaspti{\cdot}{\cdot}$, respectively.

(4) Moreover, the statements (1)--(3) hold in the deterministic case.
\end{proposition}

The second key lemma is rephrased in the following way.

\begin{proposition}\label{second-key-proposition}
Let $f,g,s,d$ denote any functions on $\nat$ and $t,u$ from $\nat^2$ to $\nat$. Assume that $u$ and $t$ are nondecreasing and that $f$ is invertible. Let $\Sigma$ be any alphabet.
Whenever execution time-bounds do not need to specify, we should ignore $u$ and $t$.

(1) Let $K\subseteq\Sigma^*$. If $K\in \mathrm{NSPACE,\!TIME}(s(|x|),t(f^{-1}(|x|),|x|))/d(|x|)$ for inputs $x$, then the family $\LL_K =\{(L_n^{(+)},L_n^{(-)})\}_{n\in\nat}$ belongs to  $O(r(n,|x|))\mathrm{time}\mbox{-}2^{\log{e(\mathrm{lin\twon})}}/f(n)$ for inputs $x$, where $e(n)=2^{O(s\circ f(n))} + O(d\circ f(n))$,  $r(n,|x|)= t(n,|x|)$, and $\LL_K$ satisfies that, for every $n\in\nat$,  $L_n^{(+)} = \{x\mid |x| = f(n), x\in K\}$ and $L_n^{(-)} = \{x\mid |x| =  f(n),x\in \overline{K}\}$.

(2) Let $\LL=\{(L_n^{(+)},L_n^{(-)})\}_{n\in\nat}$ denote any family of promise problems over an alphabet $\Sigma$. If $\LL\in u(n,|x|)\mathrm{time}\mbox{-}2^{\log{g(\mathrm{lin}\twon)}}/f(n)$ witnessed by a family $\{N_n\}_{n\in\nat}$ for inputs $x$, then $K_{\LL}\in \mathrm{NSPACE,\!TIME}(O(\log{e'(|x|)}),O(r'(f^{-1}(|x|),|x|)))/O(e'(|x|)^2)$ for inputs $x$, where $e'(|x|)=g\circ f^{-1}(|x|)$, $r'(n,|x|)= u(n,|x|)(\log|x|+\log{e'(|x|)})$,  and $K_{\LL}=\{ x\mid n\in\nat,|x| = f(n),  x\in L(N_n)\}$.

(3) It is possible to replace $\mathrm{lin}\twon$ and $\mathrm{NSPACE,\!TIME}(\cdot,\cdot)$ in (1)--(2) by $\mathrm{lin}\twonpd$ and $\nauxpdaspti{\cdot}{\cdot}$, respectively.

(4) Moreover, the statements (1)--(3) hold in deterministic case.
\end{proposition}

In Sections \ref{sec:poly-ceilings}--\ref{sec:exp-ceilings}, we will apply the above two propositions to derive various collapses/separations of nonuniform complexity classes.

We remark that, when $f$ is a linear function (i.e., $f(n)=an+b$ for constants $a,b>0$), we can derive simpler conclusions from Propositions \ref{key-proposition}-\ref{second-key-proposition}. Such conclusions will be exemplified as Theorem \ref{dspace-2nfa-relation} and Corollary \ref{etime-equivalence}.

\section{Promise Problem Families of Polynomial Ceilings}\label{sec:poly-ceilings}

\begin{figure}[t]
\centering
\includegraphics*[height=5.1cm]{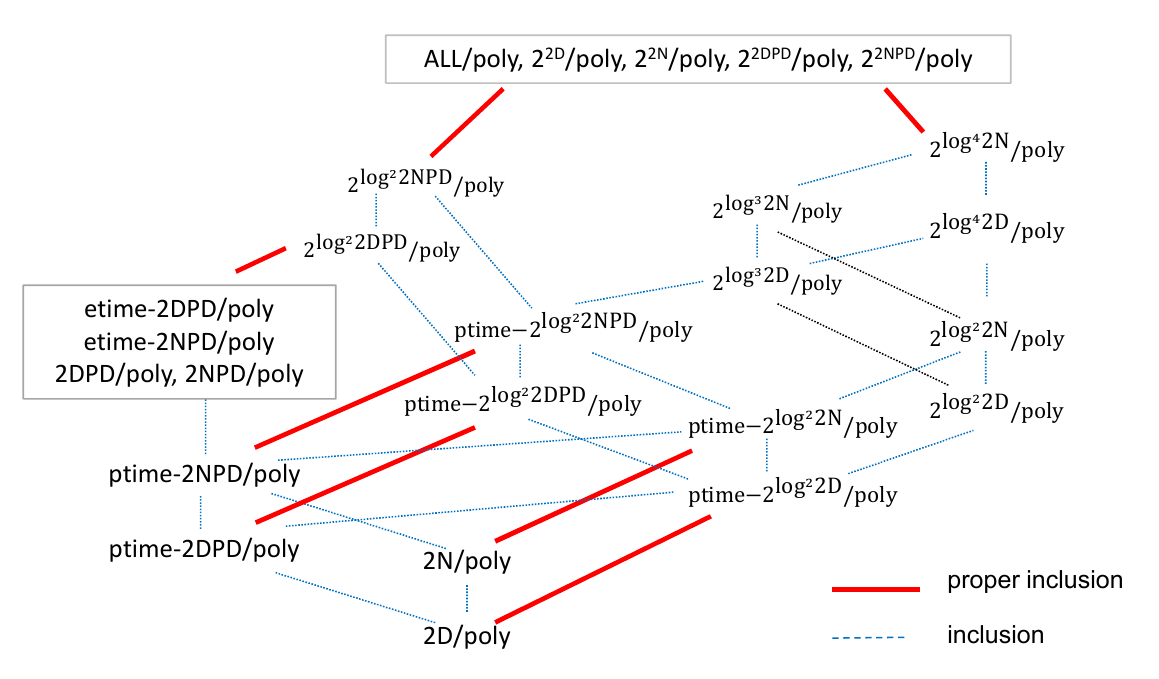}
\caption{Inclusion relationships among lower nonuniform complexity classes with polynomial ceilings. 
Solid red lines indicate ``proper'' inclusions.}\label{fig:hierarchy-poly}
\end{figure}

Let us first recall the purpose of this work, which is to study the effect of various sizes, ceilings, and execution time onto  the overall complexity of families of promise problems solvable by nonuniform families of finite and pushdown automata.
Throughout this section, nonetheless, we wish to study the computational complexity of nonuniform automata families particularly having polynomial ceilings.
We split  the section into two parts, each of which targets polynomial and exponential sizes.
Most of our results of the section are illustrated in Fig.~\ref{fig:hierarchy-poly}.

\subsection{Complexity of Polynomial-Size Automata Families}\label{sec:poly-size-automata}

Sudborough \cite{Sud78} demonstrated that all languages in $\logcfl$ can be simulated by naux-pda's running in polynomial time using $O(\log{n})$ space.
Ruzzo \cite{Ruz80} proved that such a language can be further accepted by an appropriate daux-pda running in $O(\log{n})$ space and $2^{O(\log^2n)}$ time. 
These results can relativize. Therefore, by a suitable application of Propositions \ref{key-proposition}--\ref{second-key-proposition},  these facts instantly imply that $\ptime\twonpd \subseteq \etime\twodpd$ with no ceiling bound.
From a different perspective, we here wish to discuss stronger upper bounds of $\twodpd$ and $\twonpd$ in terms of $\twod$ and $\twon$ when all promise problem families are particularly limited  to ``polynomial'' ceiling.

\begin{theorem}\label{Cook-proof}
\renewcommand{\labelitemi}{$\circ$}
\begin{enumerate}\vs{-1}
  \setlength{\topsep}{-2mm}%
  \setlength{\itemsep}{1mm}%
  \setlength{\parskip}{0cm}%

\item $\twod/\poly \subseteq \ptime\twodpd/\poly \subseteq \ptime2^{\log^2\twod}/\poly$.

\item $\twon/\poly \subseteq \ptime\twonpd/\poly \subseteq \ptime2^{\log^2\twon}/\poly$.
\end{enumerate}
\end{theorem}

With no ceiling bound, on the contrary, it is still unknown whether $\ptime\twodpd\subseteq 2^{\log^2\twod}$ holds or even $\ptime\twodpd\subseteq 2^{\lintwod}$ holds. This fact also exemplifies the significance of the use of ceiling bounds in the analysis of nonuniform automata families.

Let us return to the proof of Theorem \ref{Cook-proof} and give its proof with the help of Corollaries  \ref{twod-poly-twodpd} and \ref{log-k-2D-poly}.

For this purpose, we first recall from computational complexity theory the notion of \emph{polynomial-time many-one reducibility} (or p-m-reducibility, for short). See textbooks, e.g., \cite{DK00} for its properties.
Given a language class $\CC$, the notation $\leq^{\p}_m\!(\CC)$ denotes the closure under  p-m-reductions applied to any languages in $\CC$.
We also recall from Section \ref{sec:translation} the notation $[\Sigma,\#]$, which indicates the set $\{yx\mid y\in\Sigma_{\#}^*\#\cup\{\lambda\}, x\in\Sigma^*\}$, provided that $\#\notin\Sigma$.

\vs{-2}
\begin{proofof}{Theorem \ref{Cook-proof}}
(1) Since $\twod/\poly=\ptime\twod/\poly$, it is obvious that $\twod/\poly \subseteq \ptime\twodpd/\poly$.
In what follows, we intend to prove that $\ptime\twodpd/\poly \subseteq \ptime2^{\log^2\twod}/\poly$.
Earlier, Cook \cite{Coo71} proved that
all deterministic context-free languages are in $\mathrm{SC}^2$, namely, $\dcfl\subseteq \mathrm{SC}^2$ (see also \cite{BCM+83}).
From Cook's result, we obtain $\leq^{\p}_m\!(\dcfl)\subseteq \; \leq^{\p}_m\!(\mathrm{SC}^2)$, which is equivalent to $\dauxpdaspti{O(\log{n})}{n^{O(1)}} \subseteq \mathrm{DSPACE,\!TIME}(O(\log^2n),n^{O(1)})$ by Sudborough's daux-pda characterization \cite{Sud78}.
This argument works even with the use of polynomial-size advice. Therefore, we can conclude that $\dauxpdaspti{O(\log{n})}{n^{O(1)}}/\poly$ is included in $\mathrm{DSPACE,\!TIME}(O(\log^2n),n^{O(1)}/\poly)$.
What is remaining undone is to verify that, if $\dauxpdaspti{O(\log{n})}{n^{O(1)}}/\poly \subseteq \mathrm{DSPACE,\!TIME}(O(\log^2n),n^{O(1)}/\poly)$, then $\ptime\twodpd/\poly \subseteq \ptime2^{\log^2\twod}/\poly$ follows.

We begin with an arbitrary family $\LL$ of promise problems in $\LL\in\ptime\twodpd/\poly$. Now, we take suitable functions $g(n)=n^a+a$, $f(n)=n^b+b$, and $u(n,|x|)=(n|x|)^a+a$ for  constants $a,b>0$ so that $\LL$ belongs to $u(n,|x|)$time-$2^{\log{g(\mathrm{lin}\twodpd)}}/f(n)$, witnessed by a nonuniform family $\{N_n\}_{n\in\nat}$ of corresponding 2dpda's. Recall from Proposition \ref{key-proposition} that
$K_{\LL}  =\{w\mid w=yx\in[\Sigma,\#],  \exists n\in\nat[ |w|=f(n)\wedge  x\in L(N_n)] \}$.
By Proposition \ref{key-proposition}(1), the language $K_{\LL}$ falls in $\nauxpdaspti{O(\log(e(|w|)))}{O(r(|w|,|x|))}/O(e(|w|)^2)$ for inputs $w$ of the form $yx$ in $[\Sigma,\#]$, where $e(|w|)=g\circ f^{-1}(|w|)$ and $r(|w|,|x|)= u(f^{-1}(|w|),|x|)\log{e(|w|)}$.
Since $e(|w|)=|w|^{O(1)}$ and $r(|w|,|x|)=(|x||w|)^{O(1)}$, $K_{\LL}$ is in $\nauxpdaspti{O(\log{n})}{n^{O(1)}}/n^{O(1)}$, which is included in $\mathrm{DSPACE,\!TIME}(O(\log^2n),n^{O(1)})/n^{O(1)}$.
We then take $s(|w|)=c\log^2|w|+c$, $t(n,|x|)=(n|x|)^c+c$, and $d(|w|)=|w|^c+c$ satisfying  $K_{\LL}\in \mathrm{DSPACE,\!TIME}(s(n),t(f^{-1}(|w|),|x|))/d(|w|)$ for inputs $w=yx\in [\Sigma,\#]$.
Proposition \ref{key-proposition}(2) then implies that $\LL$ is in $O(r'(n,|x|))$time-$2^{\log{e'(\mathrm{lin}\twod)}}/f(n)$ for inputs $x$, where $e'(n)=2^{O(s\circ f(n))}+O(d\circ f(n)) = 2^{O(\log^2n)}$ and $r'(n,|x|)=t(n,|x|)$. Since $r'(n,|x|)$ is a polynomial in $(n,|x|)$,  $\LL$ belongs to $\ptime2^{\log^2\twod}/\poly$.
Therefore, we obtain $\ptime\twodpd/\poly \subseteq \ptime2^{\log^2\twod}/\poly$.

(2) Earlier, Ruzzo \cite[Corolalry 7]{Ruz80} argued that, for each 1npda,  the maximal stack height on each of its computation paths is upper-bounded by $O(\log^2|x|)$ for any input $x$.
A similar argument shows that a 1npda $M_n$ having $n$ inner states can be simulated by another suitable 1npda $N_n$ having $n^{O(1)}$ inner states using only $O(\log^2|x|)$ stack height for all inputs $x$. When all ``valid'' inputs of $N_n$ are of size at most $n^k$ for a certain fixed constant $k>0$, the stack height is upper-bounded by $O(k^2\log^2n)$. Thus, we can encode all possible stack contents of height $O(k^2\log^2n)$  into inner states. The state complexity of $N_n$ thus becomes $2^{O(\log^2n)}$. From this fact, we can simulate $N_n$ by an appropriate 2nfa with $2^{O(\log^2n)}$ inner states. This argument leads to the conclusion that $\onenpd/\poly \subseteq \ptime2^{\log^2\twon}/\poly$. By taking the closure under $\twof$-m-reductions, we obtain $\leq^{\twof}_m\!(\onedpd/\poly) \subseteq \;  \leq^{\twof}_m\!(\ptime2^{\log^2\twon}/\poly)$.
By Corollaries \ref{twod-poly-twodpd} and \ref{log-k-2D-poly},
this inclusion is equivalent to $\ptime\twonpd/\poly \subseteq \ptime2^{\log^2\twon}/\poly$, as requested.
\end{proofof}

Theorem \ref{Cook-proof}(2) yields $\twon/\poly \subseteq \ptime\twonpd/\poly$; however, it is not yet known whether the opposite inclusion holds.  This is in fact boiled down to the $\onenpd/\poly \subseteq?\twon$ question, as shown below.

\begin{proposition}
The following three statements are all logically equivalent. (1) $\twon/\poly = \ptime\twonpd/\poly$, (2) $\onenpd/\poly \subseteq \twon/\poly$, and (3) $\logcfl\subseteq \nl/\poly$.
\end{proposition}

\begin{proof}
(1 $\Rightarrow$ 2) Since  $\onenpd/\poly$ is included in $\ptime\twonpd/\poly$, $\twon/\poly \subseteq \ptime\twonpd/\poly$ implies $\onenpd\subseteq \twon/\poly$.

(2 $\Rightarrow$ 1)
If $\onenpd/\poly \subseteq \twon/\poly$, then $\leq^{\twof}_{m}\!(\onedpd/\poly) \subseteq  \: \leq^{\twof}_{m}\!(\twon/\poly)$, which implies that $\ptime\twonpd/\poly \subseteq \twon/\poly$ by Corollary \ref{twod-poly-twodpd}. Since $\twon/\poly \subseteq \ptime\twonpd/\poly$ is obvious, it thus follows that $\twon/\poly = \ptime\twonpd/\poly$.

(3 $\Rightarrow$ 1) Assume that $\logcfl\subseteq \nl/\poly$. By definition, this is equivalent to $\nauxpdaspti{O(\log{n})}{n^{O(1)}}/\poly \subseteq \mathrm{NSPACE,\!TIME}(O(\log{n}),n^{O(1)})/\poly$. Let $\LL=\{(L_n^{(+)},L_n^{(-)})\}_{n\in\nat}$ denote any promise problem family  in $\ptime\twonpd/\poly$.
Let  $g(n)=n^a+a$, $u(n,|x|)=(n|x|)^a+a$, and $f(n)=n^b+b$ for suitable constants $a,b>0$ and assume that the family $\LL$ belongs to $u(n,|x|)$time-$2^{g(\mathrm{lin}\twonpd)}/f(n)$ for inputs $x$.
By Proposition \ref{key-proposition}(1), the language $K_{\LL}$ falls in $\nauxpdaspti{O(\log{e(|w|)})}{O(r(|w|,|x|))}/O(e(|w|)^2)$ for inputs $w=yx$ in $[\Sigma,\#]$,
where $e(|w|)=g\circ f^{-1}(|w|)=|w|^{O(1)}$, $r(|w|,|x|)=u(f^{-1}(|w|),|x|)\log{e(|w|)} =|w|^{O(1)}$. Since $K_{\LL}$ belongs to $\nauxpdaspti{O(\log\log{n})}{n^{O(1)}}/\poly$. Our assumption ensures that $K_L$ falls in $\mathrm{DSPACE,\!TIME}(O(\log{n}),n^{O(1)})/\poly$.
Now, we take functions $s(|w|)=a\log{|w|}+a$, $t(n,|x|)=(n|x|)^b+b$, and $d(|w|)=|w|^b+b$ for positive constants $a,b$ satisfying that $\LL_K\in \mathrm{DSPACE,\!TIME}(s(|w|),t(f^{-1}(|w|),|x|))/d(|w|)$.
Notice that $\LL$ coincides with $\LL_K$.   By Proposition \ref{key-proposition}(2),  we conclude that $\LL$ is in $O(r'(n,|x|))$time-$2^{\log{e'(\mathrm{lin}\twonpd)}}/f(n)$ for inputs $x$, where $e'(n)=2^{O(s\circ f(n))}+O(d\circ f(n))$ and  $r'(n,|x|)=t(n,|x|)$. This class is included in $\ptime\twonpd/\poly$ because of  $e'(n)=n^{O(1)}$.   

(1 $\Rightarrow$ 3) Assume that $\twon/\poly = \ptime\twonpd/\poly$.
Let $K$ be a language in $\logcfl$, which is included in $\nauxpdaspti{O(\log{n})}{n^{O(1)}}/n^{O(1)}$. Let $f(n)=n$ for all $n\in\nat$.
We thus take functions $s(|x|)=a\log{|x|}+a$, $t(n,|x|)=(n|x|)^a+a$, and $d(|x|)=|x|^a+a$ so that  $K$ falls in $\nauxpdaspti{s(|x|)}{t(f^{-1}(|x|),|x|)}/d(|x|)$ for inputs $x$, where $a$ is an appropriate positive  constant.
By Proposition \ref{second-key-proposition}(1), $\LL_{K}$ belongs to $O(r(n,|x|))$time-$2^{\log{e(\mathrm{lin}\twonpd)}}/f(n)$ with $e(n)= 2^{s\circ f(n)}+O(d\circ f(n))= n^{O(1)}$ and $r(n,|x|)= t(n,|x|)$. Thus, $\LL_K$ belongs to $\ptime\twonpd/\poly$. Our assumption then ensures that $\LL_K$ is also in $\twon/\poly$. Now, we take a function $g(n)=n^b+b$ for a constant $b>0$ and assume that $\LL_K$ falls in $2^{\log{g(\mathrm{lin}\twon)}}/f(n)$. Proposition \ref{second-key-proposition}(2) then implies that $\LL_K$ is also  in $\mathrm{NSPACE,\!TIME}(O(\log{e'(|x|)}),\infty)/O(e'(|x|)^2)$ for inputs $x$, where $e'(|x|)= g\circ f^{-1}(|x|)=|x|^{O(1)}$.
Since $K=K_{\LL}$, $K$ belongs to $\mathrm{NSPACE,\!TIME}(O(\log{n}),\infty)/\poly$, which is included in $\nl/\poly$.
\end{proof}


We briefly discuss a fundamental question associated with a relationship between $\twon$ and $\twonpd$. One may ask whether or not $\twon$ and $\twonpd$ are the same. The following proposition provides a partial answer to this question when underlying machine families satisfy the polynomial ceiling constraint.

\begin{proposition}\label{unary-poly-etime}
$\twon/\poly = \twonpd/\poly$ if and only if  $\p\subseteq \nl/\poly$.
\end{proposition}

\begin{proof}
(If--part)
We assume that $\p\subseteq\nl/\poly$. This is logically equivalent to $\nl/\poly = \p/\poly$, which is the same as $\mathrm{NSPACE,\!TIME}(O(\log{n}),n^{O(1)})/n^{O(1)} = \mathrm{DSPACE,\!TIME}(n^{O(1)},n^{O(1)})/n^{O(1)}$ by definition.
Let us take an arbitrary promise problem family $\LL$ in $\twonpd/\poly$.
There are functions  $g(n)=n^{a}+a$ and $f(n)=n^{a}+a$ such that $\LL$ falls in   $2^{\log{g(\mathrm{lin}\twonpd)}}/f(n)$  for an appropriate constant $a>0$.   We apply Proposition \ref{key-proposition}(1) and conclude that the language $K_{\LL}$ belongs to  $\nauxpdaspti{O(\log{e(|w|)})}{\infty}/O(e(|w|)^2)$ for inputs $w$ in $[\Sigma,\#]$, where $e(|w|)=g\circ f^{-1}(|w|)$. This nonuniform class is included in $\nauxpdaspti{O(\log{n})}{\infty}/\poly$ since $e(|w|)=|w|^{O(1)}$.

At this point, it is important to remark that, by Cook \cite{Coo71}
and Chandra, Kozen, and Stockmeyer \cite{CKS81},
$\nauxpdaspti{O(\log{n})}{\infty}$ equals $\p$.
By appending polynomial-size advice to underlying naux-pda's in the above equality,
$\nauxpdaspti{O(\log{n})}{\infty}/\poly$ coincides with $\p/\poly$.
Therefore, by our assumption, $K_{\LL}$ further falls in $\nl/\poly$, which equals  $\mathrm{NSPACE,\!TIME}(O(\log{n}),\infty)/n^{O(1)}$.
Now, we take functions  $s(n)=b\log{n}+b$ and $d(n)=n^{b}+b$ for an appropriate constant $b>0$ and assume that $K_{\LL}$ is an element of $\mathrm{NSPACE,\!TIME}(s(|w|),\infty)/d(|w|)$.  We apply Proposition \ref{key-proposition}(2) and immediately obtain $\LL_K\in 2^{\log{e'(\mathrm{lin}\twon)}}/f(n)$, where $e'(n)=2^{O(s\circ f(n))}+O(d\circ f(n))$. Since $\LL = \LL_K$ and $e'(n)=n^{O(1)}$, $\LL$ belongs to  $\twon/\poly$. Therefore, we conclude that $\twonpd/\poly \subseteq \twon/\poly$. The opposite inclusion is obvious.

(Only if--part)
On the contrary, we assume that $\twon/\poly = \twonpd/\poly$.
We start with an arbitrary  language $K$  taken from $\p/\poly$. Our goal is to prove that $K$ belongs to $\nl/\poly$.  As noted above, $K$ is in $\nauxpdaspti{O(\log{n})}{\infty}/n^{O(1)}$.  Let $f(n)=n$ for all $n\in\nat$. We then take functions $s(|x|)=a\log{|x|}+a$ and $d(|w|)=n^a+a$ ensuring that $K$ is in $\nauxpdaspti{s(|x|)}{\infty}/d(|x|)$ for inputs $x$,  where $a$ is  an appropriate positive constant. Proposition \ref{second-key-proposition}(1) then forces $\LL_K$ to fall in $2^{\log{e(\mathrm{lin}\twonpd)}}/f(n)$ for $e(n)=2^{O(s\circ f(n))}+O(d\circ f(n))$. Since $e(n)=n^{O(1)}$, $\LL$ is contained in $\twonpd/\poly$, and thus   $\LL_K$ falls in $\twon/\poly$ by our assumption.

Now, there is a function $g(n)=n^b+b$ for which $\LL_K$ belongs to  $2^{\log{g(\mathrm{lin}\twon)}}/f(n)$ for a suitable constant $b>0$. By Proposition \ref{second-key-proposition}(2),  we conclude that $K$ is in $\mathrm{NSPACE,\!TIME}(O(\log{e'(|x|)}),\infty)/O(e'(|x|)^2)$, where $e'(|x|)=g\circ f^{-1}(|x|)$. Since $e'(|x|)=|x|^{O(1)}$, $K$ falls in $\mathrm{NSPACE,\!TIME}(O(\log{n}),\infty)/n^{O(1)}$, which is the same as $\nl/\poly$. The inclusion $\p/\poly \subseteq \nl/\poly$ thus follows.
\end{proof}


For the time-bounded complexity classes $\ptime\twodpd/\poly$ and $\ptime\twonpd/\poly$, it was proven in \cite{Yam21} that $\ptime\twodpd/\poly = \ptime\twonpd/\poly$ if and only if $\logdcfl/\poly = \logcfl/\poly$.
However, when underlying two-way pushdown automata run in exponential time, the situation looks completely different.

\begin{proposition}\label{twonpd-with-poly}
$\twonpd/\poly = \twodpd/\poly = \etime\twodpd/\poly = \etime\twonpd/\poly$.
\end{proposition}

The existence of a ceiling bound plays an essential role in the following proof. In sharp contrast to this proposition, it is not yet known whether $\twodpd = \twonpd$ holds.

 \vs{-2}
\begin{proofof}{Proposition \ref{twonpd-with-poly}}
It is obvious that  $\etime\twodpd/\poly \subseteq \etime\twonpd/\poly \subseteq \twonpd/\poly$ and $\etime\twodpd/\poly \subseteq \twodpd/\poly \subseteq \twonpd/\poly$.
It thus suffices to verify that $\twonpd/\poly \subseteq \etime\twodpd/\poly$.
In what follows,  we aim at proving this claim.

For any promise problem family $\LL=\{(L_n^{(+)},L_n^{(-)})\}_{n\in\nat}$ in $\twonpd/\poly$, we take functions $g(n)=n^a+a$ and $f(n)=n^b+b$ for suitable constants $a,b>0$ and a family $\{N_n\}_{n\in\nat}$ of $g(n)$-size 2npda's solving $\LL$.
The function $f$ is assumed to satisfy $L_n^{(+)} \cup L_n^{(-)}\subseteq \Sigma^{\leq f(n)}$ for all $n\in\nat$.
We then define  $e(|w|)=g\circ f^{-1}(|w|)$, which equals $|w|^{O(1)}$ for inputs $w$ of the form $yx$ in $[\Sigma,\#]$.
By Proposition \ref{key-proposition}(1), the language
$K_{\LL}  =\{w=yx\mid \exists n\in\nat[ |w|=f(n)\wedge  y\in\Sigma_{\#}^*\#\cup\{\lambda\} \wedge x\in L(N_n)] \}$ belongs to $\nauxpdaspti{O(\log{e(|w|)})}{\infty}/O(e(|w|)^2)$,
and thus it is in $\nauxpdaspti{O(\log{n})}{\infty}/\poly$.

By an advised version of Cook's result \cite{Coo71}, $K_{\LL}$ also belongs to $\dauxpdaspti{O(\log{n})}{\infty}/\poly$.
Ruzzo \cite{Ruz80} claimed that the time complexity of underlying daux-pda's can be reduced to exponentials (i.e., $2^{n^{O(1)}}$) without changing this complexity classes.
There are functions  $s(|w|)=b\log{|w|}+b$, $t(n,|x|)=2^{(n|x|)^c+c}$, and $d(|w|)=|w|^b+b$ for suitable constants $b,c>0$ such that $K_{\LL}$ falls in $\dauxpdaspti{s(|w|)}{t(f^{-1}(|w|),|x|)}/d(|w|)$ for inputs $w=yx \in [\Sigma,\#]$.
Since  $s\circ f(n) = O(\log{n})$ and $d\circ f(n) = n^{O(1)}$,  $e'(n)=2^{O(s\circ f(n))}+O(d\circ f(n))$ equals $n^{O(1)}$ and $r'(n,|x|)=t(n,|x|)$ equals $2^{(n|x|)^{O(1)}}$.
It is remarked that $L_n^{(+)}$ and $L_n^{(-)}$ satisfy the following relationships:
$L_n^{(+)}=\{x\mid [ |x| <  f(n) \wedge 1^{f(n)-|x|-1}\# x\in K_{\LL}] \vee [ |x|=f(n)\wedge x\in K_{\LL}] \}$ and $L_n^{(-)}=\{x\mid [ |x| < f(n) \wedge 1^{f(n)-|x|-1}\# x\notin K_{\LL}] \vee [|x|=f(n)\wedge x\notin K_{\LL}] \}$.   
Proposition \ref{key-proposition}(2) then implies that $\LL$ belongs to $O(r'(n,|x|))$time-$2^{\log{e'(\mathrm{lin}\twodpd)}}/f(n)$, which is included in $\etime\twodpd/\poly$.
Therefore, $\LL$ falls in $\etime\twodpd/\poly$.
\end{proofof}

Proposition \ref{twonpd-with-poly}  instantly yields the following equalities.

\begin{corollary}
$\twoupd/\poly = \twonpd/\poly = \co \twonpd/\poly$, where $\twoupd$ is the ``unambiguous'' version of $\twonpd$.
\end{corollary}

\subsection{Complexity of Superpolynomial-Size Automata Families}\label{sec:exp-size-automata}

We have studied in Section \ref{sec:poly-size-automata} descriptional/computational complexity of promise problem families solvable by ``polynomial-size'' automata families.
In contrast   here, we wish to explore the roles of various nonuniform automata families of ``superpolynomial size''.

We start with the following easy claim. For a language $L$ over an alphabet $\Sigma$ and a string $x$ in $\Sigma^*$, the notation $L(x)$ denotes either the acceptance (1) or the rejection (0) of $x$ to $L$.

\begin{proposition}\label{ALL-equivalent}
(1) $\mathrm{ALL}/\poly = 2^{\twod}/\poly = 2^{\twon}/\poly = 2^{\twodpd}/\poly = 2^{\twonpd}/\poly$.

(2) $\mathrm{ALL}/\poly = \ptime2^{\twod}/\poly =  \ptime2^{\twodpd}/\poly$. 
\end{proposition}

\begin{proof}
It suffices to prove that $\mathrm{ALL}/\poly \subseteq \ptime2^{\twod}/\poly$ because $\ptime2^{\twod}/\poly$ is the lowest class in complexity than the other complexity classes listed in the proposition.
We take any promise problem family $\LL=\{(L_n^{(+)},L_n^{(-)})\}_{n\in\nat}$ in $\mathrm{ALL}/\poly$ over an alphabet $\Sigma$. There is a polynomial $p$ satisfying  $L_n^{(+)}\cup L_n^{(-)}\subseteq \Sigma^{\leq p(n)}$ for all $n\in\nat$.
For each index $n\in\nat$, we wish to define a 2dfa $M_n$ to solve $(L_n^{(+)},L_n^{(-)})$.   The set $Q_n$ of $M_n$'s inner states consists of all tuples of the from $(j,s_1s_2 \cdots s_j,b)$ with $j\in[p(n)]$, $s_1,\ldots,s_j\in\Sigma$, and $b=L_n^{(+)}(s_1s_2\cdots s_j)$.
On valid input $x$ of the form $s_1s_2\cdots s_m$ with $s_i\in\Sigma$ for all $i\in[m]$, we process the input symbols $s_1,s_2,\ldots,s_m$ one by one as we change inner states as $(1,s_1,b_1),(2,s_1s_2,b_2),\ldots,(m,s_1s_2\cdots s_m,b_m)$ in this order. If we read $x$ completely, then we output $b_m$ and halt. Otherwise, we immediately reject $x$. Obviously, $M_n$ can solve $(L_n^{(+)},L_n^{(-)})$.
The runtime of $M_n$ on $x$ is proportional to $n|x|$.
Since  $|Q_n|\leq 2^{O(p(n))}$, we conclude that  $\LL$ belongs to $\ptime2^{\twod}/\poly$.
\end{proof}

Proposition \ref{ALL-equivalent} suggests that exponential-size automata families turn out to be too powerful.  This fact forces us to look into automata families of ``sub-exponential'' size.


Theorem \ref{Cook-proof} is extendable to $2^{O(\log^k{n})}$-size automata families for various values of $k\geq2$.

\begin{theorem}\label{log-hierarchy-relation}
For any integer $k\geq2$, $\ptime2^{\log^k\twon}/\poly \subseteq \ptime2^{\log^k\twonpd}/\poly \subseteq 2^{\log^{k+1}\twod}/\poly$.
\end{theorem}

\begin{proof}
The first inclusion is obvious. Thus, we focus on the second inclusion between  $\ptime2^{\log^k\twonpd}/\poly$ and $2^{\log^{k+1}\twod}/\poly$.
To show this, we intend to verify that the following two statements. (1) If $\nauxpdaspti{O(\log^kn)}{n^{O(1)}}/2^{O(\log^kn)} \subseteq \mathrm{DSPACE,\!TIME}(O(\log^{k+1}n),\infty)/2^{O(\log^kn)}$, then $\ptime2^{\log^k\twonpd}/\poly \subseteq 2^{\log^{k+1}\twod}/\poly$. (2) $\nauxpdaspti{O(\log^kn)}{n^{O(1)}}/2^{O(\log^kn)}$ is included in $\mathrm{DSPACE,\!TIME}(O(\log^{k+1}n),\infty)/2^{O(\log^kn)}$.
These statements together lead to the desired inclusion.

(1) Now, we assume that  $\nauxpdaspti{O(\log^kn)}{n^{O(1)}}/2^{O(\log^kn)}$ is included in $\mathrm{DSPACE,\!TIME}(O(\log^{k+1}n),\infty)/2^{O(\log^kn)}$.
Take an arbitrary promise problem family $\LL=\{(L_n^{(+)},L_n^{(-)})\}_{n\in\nat}$ in $\ptime2^{\log^k\twonpd}/\poly$ over an alphabet $\Sigma$. We also assume that $\LL$ is of $(n^a+a)$-ceiling and also we set  $f(n)=n^a+a$ for all $n\in\nat$.
There are two functions  $g(n)=2^{c\log^kn+c}$ and $u(n,|x|)= (n|x|)^a+a$ for  
constants $a,c>0$ and a family $\{N_n\}_{nb\in\nat}$ of $g(n)$-size 2npda's solving $\LL$ in time $u(n,|x|)$ on inputs $x$.
We recall the language
$K_{\LL}  =\{w\mid w=yx\in[\Sigma,\#], \exists n\in\nat[ |w|=f(n)\wedge x\in L(N_n)] \}$.
Note that $e(|w|)=g\circ f^{-1}(|w|) = 2^{O(\log^k|w|)}$ and $r(|w|,|x|)=u(f^{-1}(|w|),|x|)\log{e(|w|)}=(|x||w|)^{O(1)}$.
Proposition \ref{key-proposition}(1) then provides an naux-pda $N$ for $K_{\LL}$ running in
time $O(r(|w|,|x|))$ using space $O(\log{e(|w|)})$ with $O(e(|w|)^2)$-size advice for inputs $w$ of the form $yx\in [\Sigma,\#]$.
This implies that $K_{\LL}$ falls in $\nauxpdaspti{O(\log^kn)}{n^{O(1)}}/2^{O(\log^kn)}$.

By our assumption, we immediately obtain $K_{\LL}\in \mathrm{DSPACE,\!TIME}(O(\log^{k+1}n),\infty)/2^{O(\log^kn)}$. Hence, there exist two functions $s(n)=c\log^{k+1}n+c$  and $d(n)=2^{a\log^kn+a}$ for constants $a,c>0$ and  a DTM $M$ that recognizes $K_{\LL}$ using space $s(|w|)$ with the help of $d(|w|)$-size advice.
Note that $e'(n)=2^{O(s\circ f(n))}+O(d\circ f(n)) =2^{O(\log^{k+1}n)}$ because $s\circ f(n)=O(\log^{k+1}n)$, and $d\circ f(n)=2^{O(\log^kn)}$. We then apply Proposition \ref{key-proposition}(2) to conclude that $\LL$ belongs to $2^{g(\mathrm{lin}\twod)}/f(n)$, which is included in $2^{\log^{k+1}\twod}/\poly$.

(2)
As Ruzzo \cite{Ruz80} demonstrated, $\nauxpdaspti{O(\log^kn)}{n^{O(1)}}$ is included in $\mathrm{DSPACE,\!TIME}(O(\log^{k+1}n),\infty)$. We remark that his proof can be carried out even in the presence of $2^{O(\log^kn)}$-size advice since all accesses by  an naux-pda to advice can be simulated by an appropriate DTM.
\end{proof}

Theorem \ref{log-hierarchy-relation} indicates that the complexity of $\ptime2^{\log^k\twod}/\poly$ is relatively high. How high is it? Is it higher than, say, $\twodpd/\poly$? As an  answer to this question, we demonstrate that  $\ptime2^{\log^k\twod}/\poly$ is not ``small'' enough to be included in $\twonpd/\poly$.

\begin{theorem}\label{ptime-log-k-subset}
$\ptime2^{\log^2\twod}/\poly \nsubseteq \twonpd/\poly$.
\end{theorem}

\begin{proof}
To verify the correctness of the theorem, we plan to  claim that (1)  $\ptime2^{\log^2\twod}/\poly \subseteq \twonpd/\poly$ implies $\mathrm{SC}^2/2^{O(\log^2{n})} \subseteq \p/\poly$ and (2) $\mathrm{SC}^2/2^{O(\log^2{n})}\nsubseteq \p/\poly$. Combining (1) and (2) immediately implies the desired theorem.

(1) Assuming that $\ptime2^{\log^2\twod}/\poly\subseteq \twonpd/\poly$, we consider   any language $K$ in $\mathrm{SC}^2/2^{O(\log^2n)}$ over an alphabet $\Sigma$.
We first define $f(n)=n$ for all $n\in\nat$.
We then take  functions $s(|w|)= a\log^2{|x|}+a$, $t(n,|x|)=(n|x|)^b+b$, and $d(|x|)=2^{c\log^2|x|+c}$ for suitable  constants $a,b,c>0$ so that  $K$ falls in $\mathrm{DSPACE,\!TIME}(s(|x|),t(f^{-1}(|x|),|x|))/d(|x|)$ for inputs $x$.
Let us consider the language $\LL_K=\{(L_n^{(+)},L_n^{(-)})\}_{n\in\nat}$ defined by
setting $L_n^{(+)}=\{x\in\Sigma^*\mid |x|=f(n), x\in K\}$ and $L_n^{(-)}=\{x\in\Sigma^*\mid |x|=f(n),x\notin K\}$.
Since $s\circ f(n)= O(\log^2n)$ and $d\circ f(n)= 2^{O(\log^2n)}$,
it follows that $e(n)=2^{O(s\circ f(n))}+O(d\circ f(n)) = 2^{O(\log^2n)}$. Moreover, let $r(n,|x|)=t(n,|x|)$. Thus,  Proposition \ref{second-key-proposition}(1) places  $\LL_K$ in
$O(r(n,|x|))$time-$2^{\log{e(\mathrm{lin}\twonpd)}}/f(n)$ for inputs $x$, and thus $K_{\LL}$ is further contained in $\ptime2^{\log^2\twod}/\poly$.

The language $\LL_K$ also falls in $\twonpd/\poly$ by our assumption. Let $g(n)=n^{k}+k$ for a constant $k>0$ and assume that $\LL_K\in 2^{g(\mathrm{lin}\twonpd)}/f(n)$.
Since  $e'(|x|)=g\circ f^{-1}(|x|) = |x|^{O(1)}$, by Proposition \ref{second-key-proposition}(2), $K$ belongs to $\nauxpdaspti{O(\log{e'(|x|)})}{\infty}/O(e'(|x|)^2)$, which is included in
$\nauxpdaspti{O(\log{n})}{\infty}/\poly$. The latter class coincides with $\p/\poly$ by an advice analogue of Cook's result \cite{Coo71}.

(2)
The following argument proceeds by using a diagonalization technique.
To obtain the desired separation of $\mathrm{SC}^2/2^{O(\log^2n)}\nsubseteq \p/\poly$,
we fixate both an input alphabet and an advice alphabet to be $\Sigma=\{0,1\}$.
Firstly, we enumerate all DTMs as  $M_1,M_2,\ldots$ and, for each $m\in\nat$, enumerate all strings $x_{1,m},x_{2,m},\ldots$ and all advice strings $a_{1,m},a_{2,m},\ldots$ over $\Sigma$ in the lexicographic order, where $|x_{j,m}|=|a_{j,m}|=m$ with $1\leq j\leq 2^m$.

Given a triplet $(i,k,l)$, we express the outcome (0 or 1) of $M_i$ on input $x$ together with advice string $a_{k,|x|^l}$ running within time $|x|^l$ as $M_i(x; a_{k,|x|^l})$ if it exists; otherwise, we automatically set $M_i(x;a_{k,|x|^l})$ to be $\lambda$, where we conventionally identify
1 and 0 with ``acceptance'' and ``rejection'', respectively.

Hereafter, we wish to define an advice function $h$ and a language $L$. We inductively define $A_{j,m}$ as follows. Let $A_{0,m}=\Sigma^{\leq m^{\log{m}}}$. For each $j\geq1$ and a bit $b\in\{0,1\}$, we set $g_{i,l}(x_{j+1,m},b) = |\{z\in A_{j,m}\mid M_i(x_{j+1,m},z)=b\}|$. We then define $\alpha_{j+1,m}$ to be $b$ if $g_{i,l}(x_{j+1,m},b)\leq g_{i,l}(x_{j+1,m},\bar{b})$, where $\bar{b}=1-b$.  We then define $A_{j+1,m}=\{z\in A_{j,m}\mid M_i(x_{j+1,m},z)=\alpha_{j+1,m}\}$. For readability, we write $\overline{A_{j+1,m}}=A_{j,m}-A_{j+1,m}$. It then follows that $|A_{j+1,m}|\leq \frac{1}{2}|A_{j,m}|$ for any$j\geq0$. Now, we take the smallest natural number $r_0$ satisfying $|A_{r_0,m}|=0$.

We first assume the existence of efficient encoding function $\pair{\cdot}:\nat^3\to\nat$.
Given each $m\in\nat$, we set $h(m)$ to be $1^m\#1^i\# 1^l\# x_{1,m}\alpha_{a,m}\# x_{2,m}\alpha_{2,m}\# \cdots \# x_{r_0,m}\alpha_{r_0,m}\# 0^c$  $m=\pair{i,l,t}$
for $i,l\leq \log{m}$ and $t\in\nat$, where $c$ is an appropriate number so that $|h(m)|=5m^{\ceilings{\log{m}}+1}$.  This comes form the fact that $|h(m)|=(m+1)+(i+1)+(l+1)+ (m+1)r_0 + c \leq 4m^{\ceilings{\log{m}}+1}+c$.
In particular, it follows that, for each $m$ and $j$, $x_{j,m}\alpha_{j,m}$ equals $x_{j,m}1$ if and only if $M_i$ accepts $x_{j,m}$  using $a_{k,m^l}$ within $m^l$ steps.

Next,  we define the example language $L$ to be the set $\{x_{j,m} \mid j,m\in\nat, \text{ $\alpha_{j,m}$ in $h(m)$ equals $0$ }\}$. We then wish to prove that $L\notin \p/\poly$.
Now, assuming otherwise, we take a number $m\in\nat^{+}$ and a pair $(i,l)$ such that $i,l\leq \log m$ and $L=L(M_i,g')$ for a suitable advice function $g'$. We can assume that $g'(m)$ always has the value of the form $a_{k_m,m^l}$ for an appropriate number $k_m$. Note that, for any $m$ and $j$, $x_{j,m}\in L$ iff $M_i(x_{j,m},a_{k_m,m^l})=1$.

Now, we choose a number $j_0$ satisfying that, for all $j<j_0$, $a_{k_m,m^l}\in \overline{A_{j,m}}$ and $a_{k_m,m^l}\in A_{j_0,m}$. Such a number always exists since $A_{j+1,m}\subseteq A_{j,m} \subseteq A_{0,m}=\Sigma^{\leq m^{\log{m}}}$ for all $j\geq0$.
By the choice of $j_0$, since  $M_i(x_{j_0,m},a_{k_m,m^l})=\alpha_{j_0,m}$, we obtain $\alpha_{j_0,m}=1$. However, by the definition of $L$, $x_{j_0,m}\in L$ implies $\alpha_{j_0,m}=0$. This is a contradiction.
\end{proof}

From Theorem \ref{ptime-log-k-subset}, we immediately obtain the following corollary.

\begin{corollary}
(1) $\ptime2^{\log^2\twon}/\poly \nsubseteq \twon/\poly$.
(2) $\ptime2^{\log^2\twod}/\poly \nsubseteq \twod/\poly$.
(3) $\ptime2^{\log^2\twonpd}/\poly \nsubseteq \ptime\twonpd/\poly$.
(4) $\ptime2^{\log^2\twodpd}/\poly \nsubseteq \ptime\twodpd/\poly$.
\end{corollary}


Theorem \ref{ptime-log-k-subset} suggests that $\ptime2^{\log^k\twon}/\poly$ is not ``small'',  compared to $\twonpd/\poly$. Now, we wonder if   $\ptime2^{\log^k\twon}/\poly$ collapses down to $\ptime2^{\log^k\twod}/\poly$.
It then turns out by the following proposition that they may not coincide because the collapse of them is related to answering the $\mathrm{NSC}^{k}\subseteq$?$\mathrm{SC}^{k}/2^{O(\log^kn)}$ question, which is widely anticipated to be false.

\begin{proposition}\label{log-k-twod-SC}
For each integer with $k\geq2$,
$\ptime2^{\log^k\twod}/\poly = \ptime2^{\log^k\twon}/\poly$ if and only if $\mathrm{NSC}^k \subseteq \mathrm{SC}^k/2^{O(\log^kn)}$.
\end{proposition}

\begin{proof}
(If--part)
Our assumption is that $\mathrm{NSC}^k\subseteq \mathrm{SC}^k/2^{O(\log^kn)}$. This inclusion relation is logically equivalent to $\mathrm{SC}^k/2^{O(\log^k{n})} = \mathrm{NSC}^k/2^{O(\log^k{n})}$, which we use in the subsequent argument.
Now, we begin with an arbitrary promise problem family
$\LL=\{(L_n^{(+)},L_n^{(-)})\}_{n\in\nat}$ in $\ptime2^{\log^k\twon}/\poly$ over an alphabet $\Sigma$.
Take a polynomial $f$ for which $\LL$ has $f(n)$-ceiling.
For a suitable constant $a\in\nat^{+}$, we set $g(n)= 2^{a\floors{\log^k{n}}+a}$ and $u(n,|x|) = (n|x|)^a+a$ for all $n\in\nat$ and assume that $\LL$ is in $u(n,|x|)$time-$2^{\log{g(\mathrm{lin}\twon)}}/f(n)$; namely,
a certain family $\{N_n\}_{n\in\nat}$ of $2^{g(n)}$-size 2dfa's solves $\LL$ in time  $u(n,|x|)$ for inputs $x$.
By Proposition \ref{key-proposition}(1),  the language
$K_{\LL}  =\{w\mid w=yx\in[\Sigma,\#], \exists n\in\nat[ |w|=f(n)\wedge  x\in L(N_n)] \}$ is solvable by an NTM $N$ in time $O(r(|w|,|x|)$
using space $O(\log{e(|w|)})$ together with advice of size $O(e(|w|)^2)$ for inputs $w=yx\in[\Sigma,\#]$, where $e(|w|)=g\circ f^{-1}(|w|)$ and $r(|w|,|x|)= u(f^{-1}(|w|),|x|)\log{e(|w|)}$.
Since $e(|w|)=2^{O(\log^k|w|)}$ and $r(|w|,|x|)= (|x||w|)^{O(1)}$,  we derive the membership of $K_{\LL}$ to $\mathrm{NSPACE,\!TIME}(O(\log^k|w|),|w|^{O(1)})/2^{O(\log^k|w|)}$ for inputs $w$. This class is included in $\mathrm{NSC}^k/2^{O(\log^kn)}$.

By our assumption, we obtain $K_{\LL}\in\mathrm{SC}^k/2^{O(\log^kn)}$, which implies that $K_{\LL}$ is in $\mathrm{DSPACE,\!TIME}(s(|w|),t(f^{-1}(|w|),|x|))/d(|w|)$ for inputs $w=yx$ in $[\Sigma,\#]$ with appropriate functions
$s(|w|)=  c\floors{\log^k|w|}+c$, $t(n,|x|)=(n|x|)^b+b$, and $d(|w|)=  2^{c\floors{\log^k|w|}+c}$, where $b,c$ are appropriate positive constants.  
It then follows by Proposition \ref{key-proposition}(2) that
$\LL$ is in $O(r'(n,|x|))$time-$2^{\log{e'(\mathrm{lin}\twod)}}/f(n)$ for inputs $x$,
where
$e'(n)=2^{O(s\circ f(n))}+ O(d\circ f(n))$ and $r'(n,|x|)=t(n,|x|)$.
Note that
$L_n^{(+)} = \{x\mid [|x|<  f(n) \wedge 1^{f(n)-|x|-1}\# x\in K] \vee [|x|=f(n) \wedge x\in K]\}$ and $L_n^{(-)} = \{x\mid [|x|<  f(n) \wedge 1^{f(n)-|x|-1}\# x\in \overline{K}]\vee [|x|=f(n)\wedge x\in \overline{K}]\}$.
Since $s\circ f(n) =  O(\log^k{n})$ and $d\circ f(n)=2^{O(\log^kn)}$, we obtain $e'(n)= 2^{O(\log^kn)}$.
Therefore, $\LL$ belongs to $\ptime2^{\log^k\twod}/\poly$, as requested.


(Only if--part)
On the contrary, assuming that $\ptime2^{\log^k\twod}/\poly = \ptime2^{\log^k\twon}/\poly$, we aim at deriving the inclusion of  $\mathrm{NSC}^k\subseteq \mathrm{SC}^k/2^{O(\log^kn)}$.  We set  $f(n)=n$ for all $n\in\nat$.
Let us consider any language $K$ in $\mathrm{NSC}^k$ over an alphabet $\Sigma$ and take an NTM $N$ for $K$ running in $(|x|f^{-1}(|x|))^b+b$ time using at most $a\log^k{|x|}+a$ space with $(|x|^a+a)$-size advice for inputs $x$, where $a,b$ are appropriate positive constants.
For convenience, we set $s(|x|)=a\floors{\log^k|x|}+a$, $t(n,|x|)=(n|x|)^b+b$, and $d(|x|)=|x|^a+a$  for all $n\in\nat^{+}$.
Proposition \ref{second-key-proposition}(1) then ensures the existence of a family $\{N_n\}_{n\in\nat}$ of $e(n)$-size $O(1)$-choice 2nfa's that solves $\LL=\{(L_n^{(+)},L_n^{(-)})\}_{n\in\nat}$ in time $r(n,|x|)$ for inputs $x$, where $L_n^{(+)}=\{x\in\Sigma^* \mid |x|=f(n), x\in K\}$, $L_n^{(-)}=\{x\in\Sigma^* \mid |x|=f(n), x\notin K\}$, and $e(n)=2^{O(s\circ f(n))}+O(d\circ f(n))$, which equals $2^{O(\log^kn)}$, and $r(n,|x|)=t(n,|x|)$.
It thus follows that the family $\LL_K=\{(L_n^{(+)},L_n^{(-)})\}_{n\in\nat}$ belongs to $\ptime2^{\log^k\twon}/\poly$.

Our assumption therefore forces  $\LL_K$ to fall in $\ptime2^{\log^k\twod}/\poly$; that is, there is a family $\{M_n\}_{n\in\nat}$ of $2^{\log{g(n)}}$-size 2dfa's solving $\LL_K$ within $u(n,|x|)$ time for inputs $x$.
Proposition \ref{second-key-proposition}(2) then yields a DTM $M$ that recognizes $K$ in $O(r'(f^{-1}(|x|),|x|))$ time using $O(\log{e'(|x|)})$ space with advice of size $O(e'(|x|)^2)$ for inputs $x$, where $e'(|x|)=g\circ f^{-1}(|x|)=2^{O(\log^k|x|)}$ and $r'(n,|x|)=u(n,|x|)\log{e'(|x|)} = (n|x|)^{O(1)}$. These facts imply that $K$ belongs to $\mathrm{SC}^k/2^{O(\log^kn)}$.
\end{proof}


Under no presence of time bounds, Savitch's theorem \cite{Sav70} shows an inclusion of $\mathrm{NSPACE,\!TIME}(O(\log^kn),\infty)$ within $\mathrm{DSPACE,\!TIME}(O(\log^{2k}n),\infty)$. An advised variant of this theorem helps us prove the following inclusion relationships.

\begin{proposition}\label{coincide-log-k-twod}
Let $k$ denote any positive integer.
(1)  $2^{\log^k\twon}/\poly \subseteq 2^{\log^{2k}\twod}/\poly$.
(2)  $2^{\log^k\twonpd}/\poly \subseteq 2^{\log^{2k}\twodpd}/\poly$.
\end{proposition}

\begin{proof}
(1) We begin with an arbitrary family $\LL=\{(L_n^{(+)},L_n^{(-)})\}_{n\in\nat}$ in $2^{\log^k\twon}/\poly$ over an alphabet $\Sigma$. Take a constant $c>0$ satisfying $L_n^{(+)}\cup L_n^{(-)}\subseteq \Sigma^{\leq n^c+c}$ for all $n\in\nat$. Take also a family $\{N_n\}_{n\in\nat}$ of $2^{a\log^kn+a}$-size 2nfa's that solves $\LL$ for a suitable constant $a>0$. Now, we set $f(n)=n^c+c$ and $g(n)=2^{a\log^kn+a}$
and define $K_{\LL}$ to be the set
$K_{\LL}  =\{w=yx\mid \exists n\in\nat[ |w|=f(n)\wedge  y\in\Sigma_{\#}^*\#\cup\{\lambda\} \wedge x\in L(N_n)] \}$.
We apply Proposition \ref{key-proposition}(1).
Note that $e(n)=g\circ f^{-1}(n)$ equals $2^{O(\log^kn)}$.
The language $K_{\LL}$ then falls in $\mathrm{NSPACE,\!TIME}(O(\log{e(|w|)}),\infty)/O(e(|w|)^2)$, which is included in
$\mathrm{NSPACE,\!TIME}(O(\log^k|w|),\infty)/2^{O(\log^k|w|)}$.
By an advised variant of Savitch's theorem \cite{Sav70},
$ \mathrm{NSPACE,\!TIME}(O(\log^k{|w|}),\infty)/2^{O(\log^k{|w|})}$ is included in $\mathrm{DSPACE,\!TIME}(O(\log^{2k}{|w|}),\infty)/2^{O(\log^k{|w|})}$,  and thus
$K_{\LL}$ also belongs to $\mathrm{DSPACE,\!TIME}(O(\log^{2k}n),\infty)/2^{O(\log^kn)}$.
Now, we take a DTM $M$ that recognizes $K_{\LL}$ using space $b\log^{2k}|w|+b$ together with advice of size $2^{a'\log^k|w|+a'}$ for inputs $w$ of the form $yx$ in $[\Sigma,\#]$, where $a',b$ are appropriate positive constants.
Let  $s(|w|)=b\log^{2k}|w|+b$ and $d(|w|)=2^{a'\log^k|w|+a'}$.
We apply Proposition \ref{key-proposition}(2).
Note that $e'(n)=2^{O(s\circ f(n))} + O(d\circ f(n))$ equals $2^{O(\log^{2k}n)}$.
We then conclude that $\LL$ falls in $2^{\log{e'(\mathrm{lin}\twod)}}/f(n)$, which is included in
$2^{\log^k\twod}/\poly$. Therefore, it follows that $2^{\log^k\twon}/\poly \subseteq 2^{\log^{2k}\twod}/\poly$.

(2)
 In the argument of (1), we need to replace $2^{\log^k\twon}$ and $\mathrm{NSPACE,\!TIME}(O(\log^kn),\infty)/2^{O(\log^kn)}$ respectively by $2^{\log^k\twonpd}/\poly$ and $\nauxpdaspti{O(\log^kn)}{\infty}/2^{O(\log^kn)}$. Similarly, $2^{\log^{2k}\twod}/\poly$ and $\mathrm{DSPACE,\!TIME}(O(\log^{2k}n),\infty)/2^{O(\log^kn)}$ are replaced by $2^{\log^{2k}\twodpd}/\poly$ and $\dauxpdaspti{O(\log^{2k}n)}{\infty}/2^{O(\log^kn)}$, respectively. We then use the advised variant of Cook's result \cite{Coo71} in place of the advised variant of Savitch's theorem.
\end{proof}

In certain circumstances, it is possible to separate two complexity classes without relying on any unproven assumptions. As such a case, the following proposition points out a gap between $2^{\log^k\twodpd}/\poly$ and $2^{\twodpd}/\poly$ with no hidden assumptions.

\begin{proposition}\label{separation-ALL}
For any integer $k\geq2$, $2^{\log^k\twodpd}/\poly \neq 2^{\twodpd}/\poly$.
\end{proposition}

\begin{proof}
We fix $k\geq2$ arbitrarily and set $\Sigma=\{0,1\}$. In addition, we set $\Sigma_{\#}=\Sigma\cup\{\#\}$. We then  enumerate  all functions of the form $\mu_a(n)=a\log^kn+a$ for an arbitrary number $a\in\nat^{+}$ and, for each pair $(a,n)$, enumerate  all 2npda's $M^{(a,n})_{i}$ satisfying  $|Q^{(n)}_{a,i}|\leq 2^{\mu_a(n)}$ for all $i\in\nat$, where $Q^{(n)}_{a,i}$ is a set of inner states of $M^{(a,n)}_{i}$.
We then define a family $\LL_k=\{(L_n^{(+)},L_n^{(-)})\}_{n\in\nat}$ of promise problems by setting $L_n^{(+) }=\{w\mid w=1^i\#1^a\#x\mid  |w|\leq n, x\in\Sigma^*, \text{ $M^{(a,n)}_{i}$ does not accept $w$ }\}$ and $L_n^{(-)}= \{w\mid w=1^i\#1^a\#x,  |w|\leq n, x\in\Sigma^*, \text{ $M^{(n)}_{a,i}$ does not reject $w$ }\}$. Since  $\LL_k$ has a polynomial ceiling, it obviously falls in $\mathrm{ALL}/\poly$, which is the same as $2^{\twodpd}/\poly$ by Proposition \ref{ALL-equivalent}(1).

Now, we wish to prove that $\LL_k\notin 2^{\log^k\twodpd}/\poly$. Toward a contradiction, we assume that $\LL_k$ is in $2^{\log^k\twodpd}/\poly$.
There exist a constant $a>0$ and a sequence $(i_1,i_2,\ldots)$ for which
the family $\{M^{(a,n)}_{i_n}\}_{n\in\nat}$ of $2^{\mu_a(n)}$-size 2dpda's solves $\LL_k$.  Note that $L_n^{(+)} \subseteq L(M^{(a,n)}_{i_n})$ and $L_n^{(-)} \subseteq \Sigma^*-L(M^{(a,n)}_{i_n})$ . Thus, it follows that, for all valid inputs $w$ of the form $1^{i_n}\#1^t\#x$, $w\notin L_n^{(+)}$ if and only if $w\in L(M^{(n)}_{a,i_n})$. This is a clear contradiction. Hence, $\LL_k$ does not belong to $2^{\log^k\twodpd}/\poly$.
\end{proof}


In the end of this subsection, we study the behavior of another types of superpolynomial-size automata families.
In a way similar to Theorem \ref{log-k-twod-SC}, we can prove the following equivalence relations.

\begin{theorem}\label{dspace-2nfa-relation}
For each $k\in\nat$ with $k\geq2$, $\ptime2^{\sqrt[k]{\mathrm{lin}\twod}}/\poly = \ptime2^{\sqrt[k]{\mathrm{lin}\twon}}/\poly$ if and only if $\mathrm{NSPACE,\!TIME}(O(n^{1/k}),n^{O(1)})\subseteq \mathrm{DSPACE,\!TIME}(O(n^{1/k}),n^{O(1)})/2^{O(n^{1/k})}$.
\end{theorem}

\begin{proof}
(If--part) Assume that $\mathrm{NSPACE,\!TIME}(O(n^{1/k}),n^{O(1)})$ is included in $\mathrm{DSPACE,\!TIME}(O(n^{1/k}),n^{O(1)})/2^{O(n^{1/k})}$.
Take an arbitrary promise problem family $\LL=\{(L_n^{(+)},L_n^{(-)})\}_{n\in\nat}$ in $\ptime2^{\sqrt[k]{\mathrm{lin}\twon}}/\poly$.
There is a suitable function $f(n)=n^p+p$ for a constant $p\geq1$ satisfying $L_n^{(+)}\cup L_n^{(-)} \subseteq \Sigma^{<  f(n)}$ for all $n\in\nat$.
Now, we take a family $\NN=\{N_n\}_{n\in\nat}$ of $g(n)$-size 2nfa's $N_n$ that solves $\LL$ in time $u(n,|x|)$, where $g(n)=2^{an^{1/k}+a}$ and $u(n,|x|)=(n|x|)^b+b$ for appropriate constants $a,b>0$.
Proposition \ref{key-proposition}(1) asserts the existence of an NTM $N$ that recognizes the language
$K_{\LL}  =\{w\mid w=yx\in[\Sigma,\#], \exists n\in\nat[ |w|=f(n)\wedge  x\in L(N_n)] \}$  in time $O(r(|w|,|x|))$ using space $O(\log{e(|w|)})$ with $O(e(|w|)^2)$-size advice for inputs $w$ of the form $yx$ in $[\Sigma,\#]$, where $e(|w|)= g\circ f^{-1}(|w|) = 2^{O(|w|^{1/kp})}\subseteq 2^{O(|w|^{1/k})}$ and $r(|w|,|x|)=u(f^{-1}(|w|),|x|)\log{e(|w|)} = (|x||w|)^{O(1)}$.
This implies that $K_{\LL}$ is in   $\mathrm{NSPACE,\!TIME}(O(|w|^{1/k}),|w|^{O(1)})/2^{O(|w|^{1/k})}$.

By our assumption, we conclude that $K_{\LL}$ belongs to  $\mathrm{DSPACE,\!TIME}(O(|w|^{1/k}),|w|^{O(1)})/2^{O(|w|^{1/k})}$. We choose $s(|w|)= b'|w|^{1/k}+b'$, $t(n,|x|) = (n|x|)^{a'}+{a'}$, and $d(|w|)=2^{b'|w|^{1/k}+b'}$ for constants $a',b'>0$ so that $K_{\LL}$ is in $\mathrm{DSPACE,\!TIME}(s(|w|),t(f^{-1}(|w|),|x|))/d(|w|)$ for inputs $w=yx$ in $[\Sigma,\#]$.
By Proposition \ref{key-proposition}(2), we take a family $\{M_n\}_{n\in\nat}$ of $2^{\log{e'(n)}}$-size 2dfa's solving $\LL$ in time $r'(n,|x|)$ for inputs $x$, where $e'(n)=2^{O(s\circ f(n))}+O(d\circ f(n))$ and $r'(n,|x|)=t(n,|x|)$.  Since $s\circ f(n)=O(|w|^{1/kp})$ and $d\circ f(n)=2^{O(|w|^{1/kp})}$, we conclude that $\LL$ is in $n^{O(1)}$time-$2^{(\mathrm{lin}\twod)^{1/kp}}/f(n)$, which is included in $\ptime2^{\sqrt[k]{\mathrm{lin}\twod}}/\poly$.

Only if--part)
We start this proof by assuming $\ptime2^{\sqrt[k]{\mathrm{lin}\twon}}/\poly = \ptime2^{\sqrt[k]{\mathrm{lin}\twod}}/\poly$.
Take any language $K$ in
$\mathrm{NSPACE,\!TIME}(O(|x|^{1/k}),|x|^{O(1)})/2^{O(|x|^{1/k})}$ over an alphabet $\Sigma$ and an NTM $N$ recognizes $K$ in time at most $|x|^{a}+a$ using space $b|x|^{1/k}+b$ with $2^{c|x|^{1/k}+c}$-size advice,
where $a,b,c$ are positive constants. Let $f(n)=n$ for all $n\in\nat$. Let $s(|x|)=b|x|^{1/k}+b$, $t(n,|x|)=n(|x|^a+a)$, and $d(|x|)=2^{c|x|^{1/k}+c}$ for all $x$.
We convert $K$ to the family $\LL_K =\{(L_n^{(+)},L_n^{(-)})\}_{n\in\nat}$ of promise problems by setting $L_n^{(+)} = \{x\in \Sigma^* \mid |x|=f(n), x\in K\}$ and $L_n^{(-)} = \{x\in\Sigma^* \mid |x|=f(n), x\notin K\}$ for each index $n\in\nat$.
By Proposition \ref{second-key-proposition}(1), there exists a family $\{N_n\}_{n\in\nat}$ of $e(n)$-size 2nfa's solving $\LL_K$ in $O(r(n,|x|))$ time, where $e(n)=2^{O(s\circ f(n))}+O(d\circ f(n)) = 2^{O(\log^{1/k}n)}$ and $r(n,|x|)=t(n,|x|)$. This means that $\LL_K$ falls in $\ptime2^{\sqrt[k]{\mathrm{lin}\twon}}/\poly$.

By our assumption, we obtain $\LL_K\in \ptime2^{\sqrt[k]{\mathrm{lin}\twod}}/\poly$.  From this, there is a family $\{M_n\}_{n\in\nat}$ of $g(n)$-size 2dfa's that solves $\LL_K$ in time  $u(n,|x|)$, where $g(n)=2^{b'n^{1/k}+b'}$ and $u(n,|x|)=(n|x|)^{a'}+a'$ for appropriate constants  $a',b'>0$.
By Proposition \ref{second-key-proposition}(2), we take a DTM $M$ that recognizes $K$ in  time $O(r'(f^{-1}(|x|),|x|))$ using space $O(\log{e'(|x|)})$ with  advice of size $O(e'(|x|)^2)$. Since $e'(|x|)= g\circ f^{-1}(|x|) = 2^{O(|x|^{1/k})}$, $r'(n,|x|)=u(n,|x|)\log{e'(|x|)} = (n|x|)^{O(1)}$,  $K$ belongs to $\mathrm{DSPACE,\!TIME}(O(|x|^{1/k}),|x|^{O(1)})/2^{O(|x|^{1/k})}$ for inputs $x$.
\end{proof}

Theorem \ref{dspace-2nfa-relation} can be further generalized by taking an arbitrary real number $\varepsilon\in[0,1)$ in place of $1/k$ used for $n^{1/k}$ in the theorem. We remark that the union $\bigcup_{\varepsilon\in[0,1)} \mathrm{DSPACE,\!TIME}(O(n^{\varepsilon}),n^{O(1)})$ was already studied in \cite{Yam17,Yam18} as the complexity class $\mathrm{PsubLIN}$ in connection to the so-called \emph{linear space hypothesis}. See also Section \ref{sec:one-way-pushdown}.

\begin{corollary}\label{etime-equivalence}
For any real number $\varepsilon$ in $[0,1)$, $\ptime2^{\mathrm{lin}\twod^{\varepsilon}}\!/\poly = \ptime2^{\mathrm{lin}\twon^{\varepsilon}}/\poly$ if and only if $\mathrm{NSPACE,\!TIME}(O(n^{\varepsilon}),n^{O(1)}) \subseteq \mathrm{DSPACE,\!TIME}(O(n^{\varepsilon}),n^{O(1)})/2^{O(n^{\varepsilon})}$.
\end{corollary}

\begin{proof}
In the proof of Theorem \ref{dspace-2nfa-relation}, we only need to replace the term $n^{1/k}$ by $n^{\varepsilon}$.
\end{proof}

\section{Promise Problem Families of Exponential Ceilings}\label{sec:exp-ceilings}

\begin{figure}[t]
\centering
\includegraphics*[height=5.0cm]{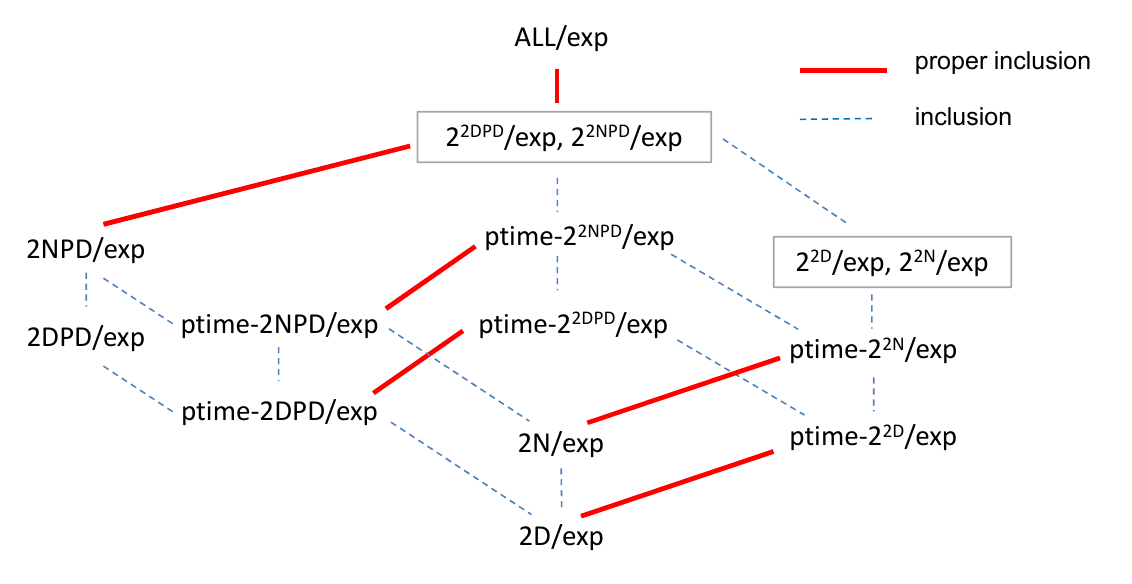}
\caption{Inclusion relationships among nonuniform complexity classes with exponential ceilings except for the separation result of Theorem \ref{separation-exp-ceiling}. Solid red lines indicate \emph{proper} inclusions.}\label{fig:hierarchy-exp}
\end{figure}

We turn our attention to families of promise problems with exponential ceilings. An early result along this line of study includes Kapoutsis's  result \cite{Kap14} on a close connection between the $\twod/\mathrm{exp} =? \twon/\mathrm{exp}$ question and the $\mathrm{NSPACE,\!TIME}(O(\log\log{n}),\infty) \subseteq? \mathrm{DSPACE,\!TIME}(O(\log\log{n}),\infty)/\polylog$ question.
In this section, we further explore other class relationships among nonuniform complexity classes of exponential ceilings.
Most of our results of this section are illustrated in Fig.~\ref{fig:hierarchy-exp}.

Concerning the presence of execution time-bounds, in particular when a ceiling bound is exponential, the polynomial runtime restriction causes a significantly different effect on underlying  machines as shown in the following proposition. To describe the proposition, we need to introduce another advised complexity class $\CC^{\log}_{\#}$, which consists of only subsets of
$\{w\mid w=yx\in[\Sigma,\#],  \exists n(f(n)=|w|)\}$ for where any arbitrary 2dft-constructible function $f$ on $\nat$, where $[\Sigma,\#]=\{yx\mid y\in\Sigma_{\#}^*\#\cup\{\lambda\}, x\in\Sigma^*\}$.
The new notation $\mathrm{SC}^{\log}_{\#}/2^{\polylog}$ denotes the class $\mathrm{DSPACE,\!TIME}(\log^{O(1)}|w|,(|x|\log|w|)^{O(1)})/2^{\log^{O(1)}|w|}$ for inputs $w$ of the form $yx$ in $[\Sigma,\#]$. Similarly, the nondeterministic variant, $\mathrm{NSC}^{\log}_{\#}/2^{\polylog}$, denotes $\mathrm{NSPACE,\!TIME}(\log^{O(1)}|w|,(|x|\log|w|)^{O(1)})/2^{\log^{O(1)}|w|}$ for inputs $w =yx$ in $[\Sigma,\#]$.

\begin{proposition}\label{exp-twod-twon}
(1) $2^{\twod}/\mathrm{exp} = 2^{\twon}/\mathrm{exp}$.

(2) If $\mathrm{SC}^{\log}_{\#}/2^{\polylog} = \mathrm{NSC}^{\log}_{\#}/2^{\polylog}$, then  $\ptime2^{\twod}/\mathrm{exp} = \ptime2^{\twon}/\mathrm{exp}$.

(3) If $\ptime2^{\twod}/\mathrm{exp} = \ptime2^{\twon}/\mathrm{exp}$, then $\mathrm{SC}/2^{\polylog} = \mathrm{NSC}/2^{\polylog}$.
\end{proposition}

\begin{proof}
(1)  It is clear that $2^{\twod}/\mathrm{exp} \subseteq 2^{\twon}/\mathrm{exp}$. In what follows, we intend to prove the opposite inclusion.
For this purpose, we first recall that $\mathrm{DSPACE,\!TIME}(\log^{O(1)}n,\infty)/2^{\mathrm{polylog}}$ equals $\mathrm{NSPACE,\!TIME}(\log^{O(1)}n,\infty)/2^{\mathrm{polylog}}$ by an advised version of Savitch's theorem \cite{Sav70}.
Thus, it suffices to prove that  $\mathrm{DSPACE,\!TIME}(\log^{O(1)}n,\infty)/2^{\polylog} = \mathrm{NSPACE,\!TIME}(\log^{O(1)}n,\infty)/2^{\polylog} $ implies $2^{\twon}/\mathrm{exp} \subseteq  2^{\twod}/\mathrm{exp}$ using Proposition \ref{key-proposition}.

Let $\LL=\{(L_n^{(+)},L_n^{(-)})\}_{n\in\nat}$ denote any family in $2^{\twon}/\mathrm{exp}$ over an alphabet $\Sigma$. Let $c$ denote a constant in $\nat^{+}$ satisfying $L_n^{(+)}\cup L_n^{(-)}\subseteq \Sigma^{\leq 2^{n^c+c}}$ for all $n\in\nat$.
By setting $f(n)=2^{n^c+c}$ and $g(n)=2^{n^b+b}$ for a suitable constant $b>0$,
we assume that $\LL$ falls in $2^{\log{g(\mathrm{lin}\twon)}}/f(n)$ witnessed by an associated family $\{N_n\}_{n\in\nat}$ of 2nfa's.
Recall the language
$K_{\LL}  =\{w\mid w=yx\in[\Sigma,\#],  \exists n\in\nat[ |w|=f(n)\wedge  x\in L(N_n)] \}$.
By Proposition \ref{key-proposition}(1), this language $K_{\LL}$ falls in $\mathrm{NSPACE,\!TIME}(O(\log{e(|w|)}),\infty)/O(e(|w|)^2)$ for inputs $w$, where $e(|w|)=g\circ f^{-1}(|w|)$. Since $e(|w|)=2^{\log^{O(1)}|w|}$, $K_{\LL}$ is also in
$\mathrm{NSPACE,\!TIME}(\log^{O(1)}|w|,\infty)/2^{\polylog}$, which coincides with   $\mathrm{DSPACE,\!TIME}(\log^{O(1)}|w|,\infty)/2^{\mathrm{polylog}}$.  

Now, we take functions $s(|w|)=a\log^k|w|+a$ and $d(|w|)=2^{a\log^k|w|+a}$ satisfying  $K_{\LL}\in \mathrm{DSPACE,\!TIME}(s(|w|),\infty)/d(|w|)$ for suitable constants $a,k\geq1$.
By Proposition \ref{key-proposition}(2), the family $\LL$ belongs to $2^{\log{e'(\mathrm{lin}\twod)}}/f(n)$,  where $e'(n)=2^{O(s\circ f(n))}+O(d\circ f(n))$. Note that $s\circ f(n)=n^{O(1)}$ and $d\circ f(n)=2^{n^{O(1)}}$. This implies that $K_{\LL}$ must belong to $2^{\twod}/\mathrm{exp}$, We therefore conclude that $2^{\twon}/\mathrm{exp}\subseteq 2^{\twod}/\mathrm{exp}$.

(2)
The proof of this statement is similar in essence to (1). Here, we only argue how the runtime restriction of underlying 2nfa's and 2dfa's affects the runtime of an NTM and a DTM,  respectively. Let $u(n,|x|)=(n|x|)^a+a$ for a suitable constant $a>0$ and assume that each $N_n$ runs in time $u(n,|x|)$ for inputs $x$. Proposition \ref{key-proposition}(1) provides $r(|w|,|x|)=u(f^{-1}(|w|),|x|)\log{e(|w|)}$, which equals $(|x|\log^{1/c}|w|)^{O(1)}\cdot \log^{O(1)}|w| = (|x| \log|w|)^{O(1)}$.
Hence, $K_{\LL}$ falls in $\mathrm{NSC}^{\log}_{\#}/2^{\polylog}$. By our assumption, $K_{\LL}$ is also in $\mathrm{SC}^{\log}_{\#}/2^{\polylog}$. Take $t(n,|w|)=n^c|x|^c+c$ for a constant $c>0$ and assume that $N$ runs in time $t(f^{-1}(|w|),|x|)$ on inputs $w=yx\in [\Sigma,\#]$.
Proposition \ref{key-proposition}(2) provides $r'(n,|x|) = t(n,|x|)$, which is thus a polynomial in $(n,|x|)$. Therefore, $\LL$ belongs to $\ptime2^{\twod}/\mathrm{exp}$.

(3)
Assume that $\ptime2^{\twod}/\mathrm{exp}= \ptime2^{\twon}/\mathrm{exp}$. We begin with an arbitrary language  $K$ in $\mathrm{NSC}/2^{\polylog}$. We define $f(n)=2^n$ for all $n\in\nat$. We then choose $s(|x|)=a\log^k|x|+a$ and $d(|x|)=2^{a\log^k|x|+a}$ for constants $a,b,k>0$ so that $K$ belongs to $\mathrm{NSPACE,\!TIME}(s(|x|),t(f^{-1}(|x|),|x|))/d(|x|)$ for inputs $x$.
We then obtain $s\circ f(n)=O(\log^kn)$ and $d\circ f(n)=2^{O(\log^kn)}$. Thus, the value  $e(n)=2^{O(s\circ f(n))}+O(d\circ f(n))$ is at most $n^{O(1)}$.
Proposition \ref{second-key-proposition}(1) places $\LL_K$ into $O(r(n,|x|))$time-$2^{\log{e(\mathrm{lin}\twon)}}/f(n)$, where $r(n,|x|)=t(n,|x|)$. This implies that $\LL_K\in \ptime2^{\twon}/\mathrm{exp}$. Our assumption further puts $\LL_K$ into $\ptime2^{\twod}/\mathrm{exp}$. There are functions $g(n)=2^{n^k+c}$  and $u(n,|x|)=n^c|x|^c+c$ for constants $c,k>0$ such that $\LL_K$ is in $u(n,|x|)$time-$2^{\log{g(\mathrm{lin}\twod)}}/f(n)$ for inputs $x$. It then follows that  $e'(|x|)=g\circ f^{-1}(|x|)=2^{O(\log^k|x|)}$ and $r'(n,|x|)=u(n,|x|)\log{e'(|x|)} = O(n^c)\cdot |x|^{O(1)}$. By Proposition \ref{second-key-proposition}(2), $K$ belongs to $\mathrm{DSPACE,\!TIME}(O(\log{e'(|x|)}),O(r'(f^{-1}(|x|),|x|)))/O(e'(|x|)^2)$ for inputs $x$, and thus $K$ falls in $\mathrm{SC}/2^{\polylog}$.
\end{proof}

As a counterpart of Proposition \ref{separation-ALL}, we can prove the following separation.

\begin{proposition}\label{ALL-separation-exp}
$\mathrm{ALL}/\mathrm{exp} \neq 2^{\twodpd}/\mathrm{exp}$.
\end{proposition}

\begin{proof}
A basic idea of this proof is the same as the proof of Proposition \ref{separation-ALL}. Let $\Sigma=\{0,1\}$ and $\Sigma_{\#}=\Sigma\cup\{\#\}$.
We first enumerate  all functions of the form $2^{n^a+a}$ as $\{\mu_a\}_{t\in\nat}$ and, for each pair $(a,n)$,  enumerate all 2dpda's $M^{(a,n)}_{i}$ satisfying $|Q^{(q,n)}_{i}|\leq 2^{n^a+a}$ for any $i\in\nat$.
Our example family $\LL=\{(L_n^{(+)},L_n^{(-)})\}_{n\in\nat}$ is defined by setting $L_n^{(+)}= \{w\mid w=1^{i}\#1^a\# x,  |w|\leq 2^n, x\in\Sigma^*, \text{ $M^{(a,n)}_{i}$ does not accept $w$ }\}$ and $L_n^{(-)}= \{w\mid w=1^{i}\#1^a\# x, |w|\leq 2^n, x\in\Sigma^*, \text{ $M^{(a,n)}_{i}$ does not reject $w$ }\}$. Clearly, $\LL$ belongs to $\mathrm{ALL}/\mathrm{exp}$.

Next, we wish to show that $2^{\twodpd}/\mathrm{exp}$ does not contain $\LL$. 
Assuming otherwise, there are a constant $a>0$, a series $(i_n)_{n\in\nat}$ of numbers, and a family $\{M^{(a,n)}_{i_n}\}_{n\in\nat}$ of $\mu_a(n)$-size 2dpda's that solves $\LL$. It then follows that, for any valid input $w=1^{i_n}\#1^a\# x$, $w\notin L_n^{(+)}$ iff $M^{(n)}_{a,i_n}$ accepts $w$. This is a contradiction.
\end{proof}

Let us consider an exponential-ceiling version of Theorem \ref{ptime-log-k-subset}, which expresses $\ptime2^{\twod}/\mathrm{exp}\nsubseteq \twonpd/\mathrm{exp}$. Unfortunately, it does not seem to verify this relationship. However, we can prove the following weaker result.

\begin{theorem}\label{separation-exp-ceiling}
For any constant $k\in\nat^{+}$, $\ptime2^{\mathrm{lin}\twod^{2k}}/\mathrm{exp} \nsubseteq \mathrm{lin}\twonpd^{k}/\mathrm{exp}$.
\end{theorem}

\begin{proof}
To prove the theorem, we intend to verify the following two statements for each number $k\geq1$:  (1) if $\ptime2^{\mathrm{lin}\twod^{2k}}/\mathrm{exp} \subseteq \mathrm{lin}\twonpd^k/\mathrm{exp}$, then $\mathrm{SC}/2^{O(\log^{2k}n)}$ is included in $\nauxpdaspti{O(\log\log{n})}{\infty}/O(\log^{2k}n)$ and (2) $\mathrm{SC}/2^{O(\log^{2k}n)} \nsubseteq \nauxpdaspti{O(\log\log{n})}{\infty}/O(\log^{2k}n)$. The statements (1)--(2) together imply the theorem. Hereafter, we fix a number $k\in\nat^{+}$.

(1) To prove the desired statement, we first assume that $\ptime2^{\mathrm{lin}\twod^{2k}}/\mathrm{exp} \subseteq \mathrm{lin}\twonpd^k/\mathrm{exp}$. To apply Proposition \ref{second-key-proposition}, we define $f(n)=2^{n}$ for all $n\in\nat$. Let $K$ denote any language in $\mathrm{SC}/2^{O(\log^{2k}n)}$. We then choose functions $s(|x|)=a\log^{2k}|x|+a$, $t(n,|x|)=(n+1)(|x|^c+c)$, and $d(|x|)=2^{c\log^{2k}|x|+c}$ using appropriate constants $a,c>0$ for which $K$ falls in $\mathrm{DSPACE,\!TIME}(s(|x|),t(f^{-1}(|x|),|x|))/d(|x|)$ for inputs $x$.
We note that $s\circ f(n) = O(n^{2k})$ and $d\circ f(n)\leq 2^{O(n^{2k})}$.
By Proposition \ref{second-key-proposition}(1), the family $\LL_K=\{(L_n^{(+)},L_n^{(-)})\}_{n\in\nat}$ is solvable by $e(n)$-size $O(1)$-choice 2dfa's in time $O(r'(n,|x|))$ for inputs $x$,  where $e(n)=2^{O(s\circ f(n))}+O(d\circ f(n)) = 2^{O(n^{2k})}$.
Thus, $\LL_K$ belongs to $\ptime2^{\mathrm{lin}\twod^{2k}}/\mathrm{exp}$.

Our assumption further places  $\LL_K$ into $\mathrm{lin}\twonpd^k/\mathrm{exp}$.
We then take a function  $g(n)=bn^k+b$  for a suitable constant $b>0$ and assume that $\LL_K\in 2^{\log{g(\mathrm{lin}\twonpd)}}/f(n)$.
Proposition \ref{second-key-proposition}(2) implies that $K$ is solvable by an naux-pda using space $O(\log{e'(|x|)})$ with the help of $O(e'(|x|)^2)$-size advice for inputs $x$, where $e'(|x|)=g\circ f^{-1}(|x|)$. Since $e'(|x|)= O(\log^{k}|x|)$,  $K$ belongs to $\nauxpdaspti{O(\log\log{n})}{\infty}/O(\log^{2k}n)$.

(2)
Next, we prove the intended key statement.   The following argument is fundamentally similar to the proof of Theorem \ref{ptime-log-k-subset}.  We set $\Sigma=\{0,1\}$ and use it as an input alphabet as well as an advice alphabet.

Let us enumerate all naux-pda's as $M_1,M_2,\ldots$ and all functions of the form $\log^kn+k$ as $\mu_1,\mu_2,\ldots$, and, for each number $m\in\nat$, enumerate all binary strings $x_{1,m},x_{2,m},\ldots$ and all binary advice strings $a_{1,m},a_{2,m},\ldots$ so that $|x_{j,m}|=|a_{j,m}|=m$ for all $(j,m)$'s.

For each $m\in\nat$, we define $A_{j,m}$ for all $j\in\nat$ as follows. Let $A_{0,m}=\Sigma^{\leq \log^{2k}m}$. Assuming that $A_{j,m}$ has been already defined, let $g_i(x_{j+1,m},b) = |\{z\in A_{j,m}\mid M_i(x_{j+1},z)=b\}|$ and set $\alpha_{j+1,m}=b$ if $g_{i}(x_{j+1,m},b)\leq g_{i}(x_{j+1,m},\bar{b})$. We finally define $A_{j+1,m}$ to be the set $\{z\in A_{j,m}\mid M_i(x_{j+1,m},z)=\alpha_{j+1,m}\}$.
Now, we take the smallest number $r_0$ such that $|A_{r_0,m}|=0$. We then define $h(m)$ to be $1^m\#1^i\# x_{1,m}\alpha_{1,m}\# x_{2,m}\alpha_{2,m}\# \cdots \# x_{r_0,m}\alpha_{r_0,m}\# 0^c$ for $m=\pair{i,t}$,
where $c$ is a number ensuring $|h(m)|=\log^{2k}m$.
$L_k=\{x_{j,m}\mid j,m\in\nat, \text{ $\alpha_{j,m}$ in $h(m)$ is $0$ }\}$.

Our goal is to prove that $L_k\notin \nauxpdaspti{O(\log\log{n})}{\infty}/O(\log^{2k}n)$. We assume otherwise and take an index $i$ such that $L_k=L(M_i,g')$, where $g'$ is a suitable  advice function of the form $g'(m)=a_{l_m,m}$ for certain numbers $l_m\in\nat$.

We choose $j_0$ such that $a_{l_m,m}\in A_{j_0,m}$ but $a_{l_m,m}\notin A_{j,m}$ for all $j<j_0$.
Note that $x_{j_0,m}\in L$ if and only if $M_i(x_{j_0,m},a_{l_m,m})=1$. By the choice of $j_0$,
$a_{l_m,m}$ satisfies $M_i(x_{j_0,m},a_{l_m,m})=\alpha_{j_0,m}$, and thus $\alpha_{j_0,m}$ must be $1$. However, by the definition of $L_k$, $\alpha_{j_0,m}$ is $0$, a contradiction.
Therefore, the intended statement is true.
\end{proof}

Theorem \ref{separation-exp-ceiling} leads to a variety of consequences.
Here, we list only four of them without proofs.

\begin{corollary}
Let $k\in\nat^{+}$.
(1) $\ptime2^{\mathrm{lin}\twon}/\mathrm{exp} \nsubseteq \mathrm{lin}\twon^k/\mathrm{exp}$.
(2) $\ptime2^{\mathrm{lin}\twod^{2k}}/\mathrm{exp} \nsubseteq \mathrm{lin}\twod^k/\mathrm{exp}$.
(3) $\ptime2^{\mathrm{lin}\twonpd^{2k}}/\mathrm{exp} \nsubseteq \ptime\mathrm{lin}\twonpd^{k}/\mathrm{exp}$.
(4) $\ptime2^{\mathrm{lin}\twodpd^{2k}}/\mathrm{exp} \nsubseteq \ptime\mathrm{lin}\twodpd^k/\mathrm{exp}$.
\end{corollary}


Cook's argument in \cite{Coo71} does not apply to aux-pda's whose auxiliary-tape space is bounded below $\log{n}$.  From this fact and the proposition below, we may not be able to determine whether or not  $\twodpd/\mathrm{exp}$ and $\twonpd/\mathrm{exp}$ coincide.  This is in sharp contrast with Proposition \ref{twonpd-with-poly}.

\begin{proposition}\label{two-exp-result}
(1) It follows that $\twodpd/\mathrm{exp} = \twonpd/\mathrm{exp}$ if and only if $\dauxpdaspti{O(\log\log{n})}{\infty}/\polylog$ coincides with $\nauxpdaspti{O(\log\log{n})}{\infty}/\polylog$.

(2) If $\ptime\twodpd/\mathrm{exp} = \ptime\twonpd/\mathrm{exp}$, then
$\dauxpdaspti{O(\log\log{n})}{n^{O(1)}}/\polylog$ coincides with $\nauxpdaspti{O(\log\log{n})}{n^{O(1)}}/\polylog$.
\end{proposition}

\begin{proof}
(1)
(If--part) Firstly, we assume that $\dauxpdaspti{O(\log\log{n})}{\infty}/\polylog$ coincides with $\nauxpdaspti{O(\log\log{n})}{\infty}/\polylog$. Since  $\twodpd/\mathrm{exp} \subseteq \twonpd/\mathrm{exp}$ is obvious, we henceforth prove the opposite inclusion.  We begin with an arbitrary promise problem family $\LL=\{(L_n^{(+)},L_n^{(-)})\}_{n\in\nat}$ in $\twonpd/\mathrm{exp}$.
By choosing $f(n)=2^{n^k}$ and $g(n)=n^{b}+b$ for appropriate constants $k,b>0$, we assume that $\LL\in 2^{\log{g(\mathrm{lin}\twonpd)}}/f(n)$.  Consider the language $K_{\LL}$ given in Proposition \ref{key-proposition}(1).
Note that $e(|w|)=g\circ f^{-1}(|w|)= O(\log^{O(1)}|w|)$.
We  apply Proposition \ref{key-proposition}(1) and then obtain $K_{\LL}\in \nauxpdaspti{O(\log{e(|w|)})}{\infty}/O(e(|w|)^2)$ for inputs $w$ in $[\Sigma,\#]$.
Thus, $K_{\LL}$  is contained in $\nauxpdaspti{O(\log\log{n})}{\infty}/\polylog$. Hence,  $K_{\LL}$ falls in $\dauxpdaspti{O(\log\log{n})}{\infty}/\polylog$  by our assumption.
Now, we take functions $s(|w|)=c\log\log{|w|}+c$ and $d(|w|)=\log^k|w|+c'$ for constants $c,c'>0$ so that $K_{\LL} \in \mathrm{DSPACE,\!TIME}(s(|w|),\infty)/d(|w|)$ for inputs $w$. Note that $e'(n)=2^{O(s\circ f(n))}+O(d\circ f(n))$.
It follows that $e'(n)=n^{O(1)}$ since $s\circ f(n)=O(\log{n})$ and $d\circ f(n)=O(\log^kn)$.
By Proposition \ref{key-proposition}(2), $\LL$ belongs to $\twodpd/\mathrm{exp}$. Therefore, we conclude that $\twonpd/\mathrm{exp}\subseteq \twodpd/\mathrm{exp}$.

(Only If--part) Assume that $\twodpd/\mathrm{exp}$ equals $\twonpd/\mathrm{exp}$. Let $K$ denote a language in  $\nauxpdaspti{O(\log\log{|x|})}{\infty}/\polylog$ for inputs $x$.
We define $f(n)=2^n$ for all $n\in\nat$.
We then take two functions $s(|x|)=a\log\log{|x|}+a$ and $d(|x|)=\log^k|x|+b$ for appropriate constants $a,b,k>0$ and assume that $K$ falls in $\nauxpdaspti{s(|x|)}{\infty}/d(|x|)$ for inputs $x$.   Recall the promise problem family $\LL_{K}$.
By Proposition \ref{second-key-proposition}(1), we conclude that $\LL_K$ belongs to  $2^{\log{g(\mathrm{lin}\twonpd)}}/f(n)$, where $e(n)=2^{O(s\circ f(n))}+O(d\circ f(n))$. Since $e(n) = n^{O(1)}$, this latter class is included in $\twonpd/\mathrm{exp}$, and thus  $\LL\in\twodpd/\mathrm{exp}$  follows by our assumption.
We next take $g(n)=n^{k'}+a$ for an appropriate constants $a,k'\geq1$ for which $\LL$ belongs to $2^{\log{g(\mathrm{lin}\twodpd)}}/f(n)$.
Since $e'(|x|)=g\circ f^{-1}(|x|)$, $K$ falls in $\nauxpdaspti{O(\log{e'(|x|)})}{\infty}/O(e'(|x|)^2)$, which is included in $\nauxpdaspti{O(\log\log{n})}{\infty}/\polylog$ since $e'(n)= = \log^{O(1)}|x|$.

(2) Since the basic argument is essentially the same as that of (1), we discuss mostly the effect of runtime bounds. We begin by assuming $\ptime\twodpd/\mathrm{exp} = \ptime\twonpd/\mathrm{exp}$. Together with functions $f,s,d$ given above,
we take $t(n,|x|)=n^c|x|^c+c$ for an appropriate constant $c>0$ so that $K$ is in $\nauxpdaspti{s(|x|)}{t(f^{-1}(|x|),|x|)}/d(|x|)$ for inputs $x$. Let $r(n,|x|)=t(n,|x|)$.
Proposition \ref{second-key-proposition}(1) then implies that $\LL_K$ falls in  $O(r(n,|x|))$time-$2^{\log{g(\mathrm{lin}\twonpd)}}/f(n) \subseteq \ptime\twonpd/\mathrm{exp}$.
Our assumption further forces $\LL_K$ to fall in $\ptime\twodpd/\mathrm{exp}$. In addition to $g$ given above, we choose $u(n,|x|)=n^b|x|^b+b$ and assume that $\LL_K$ is in $u(n,|x|)$time-$2^{\log{g(\mathrm{lin}\twodpd)}}/f(n)$ for inputs $x$.
Note that $r'(n,|x|)=u(n,|x|)\log{e'(|x|)} = n^{O(1)}|x|^{O(1)}$. Hence, by Proposition \ref{second-key-proposition}(2), $K$ belongs to $\mathrm{DSPACE,\!TIME}(O(\log\log|x|),|x|^{O(1)})/\log^{O(1)}|x|$ for inputs $x$.
\end{proof}


To compensate Proposition \ref{two-exp-result}, we intend to demonstrate the collapse of $2^{\twonpd}/\mathrm{exp}$ down to $2^{\twodpd}/\mathrm{exp}$.  This is a pushdown-automata counterpart of Proposition \ref{exp-twod-twon}(1).

\begin{proposition}
(1) $2^{\twodpd}/\mathrm{exp} = 2^{\twonpd}/\mathrm{exp}$.

(2) If $\ptime2^{\twodpd}/\mathrm{exp} = \ptime2^{\twonpd}/\mathrm{exp}$, then $\dauxpdaspti{\log^{O(1)}n}{n^{O(1)}}/2^{\polylog} = \nauxpdaspti{\log^{O(1)}n}{n^{O(1)}}/2^{\polylog}$.
\end{proposition}

\begin{proof}
(1) Let $\LL =\{(L_n^{(+)},L_n^{(-)})\}_{n\in\nat}$ be any promise problem family in $2^{\twonpd}/\mathrm{exp}$ over an alphabet $\Sigma$.  Take  a family $\{N_n\}_{n\in\nat}$ of $2^{n^a+a}$-size 2npda's that solves $\LL$ for a suitable constant $a>0$. Take another constant $c>0$ for which $\LL$ is $2^{n^c+c}$-ceiling. Now, let $f(n)=2^{n^c+c}$ and $g(n)=2^{n^a+a}$. Note that $\LL$ belongs to $2^{\log{g(\mathrm{lin}\twonpd)}}/f(n)$.
Recall the language
$K_{\LL}  =\{w=yx\mid \exists n\in\nat[ |w|=f(n)\wedge  y\in\Sigma_{\#}^*\#\cup\{\lambda\} \wedge x\in L(N_n)] \}$.
By Proposition \ref{key-proposition}(1) with $e(n) = g\circ f^{-1}(n)$,
$K_{\LL}$ belongs to $\nauxpdaspti{O(\log{e(|w|)})}{\infty}/O(e(|w|)^2)$ for inputs $w$ in $[\Sigma,\#]$.
This class is included in $\nauxpdaspti{\log^{O(1)}|w|}{\infty}/2^{\polylog}$ since $e(n)=2^{\log^{O(1)}n}$. By an advice version of Cook's result \cite{Coo71}, $K_{\LL}$ also belongs to $\dauxpdaspti{\log^{O(1)}|w|}{\infty}/2^{\polylog}$.
Consider a daux-pda that recognizes $K_{\LL}$ using space $\log^kn+k$ with advice of $2^{\log^bn+b}$ size for suitable constants $b,k\geq1$.
Let $s(n)=\log^kn+k$ and $d(n)=2^{\log^b+b}$.   Let  $e'(n)=2^{O(s\circ f(n))}+O(d\circ f(n)) = 2^{n^{O(1)}}$.
By Proposition \ref{key-proposition}(2), $\LL$ belongs to $2^{\log{e'(\mathrm{lin}\twodpd)}}/f(n)\subseteq 2^{\twodpd}/\mathrm{exp}$.

(2)
Assume that $\ptime2^{\twodpd}/\mathrm{exp} = \ptime2^{\twonpd}/\mathrm{exp}$. Let $f(n)=2^n$ for all $n\in\nat$. We then take any language $K$ in $\dauxpdaspti{s(|x|)}{t(f^{-1}(|x|),|x|)}/d(|x|)$ for inputs $x$, where $s(|x|)=b\log^a|x|+b$, $t(n,|x|)=n^b|x|^b+b$, and $d(|x|)=2^{b\log^a|x|+b}$ for appropriate constants $a,b>0$.
We remark that $t(f^{-1}(|x|),|x|)=(\log{n})^b|x|^b+b=|x|^{b}$, where $f(n)=|x|$.
By Proposition \ref{second-key-proposition}(1),  the family $\LL_K$ falls in $O(r(n,|x|))$time-$2^{\log{e(\mathrm{lin}\twonpd)}}/f(n)$ for inputs $x$,  where $e(n)=2^{O(s\circ f(n))}+O(d\circ f(n))$ and $r(n,|x|)=t(n,|x|)$.
Since $e(n)=2^{n^{O(1)}}$, $\LL_K$  is contained in $\ptime2^{\twonpd}/\mathrm{exp}$.
By our assumption, $\LL_K$ further falls in $\ptime2^{\twodpd}/\mathrm{exp}$. We then take functions $g(n)=2^{n^c+c}$ and $u(n,|x|)= (n|x|)^c+c$  with suitable constants $b,c>0$ satisfying that $\LL_K\in u(n,|x|)\mathrm{time\mbox{-}}2^{\log{g(\mathrm{lin}\twodpd)}}/f(n)$ for inputs $x$.
Proposition \ref{second-key-proposition}(2) then implies that  $K$ belongs to $\dauxpdaspti{O(\log{e'(|x|)})}{O(r'(f^{-1}(|x|),|x|))}/O(e'(|x|)^2)$ for inputs $x$, where $e'(n)=g\circ f^{-1}(n)$ and $r'(n,|x|)=u(n,|x|)\log{e'(|x|)}$.
This complexity class is included in $\dauxpdaspti{\log^{O(1)}n}{n^{O(1)}}/2^{\polylog}$ since $e'(n)=2^{\log^{O(1)}n}$ and $r'(n,|x|)=n^{O(1)}|x|^{O(1)}$.
\end{proof}

\section{Complexity of One-Way Pushdown Automata Families}\label{sec:one-way-pushdown}

Throughout Sections \ref{sec:poly-ceilings}--\ref{sec:exp-ceilings}, we have explored relationships among nonuniform complexity classes induced by families of two-way finite and pushdown automata. Except for a few cases (such as Theorems \ref{ptime-log-k-subset} and \ref{separation-exp-ceiling}), it seems difficult to prove their collapses/separations without any unproven assumptions. When we discuss similar situations associated with ``one-way'' automata families, it is sometimes (not always) possible to demonstrate collapses/separations among the complexity classes induced by these automata families.
In this section, we intend to  discuss such collapses/separations.
Later, we briefly discuss how polynomial ceilings affect collapses/separations of those complexity classes.

We first review the known results in the literature associated with ``one-way'' automata families:  $\oned\neq \onen$ \cite{SS78}, $\twobp\subseteq 2^{\oned}$, $2^{\oned}\nsubseteq \onep$, and $\onep\subsetneqq 2^{\onep}$ \cite{Yam19a} and that  $\oneu\nsubseteq \onedpd$ and $\mathrm{1t}\onedpd\nsubseteq \onep$  \cite{Yam23}.
 As a new addition to these results,
we wish to present the following separation results.

\begin{theorem}\label{onedpd-bound}
(1) $\mathrm{1t}\onedpd\nsubseteq 2^{\onen}$. (2) $\co2^{\linoner}\nsubseteq \onenpd$ (and thus $2^{\linonep}\nsubseteq \onenpd$). (3) $\co\onenpd\nsubseteq 2^{\onenpd}$. (4) $2^{\log^2\oned}\nsubseteq \onenpd$. 
\end{theorem}

Theorem \ref{onedpd-bound}(1)  should be compared to  Theorem \ref{Cook-proof}, which shows, in the two-way setting, that $\ptime\twodpd/\poly \subseteq \ptime2^{\log^2\twod}/\poly$.


To prove Theorem \ref{onedpd-bound}, we need two supporting lemmas (Lemmas \ref{pumping-1nfa}--\ref{pumping-1npda}), called the \emph{pumping lemmas for 1nfa families and 1npda families}. For the sake of the reader, we include the proofs of those two lemmas in Appendix since explicit (stack-)state complexity upper-bounds are needed.
Given a string $w$ of the form $xyz$ accepted by a 1nfa $M$, the middle substring $y$ is said to be an \emph{iterative singleton} of $w$ for $M$ if $|y|\geq1$ and $xy^iz$ is accepted by $M$ for any number $i\in\nat$.

\begin{lemma}{\rm [pumping lemma for 1nfa families]}\label{pumping-1nfa}
Let $s$ denote any bounding function on $\nat$.  Let $\MM=\{M_n\}_{n\in\nat}$ denote any family of 1nfa's whose state complexity is at most $s(n)$. For each number $n\in\nat$, the following statement is true: for any string $w$ with $|w|\geq s(n)$ accepted by $M_n$, there is a factorization $w=xyz$ for which $y$ is an iterative singleton of $w$ for $M_n$.
\end{lemma}

Next, we show another pumping lemma for 1npda families.
Similarly to iterative singletons, we define iterative pairs as follows. For any string $w=xyzuv$ accepted by a 1npda $M$,  we say that the pair $(y,u)$ of strings is an \emph{iterative pair} of $w$ for $M$ if $|yu|\geq1$ and $xy^izu^iv$ is accepted by $M$ for all $i\in\nat$.

\begin{lemma}{\rm [pumping lemma for 1npda families]}\label{pumping-1npda}
Let $s$ denote any bounding function on $\nat$. Let $\MM=\{M_n\}_{n\in\nat}$ denote any family of 1npda's whose stack-state complexity is at most $s(n)$. For each $n\in\nat$, the following statement is true: for any string $w$ with $|w|\geq s(n)^2\cdot 2^{s(n)\log{s(n)}+1}$ accepted by $M_n$, there is a factorization $w=xyzuv$ such that $(y,u)$ is an iterative pair of $w$ for $M_n$.
\end{lemma}

With the help of the above two pumping lemmas, let us describe the proof of Theorem \ref{onedpd-bound}.
In the proof, we use certain bounds of \emph{prime numbers}. For each number $x\in\nat^{+}$, the notation $p_x$ denotes the $x$-th smallest prime number. From \cite{Axl13}, it follows that the number $p_x$ sits between $x(\ln{x}+\ln\ln{x}+\ln\ln{x}/\ln{x} +2.0/\ln{x})$ and $x(\ln{x}+\ln\ln{x}+\ln\ln{x}/\ln{x} + 2.1/\ln{x})$, which implies $p_x\leq 2x\log{x}/\log{e}\leq 2x\log{x}$. It then follows that $\log{p_x}\leq \log(2x\log{x})=\log{x}+\log\log{x}+\log{2} \leq 3\log{x}$.


\vs{-2}
\begin{proofof}{Theorem \ref{onedpd-bound}}
(1) We define $L_n^{(+)}=\{a^mb^m\mid m\geq n\}$ and $L_n^{(-)}= a^*b^* - L_n^{(+)}$ for each index $n\in\nat$ over the alphabet $\Sigma=\{a,b\}$ and  then consider the family $\LL=\{(L_n^{(+)},L_n^{(-)})\}_{n\in\nat}$.
Now, we claim that $\LL$ is in $\mathrm{1t}\onedpd$ by storing $a^m$ into a stack and reading $b^m$ as popping $a^m$ from the stack one by one. Note that this machine makes only 1 (stack) turn.

Next, we show that $\LL\notin 2^{\onen}$. Assume otherwise and take a family $\MM=\{M_n\}_{n\in\nat}$ of exponential-size 1nfa's solving $\LL$.
Take a polynomial $p$ such that $|Q_n|\leq 2^{p(n)}$ for all $n\in\nat$. We then apply the  pumping lemma for 1nfa families (Lemma \ref{pumping-1nfa}) to $\MM$.
Fix $n$ arbitrarily and take a number $m$ satisfying $m>2^{p(n)}$ and consider the target  string $w=a^mb^m$. Notice that $w$ is in $L_n^{(+)}$.
Since $|w|\geq 2^{p(n)}$, this input $w$ has a factorization with three parts $xyz$ and $y$ can be pumped an arbitrary number of times by Lemma \ref{pumping-1nfa}. This is absurd because such pumped strings are not always in $L_n^{(+)}$.

(2)
We define $\mathrm{exp}_0(n)=n$ and $\mathrm{exp}_{i+1}(n)=2^{\mathrm{exp}_i(n)}$ for all $n,i\in\nat$.  As an example language $\LL'=\{(L_n^{+)},L_n^{(-)})\}_{n\in\nat}$, we take  $L_n^{(+)}=\{a^mb^mc^m \mid m\leq \mathrm{exp}_2(n)\}$ and $L_n^{(-)}=a^*b^*c^* -  L_n^{(+)}$ for each index $n\in\nat$. Firstly, we assert that $\LL'\in \co2^{\linoner}$ by designing a family of 1pfa's solving $\co\LL'$ with one-sided error probability.
On input $w$, we check if $w\in a^*b^*c^*$ with no use of stack.
Simultaneously,
we conduct the following task. Assume that $w$ has the form $a^{k}b^{l}c^{m}$ for $k,l,m\in\nat^{+}$.
We randomly choose one of the $2^{n}$ consecutive prime numbers $p_1,p_2,\ldots, p_{2^n}$ and compute $k\;\mathrm{mod}\;p_i$, $l\;\mathrm{mod}\;p_i$, and $m\;\mathrm{mod}\;p_i$. We then check if these values coincide with each other. If so, we accept $w$: otherwise, we reject $w$. It then follows that, if $w\in L_n^{(+)}$, then we successfully accept $w$ with probability $1$. Otherwise, we err with the probability less than $1/2$.
To implement this algorithm, we need to remember all
such prime numbers using inner states.
For this purpose, we need only $2^{O(n)}$ inner states since $p_x\leq 2x\log{x}$,  as noted above.
This process is possible because $M_n$ can use $2^{O(n)}$ inner states. Therefore, $\LL'$ belongs to $\co2^{\linoner}$.

To show that $\LL'\notin\onenpd$, we assume that there is a family $\{M_n\}_{n\in\nat}$ of polynomial-size 1npda's solving $\LL'$. Let $s(n)$ denote the stack-state complexity of $M_n$, which is upper-bounded by a suitable polynomial.
Now, we apply the pumping lemma for 1npda families (Lemma \ref{pumping-1npda}) to $L(M_n)$.
It is important to note that the pumping lemma constant for $L(M_n)$ is at most $s(n)^2\cdot 2^{s(n)\log{s(n)}+1}< \mathrm{exp}_2(n)$.
For any input $w$ of the form $a^mb^mc^m$ with $m=\mathrm{exp}_2(n)$, $w$ can be factorized into $xyzuv$  so that  $(y,u)$ is an iterative pair of $w$ for $M_n$. However, this violates the definition of $L_n^{(+)}$.

The second part of (2) comes from the fact that $2^{\linoner}\cup \co2^{\linoner} \subseteq 2^{\linonep}$.

(3)
We consider the family $\KK=\{(K_n^{(+)},K_n^{(-)})\}_{n\in\nat}$ with $K_n^{(-)}=\{a^kb^lc^m\mid k,l,m\in\nat, k\neq l\vee l\neq m\}$ and $K_n^{(+)}=a^*b^*c^*-K_n^{(+)}$ for each $n\in\nat$. We then claim that ($a$) $\co\KK\in \onenpd$ and ($b$) $\KK\notin2^{\onenpd}$.

($a$) Now, we wish to prove that $\KK \in \co\onenpd$, or equivalently, $\co\KK\in\onenpd$. Notice that $\co\KK=\{(K_n^{(-)},K_n^{(+)})\}_{n\in\nat}$.
Let us consider the following 1npda $N_n$. On input $w$ of the form $a^kb^lc^m$, we nondeterministically choose either $(k,l)$ or $(l,m)$. If we take $(k,l)$ (resp., $(l,m)$), then we check  whether $k\neq l$ (resp., $l\neq m$) by using a stack.  This checking  process can be done deterministically. Obviously, $\co\KK$ is solvable by the family $\{N_n\}_{n\in\nat}$, and thus it is in $\onenpd$.  This implies that $\KK\in\co\onenpd$.

($b$) We follow the argument of (2). Since $s(n)$ is an exponential in this case,  $s(n)^2\cdot 2^{s(n)\log{s(n)}+1}<\mathrm{exp}_3(n)$ follows. By the pumping lemma, we can conclude that $\KK$ does not belong to $2^{\onenpd}$.

(4)
We note that, if $2^{\log^2\onen}\subseteq \onenpd$, then $2^{\log^2\onen}/\poly \subseteq \onenpd/\poly$ follows.
By the nondeterministic versions of Corollaries \ref{twod-poly-twodpd} and \ref{log-k-2D-poly}, we obtain both $\leq^{\twof}_{m}\!(2^{\log^2\onen}/\poly) = 2^{\log^2\twon}/\poly$ and  $\leq^{\twof}_{m}\!(\onenpd/\poly) = \twonpd/\poly$.
From $2^{\log^2\onen}/\poly \subseteq \onenpd/\poly$,
we conclude that  $\ptime2^{\log^2\twon}/\poly \subseteq \twonpd/\poly$. This consequence clearly contradicts Theorem \ref{ptime-log-k-subset}, and thus $2^{\log^2\onen}\nsubseteq \onenpd$ follows.
\end{proofof}

\begin{corollary}
(1) $\onedpd\neq \onenpd \neq \co\onenpd$. (2) $2^{\onedpd}\neq 2^{\onenpd} \neq \co2^{\onenpd}$.
\end{corollary}

\begin{proof}
(1)
Recall the family $\KK$ from the proof of Theorem \ref{onedpd-bound}(3).
If $\onenpd=\co\onenpd$, then we obtain $\co\KK\in \onenpd = \co\onenpd$; that is, $\KK$ is in $\onenpd$. This is a clear contradiction with $\KK\notin\onenpd$.
Therefore, $\onenpd\neq \co\onenpd$ follows.
If $\onedpd=\onenpd$, then $\onenpd=\co\onenpd$ must hold because $\onedpd$ is closed under complementation. Thus, the separation $\onedpd\neq \onenpd$ also follows.

(2)  Note that $\co\onenpd\subseteq \co2^{\onenpd}$. If $2^{\onenpd}=\co2^{\onenpd}$, then we obtain $\co\onenpd \subseteq 2^{\onenpd}$, a contradiction with Theorem \ref{onedpd-bound}(3). Hence, we conclude that $2^{\onenpd}\neq \co2^{\onenpd}$. Moreover, since $2^{\onedpd}=2^{\onenpd}$ implies $2^{\onenpd}=\co2^{\onenpd}$, it follows that $2^{\onedpd}\neq 2^{\onenpd}$.
\end{proof}


Hereafter, we pay our attention to the effect of ceilings for one-way machines.
Let us begin with a simple case.
By analyzing the proof of Proposition \ref{ALL-equivalent}, we obtain the following corollary for one-way machines.

\begin{corollary}\label{ALL-onenpd}
$\mathrm{ALL}/\poly = 2^{\oned}/\poly = 2^{\onen}/\poly =  2^{\onedpd}/\poly = 2^{\onenpd}/\poly$.
\end{corollary}

We have known that $\oned\neq \onen$ \cite{SS78}. However, it is not yet known whether or not $\onen\subseteq\twod$.
The following proposition hints that solving the $\onen/\poly\nsubseteq? \twod/\poly$ question seems to be quite difficult.

\begin{proposition}\label{one-way-case-collapse}
\renewcommand{\labelitemi}{$\circ$}
\begin{enumerate}\vs{-2}
  \setlength{\topsep}{-2mm}%
  \setlength{\itemsep}{1mm}%
  \setlength{\parskip}{0cm}%

\item $\onen/\poly \subseteq \twod/\poly$ if and only if $\dl/\poly=\nl/\poly$.

\item If $\oned/\poly = \onen/\poly$, then $\dl/\poly = \nl/\poly$.

\item  If $\onedpd/\poly = \onenpd/\poly$, then $\logdcfl/\poly = \logcfl/\poly$.
\end{enumerate}
\end{proposition}

\begin{proof}
(1)
If $\onen/\poly \subseteq \twod/\poly$, then we obtain $\leq^{\twof}_m\!(\onen/\poly)\subseteq \;\leq^{\twof}_m\!(\twod/\poly)$, which  is equivalent to $\twon/\poly \subseteq \twod/\poly$ by Corollary \ref{twod-poly-twodpd}(1).
Obviously, $\twod/\poly = \twon/\poly$ implies $\onen/\poly \subseteq \twod/\poly$.
As shown in \cite{Kap14}, $\twod/\poly=\twon/\poly$ is logically equivalent to $\dl/\poly=\nl/\poly$.

(2)
Assume that $\oned/\poly = \onen/\poly$. From this, $\leq^{\twof}_{m}\!(\oned/\poly) = \;  \leq^{\twof}_{m}\!(\onen/\poly)$ follows.  Similarly to (1), we can conclude that $\twod/\poly = \twon/\poly$. Finally, we obtain $\dl/\poly=\nl/\poly$.

(3)
A similar argument to (2) works by an application of Corollary \ref{twod-poly-twodpd}(2). The conclusion comes from the fact that $\ptime\twodpd/\poly = \ptime\twonpd/\poly$ iff $\logdcfl/\poly = \logcfl/\poly$ \cite{Yam21}.
\end{proof}

\section{A Brief Discussion and Future Directions}\label{sec:discussion}

For the sake of the avid reader, we wish to present a short list of open problems in the topics of this work.
There are numerous unknown relationships among the computational complexity of one-way finite automata and pushdown automata families.

\renewcommand{\labelitemi}{$\circ$}
\begin{enumerate}\vs{-2}
  \setlength{\topsep}{-2mm}%
  \setlength{\itemsep}{1mm}%
  \setlength{\parskip}{0cm}%

\item Counters have proven to be a useful device in proving in Section \ref{sec:reductions} that  $\twodct_4= \leq^{\ptime\twofct_2}_{m}\!(\twodct_4)$ and $\twodpdct_3= \leq^{\ptime\twofct_2}_{m}\!(\twodpdct_3)$. However, it is not yet known that we can remove the entire use of such counters. In other words, is it true that $\twod =\: \leq^{\ptime\twofct_2}_{m}\!(\twod)$ and $\twodpd =\: \leq^{\ptime\twofct_2}_{m}\!(\twodpd)$?

\item The relationships between one-way finite automata families and their pushdown counterparts are not fully understood. We have shown that $2^{\linonep}\nsubseteq \onenpd$ in Theorem \ref{onedpd-bound}(2). Concerning a further relationship between $\onep$ and $\onenpd$, is it true that $\onep\nsubseteq \onenpd$?

\item In Section \ref{sec:poly-ceilings}, for instance, we have discussed a relationship between the $2^{\log^k\twod}$ vs. $2^{\log^k\twon}$ question and an advised version of the $\mathrm{SC}^k$ vs. $\mathrm{NSC}^k$ question. It is interesting to find exact characterizations of similar unproven questions.

\item Although the choice of $f$ is arbitrary in the proof of Propositions \ref{key-proposition}--\ref{second-key-proposition}, Sections \ref{sec:poly-ceilings}--\ref{sec:exp-ceilings} handle only the case where $f(n)$ is a polynomial or an exponential in $n$. It is thus important to study any other possible cases.

\item What kind of direct connections exist between $2^{\lintwodpd}$ (as well as $2^{\lintwonpd}$) and standard complexity classes?

\item In \cite{Yam23}, numerous variants of $\onen$ were studied. For instance, $\oneparity$ is the class of all families of promise problems solvable by polynomial-size parity automata families. What are the computational complexities of $2^{\oneparity}$ and $2^{\log^k\oneparity}$ with or without polynomial ceilings?

\item We have discussed only two ceiling bounds: polynomial and exponential.
A general study of the effect of subexponential ceiling is still missing. Such a study is expected to enrich the theory of nonuniform state complexity.

\item We have shown in Proposition \ref{coincide-log-k-twod} the inclusion of $2^{\log^k\twon}/\poly \subseteq 2^{\log^{2k}\twod}/\poly$, whose proof is founded on Savitch's theorem \cite{Sav70}. We wonder whether the exponent of $2k$ is reduced to $2k-1$ or even $3k/2$.

\item In \cite{Yam18}, the advised version of the \emph{linear space hypothesis} (LSH) was introduced and its nonuniform state complexity characterization was given in \cite{Yam18}.
The LSH concerns polynomial-time sublinear-space computation, which indicates $\mathrm{DSPACE,\!TIME}(O(n^{\varepsilon}),n^{O(1)})$.
This complexity class seems to be related to $\ptime2^{\mathrm{lin}\twod^{\varepsilon}}/\poly$.
What kind of relationships are there between  $\twon/\poly \nsubseteq \bigcup_{\varepsilon\in[0,1)} \ptime2^{\mathrm{lin}\twod^{\varepsilon}}/\poly$ and  the nonuniform version of LSH?  We further ask whether  $2^{\twon}/\poly \nsubseteq \bigcup_{\varepsilon\in[0,1)} \ptime2^{\mathrm{lin}\twod^{\varepsilon}}/\poly$ holds or not.
\end{enumerate}


\let\oldbibliography\thebibliography
\renewcommand{\thebibliography}[1]{%
  \oldbibliography{#1}%
  \setlength{\itemsep}{-1pt}%
}
\bibliographystyle{plainurl}

\bibliographystyle{alpha}


\section*{Appendix}

In the main text, we have omitted the proofs of two lemmas: Lemmas \ref{pumping-1nfa}--\ref{pumping-1npda}. 
Here, we wish to provide these omitted proofs for the sake of completeness. 


\begin{figure}[t]
\centering
\includegraphics*[height=4.8cm]{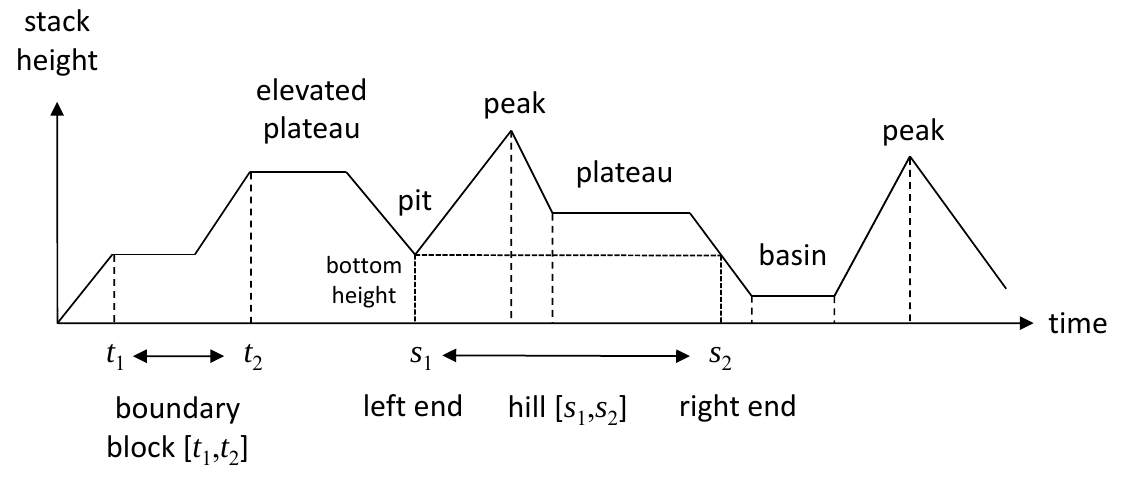}
\caption{A history of the changes of stack heights.}\label{fig:stack-height}
\end{figure}

\begin{proofof}{Lemma \ref{pumping-1nfa}}
Given a 1nfa $M_n=(Q_n,\Sigma,\{\rhd,\lhd\}, \delta_n, q_{0,n}, Q_{acc,n},Q_{rej,n})$ with $|Q_n|\leq s(n)$, we define $d_n=|Q_n-Q_{halt,n}|+1$, where $Q_{halt,n} = Q_{acc,n}\cup Q_{rej,n}$. Note that $d_n\leq s(n)$ for all $n\in\nat$ since $|Q_{halt}\neq\setempty$.  Consider a string $w$ of length $\geq s(n)$ and assume that $M_n$ accepts $w$. We then assign an inner state $q$ to each tape cell containing a symbol $\sigma$ of $w$ if $M_n$ is in inner state $q$ reading $\sigma$ on this tape cell.

Since $|w|\geq d_n$, there are two indices $i_1,i_2\in[|w|]$ that have the same inner states, say, $q$.
We then partition $w$ to $xyz$ so that $y$ is located between the $i_1$th tape cell and the $(i_2-1)$th tape cell. Consider $w_2=xyyz$. For the substring $yyz$, $M_n$ takes exactly the same inner state when $M_n$ is reading the leftmost symbol of $yyz$ and when it is reading the first symbol of the second $y$. This implies that $M_n$ makes exactly the same behavior while reading $y$. Hence, if $xyz$ is accepted, then so is $xyyz$. More generally, $M_n$ accepts $xy^iz$ for all $i\in\nat$. Hence, $y$ is indeed an iterative singleton of $w$ for $M_n$.
\end{proofof}


A standard pumping lemma for context-free languages is usually proven by arguments built up on grammar definitions of the context-free languages. The following proof of Lemma \ref{pumping-1npda}, we use an argument based on machines rather than grammars. Similar arguments were already used in \cite{Yam20} as well as \cite{Yam08,Yam16}.
To prove the lemma, we first need to recall
essential terminology from \cite{Yam20}.

The \emph{stack content} refers to a series of symbols stored in the stack (including $\bot$), which forms a ``string'' over a stack alphabet. The \emph{stack height} is then the length of the stack content.
We remark that every pushdown automaton is allowed to make its input-tape head stay still (known as an $\lambda$-move), whereas every finite automaton in this work must move its tape head to the right without making any $\lambda$-moves (in accordance with the literature).
A \emph{stack history} refers to a sequence of stack contents according to ``time''.

Now, we consider a stack history referring to a computation path on input $x$. A \emph{boundary} refers to a borderline between two adjacent tape cells. In particular, boundary $0$ is the left borderline of the tape cell indexed $0$. All boundaries are indexed from $0$ to $|\cent x\dollar|$. 
A \emph{boundary block} is a consecutive sequence of boundaries. The boundary block between boundaries $t_1$ and $t_2$ with $t_1<t_2$ is expressed as $[t_1,t_2]$, where the  numbers $t_1$ and $t_2$ are called the \emph{ends} of this boundary block and its \emph{fringe} is either the boundary $t_1-1$ or the boundary $t_2+1$ if they exist.

A \emph{peak} (resp., a \emph{pit}) is a boundary, say, $t$ for which the stack heights at the boundaries $t-1$ and $t+1$ are smaller (resp., greater) than the stack height at the boundary $t$. A \emph{plateau} is a flat boundary block and the stack height of this boundary block is called the \emph{height} of the plateau.
A \emph{basin} (resp., an \emph{elevated plateau}) $[t,t']$ is a plateau  for which the stack heights at all fringes of this plateau are greater (resp., smaller) than the height of the plateau whenever the fringes exist. 
An \emph{upward slope} means a non-flat boundary block in which the stack height never decreases. In contrast, a \emph{downward slope} is a non-flat boundary block in which no increase of the stack height occurs. A \emph{slope} refers to either an upward slope or a downward slope. 
A \emph{hill} is a boundary block $[t,t']$ such that (i) the stack height at the boundary $t$ and the stack height at the boundary $t'$ coincide, (ii) there exists either a peak or an elevated plateau, but not both, in $[t,t']$, (iii) there is no pit or basin in $[t,t']$, and (iv) one of the ends of this block is either a pit or the left edge of a basin. The lowest stack height of the hill is particularly called the \emph{bottom height}.
The \emph{height of the hill} is the difference between its bottom height and its highest stack height.
We briefly call either a peak or an elevated plateau existing in a hill by a \emph{hill top}. The opposite notion of a hill is called a \emph{valley}. 
Those concepts are illustrated in Fig.~\ref{fig:stack-height}. 

\begin{proofof}{Lemma \ref{pumping-1npda}}
Let $M_n$ denote any 1npda of the form $(Q_n,\Sigma,\{\rhd,\lhd\}, \Gamma_n, \delta_n, q_{0,n}, \bot,  Q_{acc,n},Q_{rej,n})$ with push size $e_n$. Notice that $e_n$ is at least $2$. Now, we assume that $|Q_n||\Gamma_n^{\leq e_n}|\leq s(n)$.   
To make the following argument simple, when $e_n\geq 3$, we  wish to reduce the push size to $2$. For this purpose, we modify $M_n$ to another 1npda $N_n$ as follows. Assume that $(p,w)\in \delta_n(q,\sigma,a)$ with $w=w_1w_2\cdots w_m$ and $m\leq e_n$. We then define a transition function $\delta'_n$ of $N_n$ by setting
$((p,w_{[2]}),w_1)\in \delta'_n(q,\sigma,a)$,
$\delta'_n((p,w_{[i+1]}),\lambda,w_{i+1}) = \{ ((p,w_{[i+2]}),w_{i+1}w_{i}) \}$ for all $i\in[m-1]$, and
$\delta'_n((p,w_{[m]}),\lambda,w_{m-1}) = \{ (p,w_{m}w_{m-1}) \}$, where $w_{[k]} = w_{k}w_{k+1}\cdots w_{m}$.
This makes $N_n$'s push size be $2$.   We further define $Q'_n= Q_n\cup \{(p,w)\mid p\in Q_n, w\in\Gamma^{\leq e_n}\}$ and $\Gamma'_n = \Gamma_n$. The stack-state complexity $s'(n)$ of $N_n$ is $|Q'_n||{\Gamma'}_n^{\leq 2}| = (|Q_n|+ |Q_n||\Gamma_n^{\leq e_n}|)\cdot |\Gamma_{n}^{\leq 2}| \leq 2|Q_n| |\Gamma_n^{\leq e_n}|\cdot |\Gamma_n^{\leq 2}| \leq (|Q_n||\Gamma_n^{\leq e_n}|)^2 = s(n)^2$ since $2|\Gamma_n^{\leq 2}|\leq |Q_n||\Gamma_n^{\leq e_n}|$.

\begin{figure}[t]
\centering
\includegraphics*[height=4.0cm]{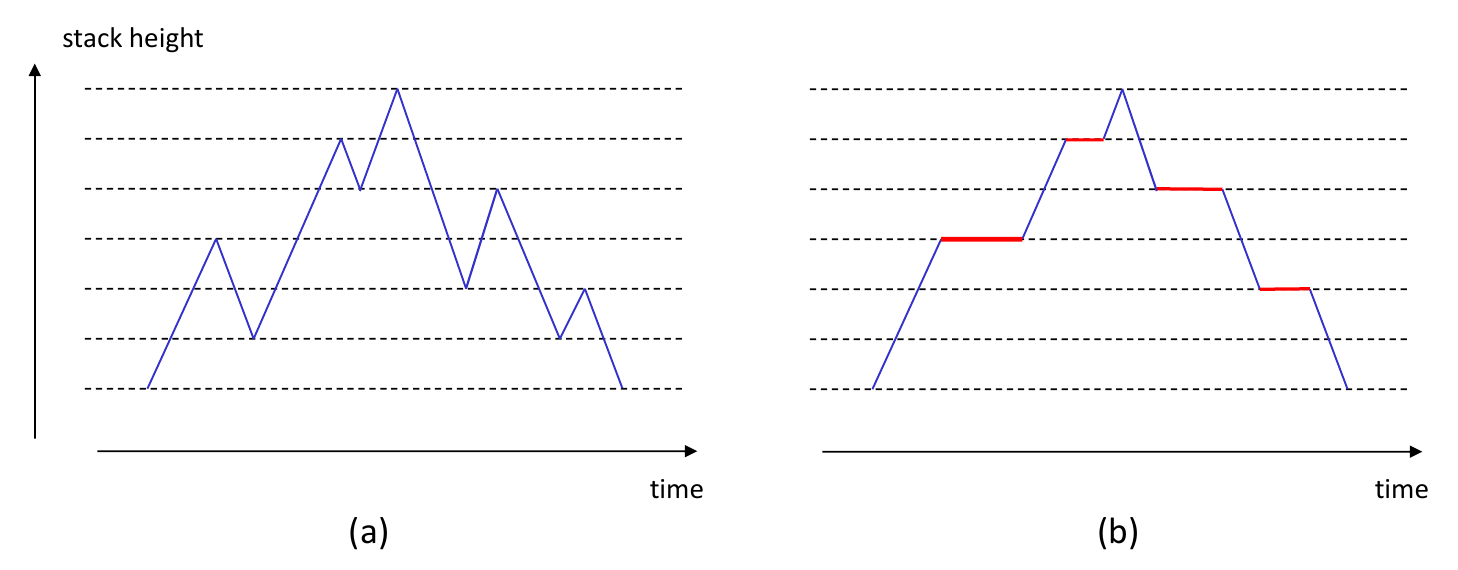}
\caption{How to trim a stack history. (a) An original stack history. (b) A trimmed stack history. Red horizontal lines show the results after trimming valleys beneath.}\label{fig:height-trimming}
\end{figure}

A \emph{state-stack pair} $(q,w)$ is an element in $Q_n\times\Gamma_n^*$, which describes a circumstance that $M_n$ is in inner state $q$ with its stack containing $w$. For convenience, we set $g(n) = s(n)^2 \cdot 2^{s(n)\log{s(n)}+1}$ for any $n\in\nat$.

To prove the lemma, we here assume that there are no iterative pairs or iterative singletons and we set our goal as to derive that the input length must be less than $g(n)$.
We describe each computation path of $M_n$ on input $x$ as a ``history'' of stack height, which is a consecutive series of stack heights according to ``time''. 

Let us consider state-stack pairs that describe $M_n$'s configurations.
The stack history of a computation path of $M_n$ on $x$ can be expressed as  a graph $T_{n,x}$, illustrated as Fig.~\ref{fig:height-trimming}(a).
For this $T_{n,x}$, we assert the following claim concerning the existence of iterative singletons and iterative pairs.

\begin{yclaim}\label{claim-iterative}
(1) If a hill has a slope (without any valleys) of length more than $s'(n)$, then there are two distinct points in the same slope, which are the same state-stack pair.
(2) If a hill has height more than $s'(n)(s'(n)-1)$, then there is a state-stack pair $(q,w)$ located at the same height in both the left slope and the right slope of the hill.
(3) If there is a consecutive series of  more than $s'(n)$ pits of the same height, then this series has two distinct pits having the same state-stack pair.
\end{yclaim}

\begin{proof}
(1) Assume that a hill has a slope of length more than $s'(n)$. In other words, the hill contains more than $s'(n)$ state-stack pairs. Since there are only $s'(n)$ different state-stack pairs, the slope contains at least two different points of the same state-stack pair.

(2) Assume that a hill has height more than $s'(n)(s'(n)-1)$. There are more than $s'(n)$ state-stack pairs in each of the left and the right slopes.
Note that the total number of two distinct state-stack pairs is exactly $s'(n)^2-s'(n)$. Thus, the hill has the same state-stack pair at the same height in both slopes.

(3) Since there are exactly $s'(n)^2-s'(n)$ distinct state-stack pairs, the current statement is obviously true.
\end{proof}

Next, we intend to modify this graph recursively from top to bottom, which we call the ``trimming'' of the graph, in the following way, depending on the following two cases.

\s
Case 1: Consider the case where there is a hill such that its peak is lower than the peak of another hill on its right-hand side. In this case, we draw the horizontal line from this peak toward the first encountered slope located in the right-hand side. We erase the subgraph forming a valley below this horizontal line.

Case 2: Consider the case where there is a hill such that its peak is lower than the peak of its right neighboring hill. We draw a horizontal line from the peak to the nearest slope located in the left-hand side. We erase the subgraph below this horizontal line.

\s
The above trimming process is depicted in Fig.~\ref{fig:height-trimming}. A graph illustrated as Fig.~\ref{fig:height-trimming}(a) is trimmed to another graph illustrated as Fig.~\ref{fig:height-trimming}(b). After this trimming process, we denote by $T'_{n,x}$ the obtained graph,  which is illustrated as Fig.~\ref{fig:stack-history-height}.

\begin{figure}[t]
\centering
\includegraphics*[height=4.0cm]{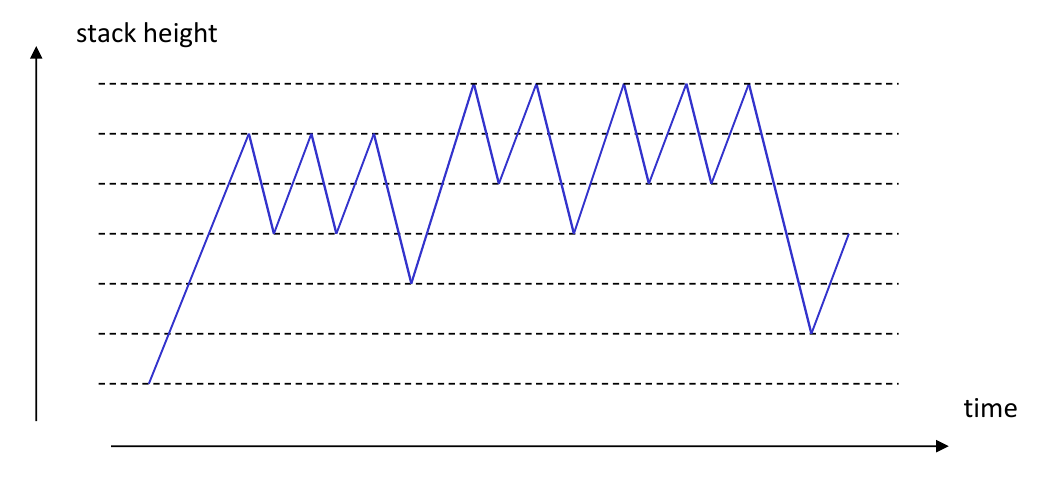}
\caption{A stack history obtained after the trimming process. }\label{fig:stack-history-height}
\end{figure}

Let $t$ denote the number of remaining hills in $T'_{n,x}$. Let $h_{max}$ denote the maximum height of the stack history $T'_{n,x}$. Note that hill's peaks have the same height as long as their bottom have the same height.
In what follows, we intend to analyze $T'_{n,x}$.

We remark the following.

(i) The height of each hill is at most $s'(n)$ because, otherwise, Claim \ref{claim-iterative} implies the existence of an iterative singleton, a contradiction.

(ii) The number of valleys whose pits have the same height is at most $s'(n)$  because, otherwise, Claim \ref{claim-iterative} implies the existence of an iterative singleton.

(iii) We focus on pits of $T'_{n,x}$. We calculate the total number of consecutive pits in $T'_{n,x}$ at height $h$ from $h=0$ to $h=h_{max}$. This number should be at most $s'(n)$ since, otherwise, Claim \ref{claim-iterative} leads to the existence of an iterative singleton.

By (i)--(iii), there are at most $\prod_{i=0}^{h_{max}}s'(n)$ points in the graph $T'_{n,x}$. Note that $\prod_{i=0}^{h_{max}}s'(n)\leq s'(n)^{h_{max}+1} \leq s'(n)^{s'(n)+1} =2^{(s'(n)+1)\log{s'(n)}} =2^{2s(n)^2\log{s(n)}}$. 

This completes the proof of the lemma. 
\end{proofof}

\end{document}